\documentclass[reqno,a4paper]{amsart}
\usepackage{comment,amssymb,latexsym,upref,enumerate,fouridx}
\usepackage{dutchcal}
\usepackage{mathrsfs,xcolor}
\usepackage{graphicx}
\graphicspath{ {./images/} }
\usepackage{subfigure}
\usepackage{placeins}
\usepackage{centernot}
\usepackage[colorlinks,linkcolor=blue,citecolor=blue]{hyperref}

\allowdisplaybreaks
\numberwithin{equation}{section}

\theoremstyle{plain}
\newtheorem{thm}{Theorem}[section]

\newtheorem{prop}[thm]{Proposition}

\theoremstyle{definition}

\theoremstyle{remark}

\newcommand{\bul}{\bullet}
\newcommand{\sgn}{\text{sgn}}
\newcommand{\wttt}{\tilde{\wtt}{}}
\newcommand{\zttt}{\tilde{\ztt}{}}
\newcommand{\ttld}{\tilde{t}{}}

\usepackage[normalem]{ulem}
 \definecolor{violet}{rgb}{0.580,0.,0.827}
 \newcommand{\corr}[3]{
 				      {\color{blue}\ifmmode\text{\,\sout{\ensuremath{{#1}}}\,}\else\sout{{#1}}\fi}
               {\color{red}#2}
               {\color{violet} #3}
 }

\input Tdef.def

\begin{document}

\title[Stability of relativistic perfect fluids with equations of state $p=K\rho$ where $1/3<K<1$]{On the stability of relativistic perfect fluids with linear equations of state $p=K\rho$ where $1/3<K<1$}

\author[E.~Marshall]{Elliot Marshall}
\address{School of Mathematics\\
9 Rainforest Walk\\
Monash University, VIC 3800\\ Australia}
\email{elliot.marshall@monash.edu}

\author[T.A.~Oliynyk]{Todd A.~Oliynyk}
\address{School of Mathematics\\
9 Rainforest Walk\\
Monash University, VIC 3800\\ Australia}
\email{todd.oliynyk@monash.edu}

\begin{abstract} 
For $1/3<K<1$, we consider the stability of two distinct families of spatially homogeneous solutions to the relativistic Euler equations with a 
linear equation of state $p=K\rho$ on exponentially expanding FLRW spacetimes.
The two families are distinguished by one being spatially isotropic while the other is not. 
We establish the future stability of nonlinear perturbations of the non-isotropic family for the full range of parameter values $1/3<K<1$, which improves a previous stability result established by the second author that required $K$ to lie in the restricted range $(1/3,1/2)$.
As a first step towards understanding the behaviour of nonlinear perturbations of the isotropic family, we construct numerical solutions to the relativistic Euler equations under a $\Tbb^2$-symmetry assumption. These solutions are generated from initial data at a fixed time that is chosen to be suitably close to the initial data of an isotropic solution. Our numerical results reveal that, for the full parameter range $1/3<K<1$, the density contrast $\frac{\del{x}\rho}{\rho}$ associated to a nonlinear
perturbation of an isotropic solution develops steep gradients near a finite number of spatial points where it becomes unbounded at future timelike infinity. This behaviour, anticipated by Rendall in \cite{Rendall:2004}, is of particular interest since it is not consistent with the standard picture for inflation in cosmology.
\end{abstract}

\maketitle

\section{Introduction\label{intro}}

Relativistic perfect fluids with a linear equation of state on a prescribed spacetime $(M,\gt)$ are governed by the relativistic Euler equations\footnote{Our indexing conventions are as follows: lower case Latin letters, e.g. $i,j,k$,
will label spacetime coordinate indices that run from $0$ to $3$ while upper case Latin letters, e.g. $I,J,K$, will label spatial coordinate indices that run from
$1$ to $3$.} 
\begin{equation}
\nablat_i \Tt^{ij}=0 \label{relEulA}
\end{equation}
where 
\begin{equation*}
\Tt^{ij} = (\rho+p)\vt^i \vt^j + p \gt^{ij}
\end{equation*}
is the stress energy tensor, $\vt^{i}$ is the fluid
four-velocity normalized by $\gt_{ij}\vt^i \vb^j=-1$, and the fluid's proper energy density $\rho$ and pressure $p$ are related by
\begin{equation*} 
p = K \rho.
\end{equation*} 
Since $K=\frac{dp}{d\rho}$ is the square of the sound speed, we will always assume\footnote{While this restriction on the sound speed is often taken for granted, it is, strictly speaking, not necessary; see \cite{Geroch:2010} for an extended discussion.} 
that $0\leq K \leq 1$ in order to ensure that the speed of sound is less than or equal to the speed of light.
We further restrict our attention to
exponentially expanding Friedmann-Lema\^{i}tre-Robertson-Walker (FLRW) spacetimes $(M,\gt)$ where 
$M = (0,1]\times \Tbb^3$
and\footnote{By introducing a change of time coordinate via $\tilde{t}=-\ln(t)$, the metric \eqref{FLRW} can be brought into the more recognizable form $\gt = -d\tilde{t}\otimes d\tilde{t} + e^{2\tilde{t}}\delta_{ij}dx^I \otimes dx^J$,
where now $\tilde{t} \in [0,\infty)$.
}
\begin{equation} \label{FLRW}
\gt = \frac{1}{t^2} g
\end{equation}
with
\begin{equation} \label{conformal}
g = -dt\otimes dt + \delta_{IJ}dx^I \otimes dx^J.
\end{equation}
 It is important to note that, due to our conventions, the future is located in the direction of \textit{decreasing} $t$ and future timelike infinity is located at $t=0$. 
Consequently, we require that $\vt^0<0$
holds in order to guarantee that the four-velocity is future directed. For use below, we find it convenient to introduce the \textit{conformal four-velocity} via
\begin{equation} \label{c-velocity}
v^i = \frac{1}{t}\vt^i.
\end{equation}

\subsection{Stability for $0\leq K\leq 1/3$} 
It can be verified by a straightforward calculation that
\begin{equation} \label{Hom-A}
(\rho_*,v_*^i) = (t^{3(1+K)}\rho_c,-\delta^i_0), \quad t\in (0,1],
\end{equation} 
defines a spatially homogeneous solution of the relativistic Euler equations \eqref{relEulA} on the exponentially expanding FLRW spacetimes $(M,\gt)$ for any choice of the parameter $0\leq K\leq 1$ and constant $\rho_c\in (0,\infty)$. From a cosmological perspective, these solutions
are, in a sense, the most natural since they are also spatially isotropic and hence do not determine a preferred direction. 

The future, nonlinear stability of the solutions \eqref{Hom-A} on the exponentially expanding FLRW spacetimes was first established in
the articles\footnote{In these articles, stability was established in the more difficult case where the fluid is coupled to Einstein's equations. However, the techniques used there also work in the simpler setting considered in this article where gravitational effects are neglected.}
articles \cite{RodnianskiSpeck:2013,Speck:2012}
for the parameter values $0<K<1/3$. Stability results for the end points $K=1/3$ and $K=0$ were established later\footnote{Again, stability was established in these articles in the more difficult case where the fluid is coupled to Einstein's equations.} in \cite{LubbeKroon:2013} and \cite{HadzicSpeck:2015},
respectively. See also \cite{Friedrich:2017,LiuOliynyk:2018b,LiuOliynyk:2018a,Oliynyk:CMP_2016} for different proofs and perspectives, the articles \cite{LeFlochWei:2021,LiuWei:2021} for related stability results for fluids with nonlinear equations of state on the exponentially expanding FLRW spacetimes, the articles \cite{FOW:2021,Speck:2013,Wei:2018} for analogous stability results on other classes of expanding cosmological spacetimes, and \cite{Ringstrom:2008} for related, early stability results for the Einstein-scalar field system. One of the important aspects of 
these works is  they demonstrate that spacetime expansion can suppress shock formation in fluids, which was first discovered in the Newtonian cosmological setting \cite{BrauerRendallReula:1994}. This is in stark contrast to fluids on Minkowski space where
arbitrary small perturbations of a class of homogeneous solutions to the relativistic Euler equations  form shocks in finite time
 \cite{Christodoulou:2007}.

A consequence of these stability proofs is that the spatial components of the conformal four-velocity $v^i$ of
small, nonlinear perturbations of the homogeneous solution \eqref{Hom-A}
decay to zero at future timelike infinity, that is,
\begin{equation*}
\lim_{t\searrow 0} v^I = 0,
\end{equation*}
for the parameter values $0\leq K < 1/3$.
This behaviour is, of course, consistent with the isotropic homogeneous solutions \eqref{Hom-A}.
On the other hand, when $K=1/3$,  the spatial components of the conformal four-velocity $v^i$ for perturbed solutions do not, in general, decay to zero at timelike infinity, and instead limit to a spatial
vector field $\xi^I$ on $\Tbb^3$, that is,
\begin{equation*} 
\lim_{t\searrow 0} v^I = \xi^I.
\end{equation*}
This behaviour is consistent with a family of \textit{non-isotropic} homogeneous solutions defined by\footnote{More generally, we could set the spatial components of the conformal four-velocity $v_\bul^I$ to be any non-zero vector in $\Rbb^3$ and determine $v_\bul^0$ via
the conditions $g_{ij}v^i_\bul v^j_\bul =-1$ and $v^0_\bul <0$. However, for simplicity, we will assume here that $v_\bul^I$ is chosen so that it is pointing in the direction of the coordinate vector field $\del{1}=\fdel{\;}{x^1}$.}
\begin{equation} \label{Hom-B}
(\rho_\bul,v_\bul^i) = (t^{3(1+K)}\rho_c,-\sqrt{1+\nu_c^2}\delta^i_0 + \nu_c\delta^i_1), \quad t\in (0,1], 
\end{equation} 
where $(\rho_c,\nu_c)\in (0,\infty)\times (0,\infty)$, 
which satisfy the relativistic Euler equations for $K=1/3$. The known stability results for $K=1/3$ imply the future stability of nonlinear perturbations of these solutions. 

Noting that solutions of the type \eqref{Hom-B} can be made arbitrarily close to solutions of the type \eqref{Hom-A} for $K=1/3$ by choosing $\nu_c$ sufficiently small, from a stability point of view there seems to
be verify little difference between the two classes of solutions for small $\nu_c$. Indeed, the future nonlinear stability of both classes of solutions, where $\nu_c$ is sufficiently small, can be achieved via a common proof. However, as will become clear, the essential difference between
these solutions is that, from an initial data point of view, stable perturbations of solutions of the type \eqref{Hom-A} are generated from initial $(\rho,v^I)|_{t=1}$ that is sufficiently close to $(\rho_*,v_*^I)|_{t=1}$ and satisfies 
\begin{equation} \label{Hom-A-idata}
\min_{x\in \Tbb^3} (g_{IJ}v^I v^J)\bigl|_{t=1}=0,
\end{equation}  
while stable  perturbations of solutions of the type \eqref{Hom-B} are generated from initial data $(\rho,v^I)|_{t=1}$ that is sufficiently close to $(\rho_\bul,v_\bul^I)|_{t=1}$ and satisfies 
\begin{equation} \label{Hom-B-idata}
\min_{x\in \Tbb^3} (g_{IJ}v^I v^J)\bigl|_{t=1}>0.
\end{equation}

\subsection{Stability for $1/3<K<1$} 
Until recently, it was not known if any solutions of the relativistic Euler equations were stable to the future for the parameter values $1/3< K< 1$. In fact, it was widely believed that
solutions to the relativistic Euler were not stable for these parameter values. This belief was due, in part, to the influential work of Rendall \cite{Rendall:2004} who used formal expansion to investigate the asymptotic behaviour of relativistic fluids on exponentially expanding FLRW spacetimes with a linear equation of state. Rendall observed that the formal expansions become inconsistent for $K$ in the range $1/3<K<1$ if the leading order term in the expansion of $g_{IJ}v^I v^J$ at $t=0$ vanished somewhere. He speculated that the inconsistent behaviour is the expansions could be due inhomogeneous
features developing in the fluid density that would ultimately result in the blow-up of the density contrast $\frac{\del{I}\rho}{\rho}$ at future timelike infinity. Speck \cite[\S 1.2.3]{Speck:2013} added further support to Rendall's arguments by presenting a heuristic analysis that suggested uninhibited growth would set in for solutions of the relativistic Euler equations for the parameter values $1/3 < K<1$ . Combined, these considerations left the stability of solutions to the relativistic Euler equations in doubt for $K$
in the range $1/3 < K <1$.

However, it was established in \cite{Oliynyk:2021} that there exists a class of non-isotropic homogeneous solutions of the relativistic Euler equations that  are stable to the future under small, nonlinear perturbations.  
This class of homogeneous solutions should be viewed as the natural continuation of the solutions \eqref{Hom-B} over the parameter range $1/3 < K <1$, and they are defined by
\begin{equation}\label{Hom-C}
\bigl(\rho_\bul,v_\bul^i) = \biggl( \frac{\rho_c t^{\frac{2(1+K)}{1-K}}}{(t^{2\mu}+ e^{2u})^{\frac{1+K}{2}}},
-t^{-\mu}\sqrt{e^{2u}+t^{2 \mu} }\delta_0^i+ t^{-\mu }e^{u}\delta_1^i\biggr), \quad t\in (0,1],
\end{equation}
where 
\begin{equation}\label{mu-def}
\mu = \frac{3K-1}{1-K}
\end{equation}
and $u=u(t)$ solves the initial value problem (IVP) 
\begin{align}
 u'\!(t) &=\frac{K\mu t^{2 \mu-1}}{t^{2
   \mu }+(1-K) e^{2 u(t)}},\quad 0<t\leq 1, \label{HomeqB.1} \\
   u(1) &= u_0. \label{HomeqB.2}
\end{align}
Existence of solutions to this IVP is guaranteed by Proposition 3.1.~from \cite{Oliynyk:2021}, which we restate here:
\begin{prop} \label{Homprop}
Suppose $1/3<K<1$, $\mu = (3K-1)/(1-K)$, and $u_0 \in \Rbb$. Then there exists a unique solution
$u \in C^\infty((0,1]) \cap C^0([0,1])$
to the initial value problem \eqref{HomeqB.1}-\eqref{HomeqB.2}
that satisfies
\begin{equation}
|u(t)-u(0)| \lesssim t^{2\mu} \AND |u'\!(t)| \lesssim t^{2\mu-1} \label{Hombounds}
\end{equation}
for all $t\in (0,1]$. Moreover, for each $\rho_c\in (0,\infty)$, the solution $u$ determines a homogeneous solution of the relativistic Euler \eqref{relEulA}
equations via \eqref{c-velocity} and \eqref{Hom-C}.
\end{prop}

The main result of  \cite{Oliynyk:2021} was a proof of the nonlinear stability to the future of the homogeneous solutions \eqref{Hom-C} for the parameter values $1/3 < K <1/2$. It was also established in \cite{Oliynyk:2021} that under a $\Tbb^2$-symmetry assumption, future stability held for the full parameter range
$1/3<K<1$. An important point that is worth emphasising is that the initial data used to generate the perturbed solutions from \cite{Oliynyk:2021} 
satisfies the condition \eqref{Hom-B-idata} at $t=1$, and furthermore, this positivity property propagates to the future in the sense that the perturbed solutions satisfy
\begin{equation*}
\inf_{x\in M}(t^{2\mu}g_{IJ}v^I v^J)>0.
\end{equation*}
It is this property of the perturbed solutions from \cite{Oliynyk:2021}
that avoids the problematic scenario identified by Rendall.  

This article has two main aims: the
first is to establish the nonlinear stability to the future of the homogeneous solutions \eqref{Hom-C} for the full parameter range $1/3 < K<1$ without the $\Tbb^2$-symmetry that was required in \cite{Oliynyk:2021}.  The second aim is to provide convincing numerical evidence that shows the density contrast blow-up scenario of Rendall is realized
if the condition \eqref{Hom-B-idata} on the initial data is violated.  

Before stating a precise version of our stability result for the homogeneous solutions \eqref{Hom-C}, we first recall two formulations
of the relativistic Euler equations from \cite{Oliynyk:2021}.   The first formulation, which was introduced in  \cite{Oliynyk:CMP_2015} and subsequently employed in \cite{Oliynyk:CMP_2016} to establish stability for the parameter range $0<K\leq 1/3$, involves representing the fluid in terms of the modified 
fluid density $\zeta$ defined via
\begin{equation}\label{mod-den}
\rho = t^{3(1+K)}\rho_c e^{(1+K)\zeta}
\end{equation}
and the spatial components $v_I$ of the conformal fluid four-covelocity\footnote{Here and in the following, all spacetime indices will be raised and lowered with the conformal metric $g_{ij}$. }  $v_i=g_{ij}v^j$. In terms of these variables, the relativistic Euler equations \eqref{relEulA} can be formulated as the following symmetric hyperbolic system:
\begin{equation} \label{relEulB}
B^k \del{k}V = \frac{1}{t}\Bc \pi V
\end{equation}
where
\begin{align}
V &= (\zeta, v_J )^{\tr} ,  \label{Vdef}\\
v_0 & = \sqrt{|v|^2 +1} , \qquad |v|^2 = \delta^{IJ}v_I v_J,  \label{v0def}\\
v^i & = \delta^{iJ}v_J - \delta^{i}_0 v_0,  \label{viupdef}\\
\Bc &= \frac{-1}{v^0}\begin{pmatrix} 1 & 0 \\ 0 & \frac{1-3K}{v_0}\delta^{JI} \end{pmatrix}, \label{Bcdef}\\
\pi &= \begin{pmatrix} 0 & 0 \\ 0 & \delta_{I}^J \end{pmatrix}, \label{pi-def} \\
L^k_I &= \delta^k_J - \frac{v_J}{v_0} \delta^k_0, \label{Ldef} \\
M_{IJ} &= \delta_{IJ} - \frac{1}{(v_0)^2}v_I v_J,  \label{Mdef}\\
B^0 &= \begin{pmatrix} K  &  \frac{K}{v^0}  L^0_M \delta^{MJ} \\ \frac{K}{v^0} \delta^{LI} L^0_L & \delta^{LI} M_{LM} \delta^{MJ} \end{pmatrix} \label{B0def}
\intertext{and}
B^K &= \frac{1}{v^0}\begin{pmatrix} Kv^K  &  K  L^K_M \delta^{MJ} \\ K \delta^{LI} L^K_L & \delta^{LI} M_{LM} \delta^{MJ} v^K \end{pmatrix}.
\label{BKdef}
\end{align}

The second formulation of the relativistic Euler equations is obtained by introducing a new density variable $\zetat$ via
\begin{equation} \label{zetatdef}
\zetat = \zeta + \ln(v_0)
\end{equation}
and decomposing the spatial components of the conformal fluid four-velocity as
\begin{align} 
v_1 &= \frac{t^{-\mu } e^{u(t)+w_1}}{\sqrt{t^{2 \mu
   }
   \left((w_2-w_3)^2+(w_2+w_3)^2\right)+1}}, \label{cov2a} \\
v_2 &= \frac{(w_2+w_3) e^{u(t)+w_1}}{\sqrt{t^{2
   \mu }
   \left((w_2-w_3)^2+(w_2+w_3)^2\right)+1}} \label{cov2b}
 \intertext{and}
v_3 &= \frac{(w_2-w_3) e^{u(t)+w_1}}{\sqrt{t^{2
   \mu }
   \left((w_2-w_3)^2+(w_2+w_3)^2\right)+1}},   \label{cov2c}
\end{align}
where $u(t)$ solves the IVP \eqref{HomeqB.1}-\eqref{HomeqB.2}. Then 
setting
\begin{align}
\wbr_1 &= u+w_1, \label{wbr1def} \\
\psi = &t^{2 \mu }+e^{2 \wbr_1}, \label{psidef}\\
\chi = &t^{2\mu }-(K-1) e^{2 \wbr_1}, \label{chidef}\\
\phi &= 2 t^{2\mu } \left(w_2^2+w_3^2\right)+1, \label{phidef} \\
\eta_\Lambda &= \left(2 w_\Lambda t^{2 \mu } 
   (w_2-w_3)+(-1)^\Lambda 1\right),  \quad \Lambda =2,3, \label{etadef}
\intertext{and}
\xi_\Lambda &= \left(2 w_\Lambda t^{2 \mu }
   (w_2+w_3)+1\right), \quad \Lambda =2,3, \label{xidef}
\end{align}
it was shown in \cite[\S 3.2]{Oliynyk:2021} that
in terms of the variables
\begin{equation} \label{Wdef}
W = (\zetat, w_1, w_2, w_3 )^{\tr}
\end{equation}
the relativistic Euler equations become 
\begin{equation} \label{relEulC}
\del{t}W + \Ac^I \del{I}W =-\frac{\mu}{t}\Pi W + t^{\mu-1}\Gc
\end{equation}
where
\begin{equation} \label{Ac1rep}
\Ac^1 = \frac{1}{\sqrt{\frac{t^{2\mu}}{e^{2\wt_1}}+1}}\begin{pmatrix}
 -\frac{1}{\sqrt{\phi}} &
   -\frac{t^{2 \mu }}{\psi \sqrt{\phi}} &
   \frac{2 t^{2\mu } w_2}{\phi^{3/2}} &
   \frac{t^{2\mu} w_3}{\phi^{3/2}}
   \\
 -\frac{K t^{2 \mu } e^{-2
   \wbr_1} \psi}{\sqrt{\phi}\chi} &
   \frac{(2 K-1)
   t^{2 \mu }+(K-1) e^{2
   \wbr_1}}{\sqrt{\phi} \chi} & -\frac{2 K
   t^{2\mu } \psi w_2}{\phi^{3/2}\chi} & -\frac{2 K t^{2\mu }
   \psi w_3}{\phi^{3/2}\chi} \\
 K t^{2\mu }w_2 e^{-2 \wbr_1}
   \sqrt{\phi} & -\frac{K
   t^{2\mu } w_2 \sqrt{\phi}}{\psi} &
   -\frac{1}{\sqrt{\phi}} & 0
   \\
  K t^{2\mu }w_3 e^{-2 \wbr_1} \sqrt{\phi} &
   -\frac{K t^{2\mu } w3 \sqrt{\phi}}{ \psi} & 0 &
  - \frac{1}{\sqrt{\phi}} \\
\end{pmatrix},
\end{equation}
\begin{equation}  \label{Ac2rep}
\Ac^2 = \frac{1}{\sqrt{\frac{t^{2\mu}}{e^{2\wt_1}}+1}}\begin{pmatrix}
 -\frac{t^{\mu } (w_3+w_2)}{\sqrt{\phi}} &
   -\frac{t^{3 \mu } (w_3+w_2)}{\psi \sqrt{\phi}} &
   \frac{t^{\mu } \eta_3}{\phi^{3/2}} &
   -\frac{t^\mu \eta_2}{\phi^{3/2}}
   \\
 -\frac{K t^{3 \mu } (w_2+w_3) e^{-2
   \wbr_1} \psi}{\sqrt{\phi}\chi} &
   \frac{t^{\mu } (w_2+w_3) \left((2 K-1)
   t^{2 \mu }+(K-1) e^{2
   \wbr_1}\right)}{\sqrt{\phi} \chi} & -\frac{K
   t^{\mu } \psi \eta_3}{\phi^{3/2}\chi} & \frac{K t^{\mu }
   \psi\eta_2}{\phi^{3/2}\chi} \\
 -\frac{1}{2} K t^{\mu } e^{-2 \wbr_1}
   \sqrt{\phi} & \frac{K
   t^{\mu } \sqrt{\phi}}{2 \psi} &
   -\frac{t^{\mu } (w_3+w_2)}{\sqrt{\phi}} & 0
   \\
 -\frac{1}{2} K t^{\mu } e^{-2 \wbr_1} \sqrt{\phi} &
   \frac{K t^{\mu } \sqrt{\phi}}{2 \psi} & 0 &
   -\frac{t^{\mu } (w_3+w_2)}{\sqrt{\phi}} \\
\end{pmatrix},
\end{equation}
\begin{equation}  \label{Ac3rep}
\Ac^3 = \frac{1}{\sqrt{\frac{t^{2\mu}}{e^{2\wt_1}}+1}}\begin{pmatrix}
 \frac{t^{\mu } (w_3-w_2)}{\sqrt{\phi}} &
   \frac{t^{3 \mu } (w_3-w_2)}{\psi \sqrt{\phi}} &
   -\frac{t^{\mu } \xi_3}{\phi^{3/2}} &
   \frac{t^\mu \xi_2}{\phi^{3/2}}
   \\
 -\frac{K t^{3 \mu } (w_2-w_3) e^{-2
   \wbr_1} \psi}{\sqrt{\phi}\chi} &
   \frac{t^{\mu } (w_2-w_3) \left((2 K-1)
   t^{2 \mu }+(K-1) e^{2
   \wbr_1}\right)}{\sqrt{\phi} \chi} & \frac{K
   t^{\mu } \psi \xi_3}{\phi^{3/2}\chi} & -\frac{K t^{\mu }
   \psi\xi_2}{\phi^{3/2}\chi} \\
 -\frac{1}{2} K t^{\mu } e^{-2 \wbr_1}
   \sqrt{\phi} & \frac{K
   t^{\mu } \sqrt{\phi}}{2 \psi} &
   \frac{t^{\mu } (w_3-w_2)}{\sqrt{\phi}} & 0
   \\
 \frac{1}{2} K t^{\mu } e^{-2 \wbr_1} \sqrt{\phi} &
   -\frac{K t^{\mu } \sqrt{\phi}}{2 \psi} & 0 &
   \frac{t^{\mu } (w_3-w_2)}{\sqrt{\phi}} \\
\end{pmatrix},
\end{equation}
\begin{equation} \label{Gcdef}
\Gc =\begin{pmatrix}
 0 \\
 -\frac{K (3 K-1) \left(e^{2 w_1}-1\right) e^{2
   u}}{\left((K-1) e^{2 u}-t^{2 \mu }\right)
   \left((K-1) e^{2 \wbr_1}-t^{2 \mu
   }\right)} \\
 0 \\
 0 
\end{pmatrix},
\end{equation}
and
\begin{equation} \label{Pidef}
\Pi = \diag(0,0,1,1).
\end{equation}
For later use, we also define
\begin{equation} \label{Piperpdef}
\Pi^\perp = \id - \Pi,
\end{equation}
and observe that $\Pi$ and $\Pi^\perp$ satisfy the relations
\begin{equation} 
\Pi^2 = \Pi, \quad (\Pi^\perp)^2 = \Pi^\perp, \quad \Pi\Pi^\perp =\Pi^\perp \Pi = 0 \AND \Pi+\Pi^\perp = \id. \label{Pirel} 
\end{equation}
An important point regarding the formulation \eqref{relEulC} is that it is symmetrizable. Indeed, as shown in \cite{Oliynyk:2021}, multiplying \eqref{relEulC} by
the positive definite, symmetric matrix
\begin{equation}  \label{A0rep}
A^0 = \begin{pmatrix}
 K & 0 & 0 & 0 \\
 0 & \frac{t^{2 \mu } e^{2 \wbr_1}-(K-1) e^{4
   \wbr_1}}{\psi^2} & 0 & 0 \\
 0 & 0 & \frac{2 e^{2 \wbr_1} \left(2 w_3^2
   t^{2 \mu }+1\right)}{\phi^2} &
   -\frac{4 w_2 w_3 t^{2 \mu } e^{2
   \wbr_1}}{\phi^2} \\
 0 & 0 & -\frac{4 w_2 w_3 t^{2 \mu } e^{2
   \wbr_1}}{\phi^2} &
   \frac{2 e^{2 \wbr_1} \left(2 w_2^2 t^{2
   \mu }+1\right)}{\phi^2} \\
\end{pmatrix}
\end{equation}
yields
\begin{equation} \label{relEulD}
A^0\del{t}W + A^I \del{I}W =-\frac{\mu}{t} A^0\Pi W + t^{\mu-1}A^0\Gc
\end{equation}
where it is straightforward to verify from \eqref{Ac1rep}-\eqref{Ac3rep} that the matrices 
\begin{equation}\label{AIdef}
A^I = A^0 \Ac^I
\end{equation}
are symmetric, that is,
\begin{equation}\label{AI-sym}
(A^I)^{\tr}=A^I.
\end{equation}

We are now in a position to state the main stability theorem of this article. The proof is presented in Section \ref{sec:proof}. Before stating the theorem, it is important to note that, due to change of variables defined via \eqref{zetatdef}-\eqref{cov2c} and \eqref{Wdef}, the homogeneous solutions \eqref{Hom-C} correspond to the trivial solution $W=0$ of \eqref{relEulC}.

\begin{thm} \label{mainthm}
Suppose $k\in\Zbb_{>3/2+1}$, $1/3<K < 1$, $\mu = (3K-1)/(1-K)$,  $\sigma > 0$, $u_0\in \Rbb$, $u \in C^\infty((0,1])\cap C^0([0,1])$ is the unique solution to the IVP \eqref{HomeqB.1}-\eqref{HomeqB.2} from Proposition \ref{Homprop} and
$\zetat_0, w^0_J \in H^{k+1}(\Tbb^3)$.
Then for $\delta>0$ small enough, there exists a unique solution
\begin{equation*}
W=(\zetat,w_J)^{\tr} \in C^0\bigl((0,1], H^{k+1}(\Tbb^3,\Rbb^4)\bigr)\cap C^1\bigl((0,1],H^{k}(\Tbb^3,\Rbb^4)\bigr)
\end{equation*}
to the initial value problem 
\begin{align}
\del{t}W + \Ac^I\del{I}W &= -\frac{\mu}{t}\Pi W + t^{\mu-1}\Gc \hspace{0.5cm}  \text{in $(0,1]\times \Tbb^3$,} \label{relEulE.1}\\
W  &= (\zetat_0, w^0_J)^{\tr} \hspace{1.65cm} \text{in $\{1\}\times \Tbb^3$,} \label{relEulE.2}
\end{align}
provided that 
\begin{equation*}
\biggl(\norm{\zetat_0}_{H^{k+1}}^2+\sum_{J=1}^3\norm{w^0_J}_{H^{k+1}}^2\biggr)^{\frac{1}{2}}\leq \delta.
\end{equation*}
Moreover, 
\begin{enumerate}[(i)]
\item $W=(\zetat,w_J)^{\tr}$ satisfies the energy estimate
\begin{equation*}
\Ec(t) + \int_t^1 \tau^{2\mu-1}\bigl(\norm{D\zetat(\tau)}_{H^k}^2+\norm{Dw_1(\tau)}_{H^k}^2\bigr)\,d\tau \lesssim \norm{\zetat_0}_{H^{k+1}}^2+\sum_{J=1}^3\norm{w^0_J}_{H^{k+1}}^2
\end{equation*}
for all $t\in (0,1]$
where\footnote{The norm $\norm{Df}_{H^k}$ is defined by  $\norm{Df}^2_{H^k}= \sum_{J=1}^3 \norm{\del{J}f}^2_{H^k}$.}
\begin{equation*}
\Ec(t)=\norm{\zetat(t)}_{H^k}^2+\norm{w_1(t)}_{H^k}^2+t^{2\mu}\Bigl(\norm{D\zetat(t)}_{H^k}^2+\norm{Dw_1(t)}_{H^k}^2+\norm{w_2(t)}_{H^{k+1}}^2+\norm{w_3(t)}_{H^{k+1}}^2\Bigr),
\end{equation*}
\item there exists functions $\zetat_*, w_1^* \in H^{k-1}(\Tbb^3)$ and $\wb_2^*,\wb_3^* \in H^{k}(\Tbb^3)$ such that the estimate
\begin{align*}
\bar{\Ec}(t) \lesssim
 t^{\mu-\sigma}
\end{align*}
holds for all $t\in (0,1]$ where
\begin{equation*}
\bar{\Ec}(t)=\norm{\zetat(t) - \zetat_*}_{H^{k-1}}+\norm{w_1(t) - w_1^*}_{H^{k-1}}
+\norm{t^\mu w_2(t) - \wb_2^*}_{H^{k}}+\norm{t^\mu w_3(t) - \wb_3^*}_{H^{k}},
\end{equation*}
\item $u$ and $W=(\zetat,w_J)^{\tr}$ determine a unique solution of the relativistic Euler equations \eqref{relEulA} on the
spacetime region $M=(0,1]\times \Tbb^3$ via the formulas 
\begin{align}
\rho &= \frac{\rho_c t^{\frac{2(1+K)}{1-K}} e^{(1+K)\zetat}}{(t^{2\mu}+ e^{2(u+w_1)})^{\frac{1+K}{2}}}, \label{relEulsol.1}\\
\vt^0 &= -t^{1-\mu}\sqrt{e^{2 (u+w_1)}+t^{2 \mu} },\label{relEulsol.2}\\
\vt^1 &=t^{1-\mu }\biggl(  \frac{e^{u+w_1}}{\sqrt{ (t^{\mu}w_2-t^{\mu}w_3)^2+(t^{\mu}w_2+t^{\mu}w_3)^2+1}} \biggr), \label{relEulsol.3} \\
\vt^2 &= t^{1-\mu }\biggl( \frac{(t^{\mu}w_2+t^{\mu}w_3) e^{u+w_1}}{\sqrt{ (t^{\mu}w_2-t^{\mu}w_3)^2+(t^{\mu}w_2+t^{\mu}w_3)^2+1}}\biggr),\label{relEulsol.4}\\
\vt^3 &= t^{1-\mu }\biggl( \frac{(t^{\mu}w_2-t^{\mu}w_3) e^{u+w_1}}{\sqrt{ (t^{\mu}w_2-t^{\mu}w_3)^2+(t^{\mu}w_2+t^{\mu}w_3)^2+1}}\biggr),  \label{relEulsol.5}
\end{align}
\item and the density contrast $\frac{\del{I}\rho}{\rho}$ satisfies
\begin{equation} \label{den-constrast-A}
\lim_{t\searrow 0} \Bigl\| \frac{\del{I}\rho}{\rho} - (1+K)\del{I}(\zetat_*-w_1^*) \Bigr\|_{H^{k-2}} = 0.
\end{equation}
\end{enumerate}
\end{thm}

\subsection{Instability for $1/3 < K< 1$}
It is essential for the stability result stated in Theorem \ref{mainthm} to hold that the initial data used to generated the nonlinear perturbations of homogeneous solutions of the type \eqref{Hom-C} satisfies the condition \eqref{Hom-B-idata}. This leaves the question of what happens when this condition is violated, which would be guaranteed to happen for some choice of initial data from any given open set
of initial data that contains initial data corresponding to an isotropic homogeneous solution \eqref{Hom-A}. To investigate this situation, we consider a $\Tbb^2$-symmetric reduction of the system \eqref{relEulB} obtained by the ansatz
\begin{align}
\tilde \zeta(t,x^1,x^2,x^3)&=\ztt(t,x^1), \label{zttt-def} \\
v_{I}(t,x^1,x^2,x^3) &=  t^{-\mu}\wtt(t,x^1) \delta_I^1, \label{wttt-def}
\end{align}
where $\zetat$ is as defined above by \eqref{zetatdef}. It is not difficult to verify via a straightforward calculation that the relativistic Euler equations \eqref{relEulB} will be satisfied provided that 
$\ztt$ and $\wtt$ solve\footnote{Here, we set $x=x^1$.} 
\begin{align}
\del{t}\ztt-\frac{\wtt}{(t^{2 \mu }+\wtt^2)^{\frac{1}{2}}}  \partial_{x}\ztt - \frac{t^{2\mu}}{(t^{2 \mu }+\wtt^2)^{\frac{3}{2}}}\partial_{x}\wtt &=0, \label{eqn:dotzeta} \\
\label{eqn:dotw}
\del{t}\wtt-\frac{Kt^{2 \mu } (t^{2 \mu }+\wtt^2)^{\frac{1}{2}}}{(t^{2 \mu }-(K-1)\wtt^2)}\partial_{x}\ztt 
+\frac{\bigl((2 K-1) t^{2 \mu }+(K-1) \wtt^2\bigr)\wtt}{(t^{2 \mu }+\wtt^2)^{\frac{1}{2}} (t^{2 \mu }-(K-1)\wtt^2)}\partial_{x}\wtt
&=\frac{t^{2 \mu -1}(-3 K+\mu +1) \wtt}{t^{2 \mu }-(K-1)\wtt^2} .
\end{align}

In Section \ref{numsol}, we numerically solve  this system for specific choices 
of initial data
\begin{equation*}
(\ztt,\wtt)|_{t={t_0}} = (\ztt_0,\wtt_0) \quad \text{in $\Tbb^1$.}
\end{equation*}
Importantly, these choices include initial data for which $\wtt_0$ crosses zero at two points in $\Tbb^1$, and as a consequence, violates the condition \eqref{Hom-B-idata}. From our numerical solutions, we observe the following behaviour: 
\begin{enumerate}[(1)]
\item For all $K\in (1/3,1)$ and all choices of initial data $(\ztt_0,\wtt_0)$ that are sufficiently close to homogeneous initial data of either family of solutions \eqref{Hom-A} and \eqref{Hom-C}, $\ztt$ and $\wtt$ remain bounded and converge pointwise as $t\searrow 0$.
\smallskip
\item For each $K\in (1/3,1)$ and each choice of initial data $(\ztt_0,\wtt_0)$ that violates \eqref{Hom-B-idata} and is sufficiently close to homogeneous initial data of the family of solutions \eqref{Hom-A}, there exists a $\ell=\ell(K)\in \Zbb_{\geq 0}$ such that \begin{equation*}
    \sup_{x\in \Tbb^1}\bigl(|\del{x}^{\ell} \ztt(t,x)|+|\del{x}^{\ell}\wtt(t,x)|\bigr) \nearrow \infty \quad \text{as $t\searrow 0$.}
\end{equation*}
This indicates an instability in the $H^\ell$-spaces for solutions of \eqref{eqn:dotzeta}-\eqref{eqn:dotw} that is not present, c.f.~Theorem \ref{mainthm}, in solutions generated from initial data satisfying \eqref{Hom-B-idata}.
We also observe that the integer $\ell$ is a monotonically decreasing function of $K$ with a minimum value of $1$. For the initial data we tested, the blow-up at $t=0$ in the derivatives occurs at a finite set of spatial points. 
\smallskip
\item For all $K\in (1/3,1)$ and all choices of initial data $(\ztt_0,\wtt_0)$ that are sufficiently close to homogeneous initial data of either family of solutions \eqref{Hom-A} and \eqref{Hom-C}, solutions to \eqref{eqn:dotzeta}-\eqref{eqn:dotw} are approximated remarkably well, for times sufficiently close to zero, by solutions to the \textit{asymptotic system}\footnote{Note this system is obtain from
 \eqref{eqn:dotzeta}-\eqref{eqn:dotw} simply by discarding the terms involving spatial derivatives.}
\begin{align}
\del{t}\zttt &=0, \label{zttt-asympt} \\
\del{t}\wttt &=\frac{t^{2 \mu -1}(-3 K+\mu +1) \wttt}{t^{2 \mu }-(K-1)\wttt^2}, \label{wttt-asympt} 
\end{align}
everywhere except, possibly, at a finite set of points where steep gradients form in $z$, which only happens for $K$ large enough and initial data violating \eqref{Hom-B-idata}.
\smallskip
\item For each $K\in (1/3,1)$ and each choice of initial data $(\ztt_0,\wtt_0)$ that violates \eqref{Hom-B-idata} and is sufficiently close to homogeneous initial data of the family of solutions \eqref{Hom-A}, the density contrast $\frac{\del{x}\rho}{\rho}$ develops steep gradients near a finite number of spatial points where it becomes unbounded as $t\searrow 0$.
This behaviour was anticipated by Rendall in \cite{Rendall:2004}, and it is \textit{not consistent} with either the standard picture for inflation in cosmology where the density contrast remains bounded as $t\searrow 0$, or with the behaviour of the density contrast of solutions generated from initial data satisfying \eqref{Hom-B-idata}, c.f.~Theorem \ref{mainthm}.
\end{enumerate}

\subsection{Stability/instability for $K=1$}
When the sound speed is equal to the speed of light, i.e. $K=1$, it is well known that the irrotational relativistic Euler equations coincide, under a change of variables, with the linear wave equation. Even though the future global existence of solutions to linear wave equations on exponentially expanding FLRW spacetimes can be inferred from standard existence results for linear wave equations, a corresponding future global existence result for the irrotational relativistic Euler equations does not automatically follow. This is because the change of variables needed to interpret a wave solution as a solution of the relativistic Euler equations requires the gradient of the wave solution to be timelike. Thus an instability in the irrotational relativistic Euler equations can still occur for $K=1$ if the gradient of the wave solution starts out timelike but becomes spacelike somewhere in finite time. This phenomena was shown in \cite{Fournodavlos:2022} to occur in the more difficult case where coupling to Einstein's equations with a positive cosmological constant was taken into account. In fact, it was shown in \cite{Fournodavlos:2022} that all wave solutions generated from initial data sets that correspond to a sufficiently small perturbation of the FLRW fluid solution (i.e.~\eqref{Hom-A} in our setting) become spacelike in finite time. This proves that the self-gravitating version of the isotropic homogeneous \eqref{Hom-A} are unstable, and in the irrotational setting at least, characterizes the cause of the instability. What is not known is if the other family of homogeneous solutions \eqref{Hom-C} or their self-gravitating versions remain stable for $K=1$.     

\subsection{Future directions}
The most natural and physically relevant generalization of the the stability result stated in Theorem \ref{mainthm} would be an analogous stability result for the coupled Einstein-Euler equations with a positive cosmological constant for
$K$ satisfying $1/3<K<1$. We expect that establishing this type of stability result is feasible by adapting the arguments from \cite{Oliynyk:CMP_2016}. This expectation is due to the behaviour of the term  $t^{-2}\rho v_i v_j$, which is the only potentially problematic term that could,
if it grew too quickly as $t\searrow 0$, prevent the use of the arguments from \cite{Oliynyk:CMP_2016}.
However, by Theorem \ref{mainthm}, we know that $\rho=\Ord\bigl(t^{\frac{2(1+K)}{1-K}}\bigr)$ and $v_i = \Ord\bigl(t^{\frac{1-3 K}{1-K}}\bigr)$ from which it follows that $t^{-2}\rho v_i v_j = \Ord(t^2)$. This shows that $t^{-2}\rho v_i v_j$ decays quickly enough as $t\searrow 0$ to expect that it should not be problematic. We are currently working on generalizing Theorem \ref{mainthm} to include coupling to Einstein's equations with a positive cosmological constant, and  we will report on  any progress in this direction in a follow-up article. We are also planning to investigate numerically, under a Gowdy symmetry assumption, if a similar behaviour, as described in Section \ref{numsol}, occurs for initial data that violates \eqref{Hom-B-idata} when coupling to Einstein's equations with a positive cosmological constant is taken into account. 

\section{Proof of Theorem \ref{mainthm}\label{sec:proof}}
\subsection{Step 1: Fuchsian formulation}
Applying the projection operator $\Pi$ to \eqref{relEulC}, while noting that $\Pi \Gc = 0$ by  \eqref{Gcdef}-\eqref{Pidef}, yields 
\begin{equation*}
\del{t}(\Pi W) + \Pi\Ac^I \del{I}W =-\frac{\mu}{t}\Pi W.
\end{equation*}
Multiplying this equation through by $t^{\mu}$ gives 
\begin{equation}\label{relEulF}
\del{t}(t^{\mu}\Pi W) + t^{\mu}\Pi\Ac^I \del{I}W = 0.
\end{equation}
Applying $\Pi^\perp$ to \eqref{relEulC}, we further observe, with the help of \eqref{Pirel}, that
\begin{equation} \label{relEulG} 
\del{t}(\Pi^\perp W) + \Pi^\perp\Ac^I \del{I}W = t^{2\mu-1}\Pi^\perp\Gc.
\end{equation}

Next, we decompose the term $\Pi^\perp\Ac^I \del{I}W$ in \eqref{relEulG} as follows
\begin{equation*}
\Pi^\perp\Ac^I \del{I}W=\Pi^\perp\Ac^I  \Pi^\perp \del{I}( \Pi^\perp W) + t^{-\mu}\Pi^\perp\Ac^I  \Pi t^{\mu}\del{I} W.
\end{equation*}
Inserting  this into \eqref{relEulG} and multiplying the resulting equation on the left by $\Pi^\perp A^0 \Pi^\perp$ gives  
\begin{equation} \label{relEulH} 
\Pi^\perp A^0 \Pi^\perp\del{t}(\Pi^\perp W) +\Pi^\perp A^0 \Pi^\perp\Ac^I  \Pi^\perp \del{I}( \Pi^\perp W) + t^{-\mu}\Pi^\perp A^0 \Pi^\perp\Ac^I  \Pi t^{\mu}\del{I} W = t^{2\mu-1}\Pi^\perp A^0\Pi^\perp\Gc.
\end{equation}
It is worth noting at this point that it is the use of this equation to control $\Pi^\perp W$ instead of \eqref{relEulG} that is responsible for the improvement of the range of the parameter values for which stability holds from $1/3<K < 1/2$ in \cite{Oliynyk:2021} to $1/3<K<1$ in this article.

Now, multiplying \eqref{relEulH} by
\begin{equation}  \label{Sdef}
S = \begin{pmatrix}
 \frac{e^{-2\wbr_1}\psi^2}{\chi} & 0 & 0 & 0 \\
 0 & \frac{\psi^2}{t^{2 \mu } e^{2 \wbr_1}-(K-1) e^{4
   \wbr_1}} & 0 & 0 \\
 0 & 0 & 0 & 0 \\
 0 & 0 & 0 & 0
\end{pmatrix}
\end{equation}
and adding the resulting equation to \eqref{relEulF} yields
\begin{align*}
\del{t}(t^{\mu}\Pi W)+S\Pi^\perp A^0 \Pi^\perp\del{t}(\Pi^\perp W) &+S\Pi^\perp A^0 \Pi^\perp\Ac^I  \Pi^\perp \del{I}( \Pi^\perp W) =-\Pi\Ac^I  t^{\mu}\del{I}W\\
&- t^{-\mu}S\Pi^\perp A^0 \Pi^\perp\Ac^I  \Pi t^{\mu}\del{I} W + t^{2\mu-1}S\Pi^\perp A^0\Pi^\perp\Gc.
\end{align*}
Setting 
\begin{equation}\label{Wbdef}
\Wb := \Pi^\perp W+t^{\mu}\Pi W = (\zetat,w_1,t^{\mu}w_2,t^{\mu}w_3)^{\tr},
\end{equation}
it is then not difficult to verify that the above equation can be expressed as 
\begin{equation}\label{relEulHa}
B^0\del{t}\Wb + B^I \del{I}\Wb = -\Pi\Ac^I  t^{\mu}\del{I}W
- t^{-\mu}S\Pi^\perp A^0 \Pi^\perp\Ac^I  \Pi t^{\mu}\del{I} W + t^{2\mu-1}S\Pi^\perp A^0\Pi^\perp\Gc
\end{equation}
where
\begin{equation}\label{B-def}
B^0 =S\Pi^\perp A^0 \Pi^\perp + \Pi \AND B^I =S\Pi^\perp A^0 \Pi^\perp \Ac^I \Pi^\perp.
\end{equation}
Noting from \eqref{Ac1rep}-\eqref{Ac3rep} that
\begin{equation*}
\Pi^\perp \Ac^I \Pi^\perp =  \frac{b^I}{\sqrt{\frac{t^{2\mu}}{e^{2\wt_1}}+1}}\begin{pmatrix}
 -\frac{1}{\sqrt{\phi}} &
   -\frac{t^{2 \mu }}{\psi \sqrt{\phi}} &0 &0
   \\
 -\frac{K t^{2 \mu } e^{-2
   \wbr_1} \psi}{\sqrt{\phi}\chi} &
   \frac{(2 K-1)
   t^{2 \mu }+(K-1) e^{2
   \wbr_1}}{\sqrt{\phi} \chi} & 0 & 0\\
0 & 0 & 0& 0\\
0 & 0& 0& 0
\end{pmatrix}
\end{equation*}
where
\begin{equation} \label{bI-def}
b^1 = 1, \quad b^2 = t^{\mu}(w_3+w_2) \AND b^3 = t^\mu(w_2-w_3),
\end{equation}
a short calculation using  \eqref{Pidef}-\eqref{Piperpdef}, \eqref{A0rep}, and \eqref{Sdef} shows that
the matrices \eqref{B-def} are given by
\begin{equation}\label{B0-form}
B^0 =  \begin{pmatrix}  \frac{ K e^{-2\wbr_1}\psi^2}{\chi} & 0 & 0 & 0 \\
 0 & 1 & 0 & \\
0 & 0 & 1 & 0 \\
0 & 0 & 0 & 1\end{pmatrix}
\end{equation}
and
\begin{equation} \label{BI-form}
B^I =  \frac{b^I}{\sqrt{\frac{t^{2\mu}}{e^{2\wt_1}}+1}}\begin{pmatrix}
 -\frac{K e^{-2\wbr_1}\psi^2}{\chi\sqrt{\phi}} &
   -\frac{K t^{2 \mu }e^{-2\wbr_1}\psi}{\sqrt{\phi}\chi} &0 &0
   \\
 -\frac{K t^{2 \mu } e^{-2
   \wbr_1} \psi}{\sqrt{\phi}\chi} &
   \frac{(2 K-1)
   t^{2 \mu }+(K-1) e^{2
   \wbr_1}}{\sqrt{\phi} \chi} & 0 & 0\\
0 & 0 & 0& 0\\
0 & 0& 0& 0
\end{pmatrix}.
\end{equation}
From these formulas, it is clear that the matrices $B^i$ are symmetric, that is,
\begin{equation} \label{B-sym}
(B^i)^{\tr}=B^i.
\end{equation}

We proceed by differentiating \eqref{relEulC} spatially to get
\begin{equation*}
\del{t}\del{J}W + \Ac^I \del{I}\del{J}W + \del{J}\Ac^I \del{I}W  = -\frac{\mu}{t}\Pi \del{J}W + t^{2\mu-1}\del{J}\Gc.
\end{equation*}
Setting
\begin{equation} \label{WbJdef}
\Wb\!_J := t^\mu \del{J}W = (t^\mu\del{J}\zetat,t^\mu \del{J}w_1,t^{\mu}\del{J}w_2,t^{\mu} \del{J} w_3)^{\tr},
\end{equation}
we can write this as
\begin{equation*} 
\del{t}\Wb\!_J + \Ac^I \del{I}\Wb\!_J + \del{J}\Ac^I \Wb_I  = \frac{\mu}{t}\Pi^\perp \Wb\!_J + t^{3\mu-1}\del{J}\Gc.
\end{equation*}
Multiplying the above equation on the left by $A^0$ and recalling the definitions \eqref{AIdef}, we find that $\Wb\!_J$ satisfies
\begin{equation}  \label{relEulI}
A^0\del{t}\Wb\!_J + A^I \del{I}\Wb\!_J = \frac{\mu}{t}A^0\Pi^\perp \Wb\!_J +  t^{3\mu-1}A^0\del{J}\Gc- A^0\del{J}\Ac^I \Wb_I.
\end{equation}
Finally, combining \eqref{relEulHa} and \eqref{relEulI} yields the Fuchsian system
\begin{equation} \label{relEulK}
\Asc^0\del{t}\Wsc + \Asc^I \del{I}\Wsc = \frac{\mu}{t}\Asc^0\Pbb \Wsc + \Fsc
\end{equation}
where
\begin{align}
\Wsc &= \begin{pmatrix} \Wb \\ \Wb\!_J \end{pmatrix}, \label{Wscdef} \\
\Asc^0 &= \begin{pmatrix} B^0 & 0 \\ 0 & A^0 \end{pmatrix}, \label{Asc0def} \\
\Asc^I &=  \begin{pmatrix} B^I & 0 \\ 0 & A^I \end{pmatrix}, \label{AscIdef} \\
\Pbb &=  \begin{pmatrix} 0 & 0 \\ 0 & \Pi^\perp \end{pmatrix}, \label{Pbbdef}
\intertext{and}
\Fsc &=\begin{pmatrix}  -\Pi\Ac^I  \Wb_I
- t^{-\mu}S\Pi^\perp A^0 \Pi^\perp\Ac^I  \Pi \Wb_I + t^{2\mu-1}S\Pi^\perp A^0\Pi^\perp\Gc \\ 
t^{3\mu-1}A^0\del{J}\Gc- A^0\del{J}\Ac^I \Wb_I\end{pmatrix}. \label{Fsc0def}
\end{align}

As will be established in Step 2 below, the Fuchsian system \eqref{relEulK} satisfies assumption needed to apply the Fuchsian global existence theory from \cite{BOOS:2021}; see, in particular,  \cite[Thm.~3.8.]{BOOS:2021} and \cite[\S 3.4.]{BOOS:2021}. This global existence theory will be used
in Step 3 of the proof to establish uniform bounds on solutions to the initial value problem \eqref{relEulE.1}-\eqref{relEulE.2} under a suitable small initial data
assumption. These bounds in conjunction with a continuation principle will then yield the existence solutions to \eqref{relEulE.1}-\eqref{relEulE.2} on $(0,1]\times \Tbb^3$
as well as decay estimates as $t\searrow 0$. 

\subsection{Step 2: Verification of the coefficient assumptions}
In order to apply Theorem 3.8.~from \cite{BOOS:2021}, see also \cite[\S 3.4.]{BOOS:2021},  to the Fuchsian system \eqref{relEulK}, we need to verify that the coefficients of this equations satisfy the assumptions from Section 3.4.~of \cite{BOOS:2021}, see also  \cite[\S 3.1.]{BOOS:2021}. To begin the verification, we set
\begin{equation} \label{bvarsdef}
\tb = t^{2\mu}, \AND \wb_\Lambda = t^\mu w_\Lambda, \quad \Lambda=2,3,
\end{equation}
and observe from \eqref{wbr1def}-\eqref{phidef}, \eqref{A0rep}, \eqref{B0-form} and  \eqref{Asc0def} that the matrix $\Asc^0$ can be treated as a map depending on the variables \eqref{bvarsdef}, that is,
\begin{equation} \label{A0smooth}
\Asc^0 = \Asc^0(\tb,\wbr_1,\wb_2,\wb_3),
\end{equation}
where for each $R>0$ there exists constants $r,\omega >0$ such that $\Asc^0$  is smooth
on the domain defined by
\begin{equation} \label{smoothdom}
(\tb,\wbr_1,\wb_2,\wb_3) \in (-r,2) \times (-R,R) \times (-R,R)\times (-R,R),
\end{equation}
and satisfies
\begin{equation} \label{A0lb}
\Asc^0(\tb,\wbr_1,0,0) \geq \omega \id 
\end{equation}
for all $(\tb,\wbr_1)\in (-r,2)\times (-R,R)$. In the following, we will always be able to choose $R>0$ and $r>0$ as needed in order to guarantee that the statements we make
are valid.

Differentiating $\Asc^0$ with respect to $t$ then shows, with the help of  \eqref{wbr1def}, \eqref{Wbdef}
and \eqref{bvarsdef}-\eqref{A0smooth}, that
\begin{align} 
\del{t}\Asc^0 &= D\Asc^0(\tb,\wbr_1,\wb_2,\wb_3) \begin{pmatrix} 2\mu t^{2\mu-1} \\ u'(t)+\del{t}w_1\\ \del{t}\wb_2\\ \del{t}\wb_3 \end{pmatrix}
\notag \\
&= D\Asc^0(\tb,\wbr_1,\wb_2,\wb_3) \left(\begin{pmatrix} 2\mu t^{2\mu-1} \\ u'(t)\\ 0 \\ 0 \end{pmatrix}
+ \Pc_1\del{t}\Wb \right) \label{dtA0}
\end{align}
where 
\begin{equation*} 
\Pc_1 = \diag(0,1,1,1),
\end{equation*}
and $\del{t}\Wb$ can be computed from from \eqref{relEulHa}, that is, 
\begin{equation} \label{dt-Wb}
\del{t}\Wb =(B^0)^{-1}\Bigl(- B^I \del{I}\Wb -\Pi\Ac^I \Wb_I
- t^{-\mu}S\Pi^\perp A^0 \Pi^\perp\Ac^I  \Pi \Wb_I + t^{2\mu-1}S\Pi^\perp A^0\Pi^\perp\Gc\Bigr).
\end{equation}
We note from \eqref{wbr1def}-\eqref{chidef}, \eqref{Sdef},  \eqref{B0-form} and \eqref{bvarsdef} that the matrices
\begin{equation}\label{B0smooth}
 S = S(\tb,\wbr_1) \AND B^0 = B^0(\tb,\wbr_1)
\end{equation}
are smooth on the domain  $(\tb,\wbr_1)\in (-r,2)\times (-R,R)$,
and  that $B^0$ is bounded below by
\begin{equation}\label{B0-lbnd}
B^0 \geq \omega \id 
\end{equation} 
for all $(\tb,\wbr_1)\in (-r,2)\times (-R,R)$ where $\omega$ can be taken as the same constant as in \eqref{A0lb}. 
We further note from  \eqref{wbr1def}-\eqref{phidef}, \eqref{A0rep}, \eqref{AIdef},  \eqref{BI-form}, \eqref{bI-def} and \eqref{bvarsdef} that the matrices
\begin{equation}\label{BIsmooth}
A^i=  A^i(\tb,\wbr_1,\wb_2,\wb_3)
\AND
B^I=  B^I(\tb,\wbr_1,\wb_2,\wb_3)
\end{equation}
are smooth on the domain  \eqref{smoothdom}, while is 
clear from \eqref{Gcdef} that
the vector-valued map
\begin{equation} \label{Gcsmooth}
\Gc = \Gc(\tb,\wbr_1,w_1)
\end{equation}
is smooth on the domain $(\tb,\wbr_1,w_1)\in (-r,2)\times (-R,R) \times (-R,R)$.

Next, setting
\begin{equation} \label{wh1def}
\wh_1 = t^\mu e^{-2 \wbr_1}, 
\end{equation}
it follows from \eqref{wbr1def}-\eqref{xidef}, \eqref{Ac1rep}-\eqref{Ac3rep} and \eqref{bvarsdef} that  the matrices $\Ac^I$  can be expanded as
\begin{equation} \label{AcIsmooth}
\Ac^I = \Ac^I_1(\wh_1,\wb_2,\wb_3)+ t^\mu \Ac^I_2(\tb,\wbr_1,\wb_2,\wb_3)+ t^{2\mu} \Ac^I_3(\tb,\wbr_1,\wb_2,\wb_3)
\end{equation}
where the $\Ac^I_2$, $\At^I_3$ are smooth on the domain \eqref{smoothdom} and the $\Ac^I_1$ are smooth on the domain
defined by
\begin{equation*}
(\wh_1,\wb_2,\wb_3) \in (-R,R)\times (-R,R)\times (-R,R).
\end{equation*}
It is also not difficult to verify from \eqref{Ac1rep}-\eqref{Ac3rep} that the $\Ac^I_1$
satisfy
\begin{equation} \label{PiperpAcIPi}
 \Pi^\perp \Ac^I_1 \Pi = 0.
\end{equation}
Differentiating the matrices $\Ac^I$ spatially, we have by \eqref{wbr1def}, \eqref{WbJdef}, \eqref{bvarsdef}, \eqref{wh1def} and
\eqref{AcIsmooth} that
\begin{align}
\del{J}\Ac^I &= D\Ac^I_1(\wh_1,\wb_2,\wb_3)\begin{pmatrix}-2 e^{-2 \wbr_1} t^\mu \del{J}w_1 \\ t^\mu \del{J}w_2 \\
 t^\mu \del{J}w_2
\end{pmatrix}\notag \\
&+ t^\mu D\Ac^I_2(\tb,\wbr_1,\wb_2,\wb_3)\begin{pmatrix} 0 \\ \del{J}w_1\\ t^\mu \del{J}w_2
\\ t^\mu\del{J} w_3 \end{pmatrix} +t^{2\mu} D\Ac^I_3(\tb,\wbr_1,\wb_2,\wb_3)\begin{pmatrix} 0 \\ \del{J}w_1\\ t^\mu \del{J}w_2
\\ t^\mu\del{J} w_3 \end{pmatrix} \notag \\
& =  \Bigl(D\Ac^I_1(\wh_1,\wb_2,\wb_3)\Pc_2 + D\Ac^I_2(\tb,\wbr_1,\wb_2,\wb_3)\Pc_3+t^\mu  D\Ac^I_2(\tb,\wbr_1,\wb_2,\wb_3)\Pc_3\Bigr) \Wb\!_J, \label{dJAcI}
\end{align}
where
\begin{equation*}
\Pc_2 = \begin{pmatrix} 0 & -2 e^{-2 \wbr_1} & 0 & 0 \\ 0 & 0 & 1 & 0 \\ 0 & 0 & 0 & 1 \end{pmatrix} \AND
\Pc_3 = \begin{pmatrix} 0 & 0 & 0 & 0 \\ 0 & 1 & 0 & 0 \\ 0 & 0 & t^\mu & 0 \\ 0 & 0 & 0 & t^\mu \end{pmatrix}.
\end{equation*}

By  \eqref{Pidef}-\eqref{Piperpdef} and \eqref{A0rep}, we note that the matrix $A^0$ satisfies $[\Pi^\perp,A^0] = 0$
and $\Pi^\perp A^0 \Pi = \Pi A^0 \Pi^\perp = 0$.
Using these identities, it is then follows from the definitions \eqref{Asc0def} and \eqref{Pbbdef} that $\Asc^0$ satisfies
 \begin{equation} \label{Asc0Pbbcom}
[\Pbb,\Asc^0] = 0
\end{equation}
and
\begin{equation} \label{PbbAsc0Pbbperp}
\Pbb^\perp \Asc^0 \Pbb = \Pbb \Asc^0 \Pbb^\perp = 0,
\end{equation}
where 
\begin{equation} \label{Pbbperpdef}
\Pbb^\perp = \id -\Pbb.
\end{equation}
Additionally,  by \eqref{Pidef}-\eqref{Pirel}, we observe that $\Pbb$ satisfies
\begin{equation} \label{Pbbrel}
\Pbb^2 = \Pbb,  \quad \Pbb^{\tr} = \Pbb,  \quad \del{t}\Pbb = 0 \AND \del{I} \Pbb = 0,
\end{equation}
while the symmetry of the matrices $\Asc^i$,  that is,
\begin{equation} \label{Ascisym}
(\Asc^i)^{\tr} = \Asc^i,
\end{equation}
is obvious from the definitions \eqref{A0rep} and \eqref{Asc0def}-\eqref{AscIdef}, and the relations \eqref{AI-sym} and \eqref{B-sym}.

Now, from the definitions  \eqref{wbr1def}, \eqref{Wbdef}, \eqref{WbJdef}, \eqref{Wscdef}, \eqref{bvarsdef}
and \eqref{wh1def}, the formulas \eqref{dtA0} and  \eqref{dt-Wb}, the estimates \eqref{Hombounds} for $u(t)$ and $u'(t)$, the smoothness properties \eqref{A0smooth}, \eqref{B0smooth}, \eqref{BIsmooth}, \eqref{Gcsmooth} and \eqref{AcIsmooth} of the matrices $\Asc^0$, $S$, $A^0$, $B^0$, $A^I$, $B^I$, $\Ac^I$ and the source term $\Gc$, the lower bound \eqref{B0-lbnd} on $B^0$, and the identity
\eqref{PiperpAcIPi}, it is not difficult to verify that for each $\mu \in (0,\infty)$ that there exists a constant $\theta>0$ such that 
\begin{equation}\label{dtAsc-bnd}
|\del{t}\Asc^0| \leq \theta (t^{2\mu-1}+1)
\end{equation}
for all $(t,\Wsc,D\Wsc)\in [0,1]\times B_R(\Rbb^{16})\times B_{R}(\Rbb^{16\times 3})$, where $D\Wsc = (\del{I}\Wsc)$. From \eqref{Fsc0def} and similar considerations, it is also not difficult to verify
\begin{equation} \label{Fsc-bnd}
|\Fsc| \lesssim (t^{2\mu-1}+1)|\Wsc| 
\end{equation}
for all $(t,\Wsc)\in [0,1]\times B_R(\Rbb^{16})$. It is also clear that we can view \eqref{relEulK} as an equation for the variables $\Wsc=(\Wb,\Wb_J)$, with $\Wb=(\zetat,w_1,\wb_2,\wb_3)$ and $\Wb_J=(\zetat_J, w_{1J},\wb_{2J},\wb_{3J})$, where the maps $\Asc^i$ and $\Fsc$ depend on the variables $(t,\Wb)$ and $(t,\Wsc)$, respectively.

Taken together, \textbf{(i)} the variable definitions \eqref{wbr1def}, \eqref{Wbdef}, \eqref{WbJdef}, \eqref{bvarsdef}
and \eqref{wh1def}, \textbf{(ii)} the smoothness properties \eqref{A0smooth}, \eqref{B0smooth}, \eqref{BIsmooth}, \eqref{Gcsmooth} and \eqref{AcIsmooth} of the matrices $\Asc^0$, $S$, $A^0$, $B^0$, $A^I$, $B^I$, $\Ac^I$ and the source term $\Gc$,
\textbf{(iii)} the identities \eqref{Asc0Pbbcom}-\eqref{PbbAsc0Pbbperp} and the lower bound \eqref{A0lb} satisfied by matrix $\Asc^0$, 
\textbf{(iv)} the definitions \eqref{AscIdef} and \eqref{Fsc0def} of the matrices $\Asc^I$ and the source term $\Fsc$, 
\textbf{(v)} the properties \eqref{Pbbrel} of the projection map $\Pbb$, and \textbf{(vi)} the bounds \eqref{dtAsc-bnd} and \eqref{Fsc-bnd}  on
$\del{t}\Asc^0$ and $\Fsc$, respectively, imply that for any\footnote{By \eqref{mu-def}, $\mu\in (0,\infty)$ corresponds to $1/3<K<1$.} $\mu \in (0,\infty)$ and $R>0$ chosen sufficiently small, 
there exist constants $\theta,\gamma_1=\gammat_1,\gamma_2=\gammat_2>0$ such that the Fuchsian system \eqref{relEulK} satisfies
satisfies all the assumptions from Section 3.4 of \cite{BOOS:2021}  for following choice of constants: $\kappa=\kappat=\mu$, $\beta_\ell=0$, $1\leq \ell \leq 7$,
\begin{gather*}
 \quad p=\begin{cases}2\mu & \text{if $0<\mu\leq 1/2$}\\
1 & \text{if $\mu > 1$} \end{cases} 
\end{gather*}
and $\lambda_1=\lambda_2=\lambda_3= \alpha=0$. 
As discussed in \cite[\S 3.4]{BOOS:2021}, under the time transformation\footnote{Note that our time variable $t$ is assumed to be positive as opposed to \cite{BOOS:2021}, where it is taken to
be negative. This
causes no difficulties as we can change between these two conventions by using the simple time transformation $t\rightarrow -t$.} 
$t \mapsto t^p$, the transformed version of \eqref{relEulK} will satisfy all of the assumptions from Section 3.1 of \cite{BOOS:2021}.  Moreover, since the matrices $\Asc^I$ have a regular limit as $t\searrow 0$, the constants $\btt$ and $\tilde{\btt}$ from Theorem 3.8 of \cite{BOOS:2021} vanish. This fact together with $\beta_1=0$ and $\kappa=\kappat=\mu$ implies that the constant\footnote{This constant is denoted by $\zeta$ in the article \cite{BOOS:2021}. We denote it here by $\mathfrak{z}$ because $\zeta$ is already
being used denote the modified density.}
$\mathfrak{z}$ from Theorem 3.8 of \cite{BOOS:2021} that determines the decay rate is given by
$\mathfrak{z}= \mu$.

\subsection{Step 3: Existence and uniqueness}
By \eqref{A0rep} and \eqref{AI-sym}, we know that the matrices $A^i$ are symmetric. Furthermore, from the analysis carried out in Step 2 above, we know that
the matrices  $A^i$ and the source term $A^0 \Gsc$  depend smoothly on the variables $(t,w_J)$ for $t\in (0,1]$ and $w_J$ in an open neighbourhood of zero, and that the matrix $A^0$ is
positive definite on this neighbourhood. As a consequence, the system \eqref{relEulE.1} is symmetrizable and can be put in the symmetric hyperbolic form \eqref{relEulD} by multiplying it on the left by the matrix $A^0$.  Since $k\in\Zbb_{>3/2+1}$ and 
$W_0 :=(\zetat_0, w^0_J)^{\tr}\in H^{k+1}(\Tbb^3,\Rbb^4)$,
we obtain from an application of standard local-in-time existence and uniqueness theorems and the continuation principle for symmetric hyperbolic systems, see Propositions 1.4, 1.5 and 2.1 from \cite[Ch.~16]{TaylorIII:1996}, the existence of a unique solution
\begin{equation*}
W=(\zetat,w_J) \in C^0\bigl((T_*,1], H^{k+1}(\Tbb^3,\Rbb^4)\bigr)\cap C^1\bigl((T_*,1],H^{k}(\Tbb^3,\Rbb^4)\bigr)
\end{equation*}
to  IVP \eqref{relEulE.1}-\eqref{relEulE.2} where $T_*\in [0,1)$ is the maximal time of existence. 
From the computations carried out in Step 1 of the proof,  this solution determines via \eqref{Wbdef} and \eqref{WbJdef}
a solution
\begin{equation} \label{Wscvar}
\Wsc = (\Wb,\Wb\!_J)  \in C^0\bigl((T_*,1], H^{k}(\Tbb^3,\Rbb^{16})\bigr)\cap C^1\bigl((T_*,1],H^{k-1}(\Tbb^3,\Rbb^{16})\bigr)
\end{equation}
of the IVP
\begin{align} 
\Asc^0\del{t}\Wsc + \Asc^I \del{I}\Wsc &= \frac{\mu}{t}\Asc^0\Pbb \Wsc + \Fsc\hspace{1.20cm}
\text{in $(T_*,1]\times \Tbb^3$,} 
\label{relEulM1} \\
\Wsc &= \Wsc_0 := (W_0,\del{J}W_0) \hspace{0.5cm}\text{in $\{1\}\times \Tbb^3$,} \label{relEulM2}
\end{align}
where we observe that
\begin{equation} \label{Wsc-idata}
\norm{\Wsc_0}_{H^k} \lesssim \norm{W_0}_{H^{k+1}} \leq \delta.
\end{equation}

On the other hand, by Step 2 we can apply\footnote{It it is important to note that the regularity $k\in \Zbb_{>3/2+1}$ of the initial data \eqref{Wsc-idata} is less than what
is required to apply Theorem 3.8.~ \cite{BOOS:2021} to the Fuchsian system \eqref{relEulK}. The reason that we can still apply this theorem is that the matrices $\Asc^I$
in \eqref{relEulK} do not have any $1/t$ singular terms; see  Remark A.3.(ii) from \cite{BeyerOliynyk:2020}.}
Theorem 3.8.~from \cite{BOOS:2021}
to the time transformed version of \eqref{relEulK} as described in \cite[Section 3.4]{BOOS:2021} to deduce, for $\delta>0$ chosen small enough and
the initial data satisfying \eqref{Wsc-idata}, the existence of a unique
solution
\begin{equation*}
\Wsc^* \in C^0\bigl((0,1],H^k(\Tbb^3,\Rbb^{16})\bigr)\cap L^\infty\bigl((0,1],H^k(\Tbb^3,\Rbb^{16}))\bigr)\cap 
C^1\bigl((0,1],H^{k-1}(\Tbb^3,\Rbb^{16})\bigr)
\end{equation*}
to the IVP \eqref{relEulM1}-\eqref{relEulM2} with $T_*=0$ that satisfies the following properties:
\begin{enumerate}[(1)]
\item The limit of $\Pbb^\perp \Wsc^*$ as $t\searrow 0$, denoted $\Pbb^\perp \Wsc^*(0)$, exists in $H^{k-1}(\Tbb^3,\Rbb^{16})$.
\item The solution $\Wsc^*$ is bounded by the energy estimate
\begin{equation}\label{eestA}
\norm{\Wsc^*(t)}_{H^k}^2 + \int_{t}^1 \frac{1}{\tau} \norm{\Pbb \Wsc^*(\tau)}_{H^k}^2\, d\tau   \lesssim \norm{\Wsc_0}_{H^k}^2
\end{equation}
for all $t\in (0,1]$, where the implied constant depends on $\delta$.
\item For any given $\sigma>0$, the solution  $\Wsc^*$ satisfies the decay estimate
\begin{gather}
\norm{\Pbb \Wsc^*(t)}_{H^{k-1}} \lesssim 
t^{\mu-\sigma} 
\AND
\norm{\Pbb^\perp \Wsc^*(t) - \Pbb^\perp \Wsc^*(0)}_{H^{k-1}} \lesssim
 t^{\mu-\sigma} \label{decayA2}
\end{gather}
for all $t\in (0,1]$, where the implied constants depend on $\delta$ and $\sigma$.
\end{enumerate}

By uniqueness, the two solutions $\Wsc$ and $\Wsc^*$ to the IVP \eqref{relEulM1}-\eqref{relEulM2} must coincide on their common
domain of definition, and consequently, $\Wsc(t)=\Wsc^*(t)$ for all $t\in (T_*,1]$.
But this implies by \eqref{Wscvar}, the energy estimate \eqref{eestA}, and Sobolev's inequality \cite[Thm.~6.2.1]{Rauch:2012}  that
\begin{equation*}
\norm{\Wb(t)}_{W^{1,\infty}} \lesssim \norm{\Wb(t)}_{H^k} \leq \norm{\Wsc(t)}_{H^{k-1}} \lesssim \norm{\Wsc_0},
\quad T^*<t\leq 1.
\end{equation*}
By shrinking $\delta$ if necessary, we can, by \eqref{Wsc-idata}, make $\norm{\Wsc_0}_{H^k}$ as small as we like, which in turn, implies via 
the above estimate that we can bound $\Wb$ by
$\norm{\Wb(t)}_{W^{1,\infty}} \leq \frac{R}{2}$ for all $t\in (T^*,1]$,
where $R>0$ is as determined in Step 2 of the proof. This bound is sufficient to guarantee that the matrices $A^i$ and the source term $A^0\Gsc$ from
the symmetric hyperbolic system \eqref{relEulD} remain well defined and that the matrix $A^0$ continues to be positive definite. 
By the continuation principle and the maximality of $T_*$, we deduce that $T_*=0$, and hence that
$\Wsc(t)=\Wsc^*(t)$ for all $t\in (0,1]$. From this and the energy estimate \eqref{eestA}, it then follows with the help of the definitions
 \eqref{Pidef}-\eqref{Piperpdef}, \eqref{Wbdef}, \eqref{WbJdef},  \eqref{Pbbdef} and \eqref{Wscvar} that
\begin{equation*}
\Ec(t) + \int_t^1 \tau^{2\mu-1}\bigl(\norm{D\zetat(\tau)}_{H^k}^2+\norm{Dw_1(\tau)}_{H^k}^2\bigr)\,d\tau \lesssim \norm{W_0}_{H^k}^2,
\quad 0<t\leq 1,
\end{equation*}
where
\begin{equation*}
\Ec(t)=\norm{\zetat(t)}_{H^k}^2+\norm{w_1(t)}_{H^k}^2+t^{2\mu}\Bigl(\norm{D\zetat(t)}_{H^k}^2+\norm{Dw_1(t)}_{H^k}^2+\norm{w_2(t)}_{H^{k+1}}^2+\norm{w_3(t)}_{H^{k+1}}^2\Bigr).
\end{equation*}
We further obtain from the decay estimate \eqref{decayA2} and \eqref{Pbbperpdef} the existence of functions $\zetat_*, w_1^* \in H^{k-1}(\Tbb^3)$ and $\wb_2^*,\wb_3^* \in H^{k}(\Tbb^3)$ such that
\begin{equation} \label{Ecb-bnd}
\bar{\Ec}(t) \lesssim
 t^{\mu-\sigma}
 \end{equation}
for all $t\in (0,1]$ where
\begin{equation*} 
\bar{\Ec}(t)=\norm{\zetat(t) - \zetat_*}_{H^{k-1}}+\norm{w_1(t) - w_1^*}_{H^{k-1}}
+\norm{t^\mu w_2(t) - \wb_2^*}_{H^{k}}+\norm{t^\mu w_3(t) - \wb_3^*}_{H^{k}}.
\end{equation*}
We also note by \eqref{c-velocity},  \eqref{mod-den}, \eqref{v0def} and \eqref{zetatdef}-\eqref{cov2c} that $u$ and
$W=(\zetat,w_J)^{\tr}$ determine a solution of the relativistic Euler equations \eqref{relEulA} on the spacetime region $M=(0,1]\times \Tbb^3$
via the formulas  \eqref{relEulsol.1}-\eqref{relEulsol.5}.

To complete the proof, we find from differentiating \eqref{relEulsol.1} that the density contrast can be expressed as
\begin{equation} \label{den-constrast-B}
\frac{\del{I}\rho}{\rho} = \frac{(1+K)(t^{2\mu}+ e^{2(u+w_1)})^{\frac{1+K}{2}}\del{I}\zetat -(1+K)(t^{2\mu}+ e^{2(u+w_1)})^{\frac{K-1}{2}}e^{2(u+w_1)}\del{I}w_1 }{(t^{2\mu}+ e^{2(u+w_1)})^{\frac{1+K}{2}}}.
\end{equation}
Since $\mu>0$, we can choose $\sigma>0$ small enough so that $\mu-\sigma >0$. Doing so then implies by \eqref{Ecb-bnd} that
$\zetat$ and $w_1$ converge in $H^{k-1}(\Tbb^3)$ to $\zetat_*$ and $w_1^*$ as $t\searrow 0$. Since $u(t)$ converges as well by Proposition \ref{Homprop}, it is then not difficult to verify from \eqref{den-constrast-B} and the Sobolev and Moser inequalities \cite[Thms.~6.2.1 \& 6.4.1]{Rauch:2012}  that
\begin{equation*} 
\lim_{t\searrow 0} \Bigl\| \frac{\del{I}\rho}{\rho} - (1+K)\del{I}(\zetat_*-w_1^*) \Bigr\|_{H^{k-2}} = 0,
\end{equation*}
which completes the proof.

\section{Numerical solutions\label{numsol}}

\subsection{Numerical setup}
In the numerical setup that we use to solve the system \eqref{eqn:dotzeta}-\eqref{eqn:dotw}, the computational domain is $[0,2\pi]$ with periodic boundary condition, the variables $\texttt{z}$ and $\texttt{w}$ are discretised in space using $2^{\text{nd}}$ order central finite differences, and time integration is performed using a standard $2^{\text{nd}}$ order Runge-Kutta method (\textit{Heun's Method}). As a consequence, our code is second order accurate\footnote{Strictly speaking one also needs to enforce the CFL condition to ensure convergence. In this case we have used the tightened 4/3 CFL condition for Heun's Method which is discussed in \cite{schneider:hal-01307287}.}. 

\subsubsection{Convergence tests}
We have verified the second order accuracy of our code with convergence tests involving perturbations of both types of homogeneous solutions \eqref{Hom-A} and \eqref{Hom-C}. 
In our convergence tests, we have evolved the system \eqref{eqn:dotzeta}-\eqref{eqn:dotw} 
staring from the 
the two initial data sets
\begin{align}
\label{eqn:numericalID_A}
(\ztt_{0},\wtt_{0}) &= (0,0.1\sin(x))
\intertext{and}
\label{eqn:numericalID_B}
(\ztt_{0},\wtt_{0}) &= (0,0.1\sin(x)+0.15)
\end{align}
using resolutions of $N =$ $200$, $400$, $800$, $1600$, $3200$, and $6400$ grid points. The initial data \eqref{eqn:numericalID_A} and \eqref{eqn:numericalID_B} satisfy the conditions \eqref{Hom-A-idata} and \eqref{Hom-B-idata}, respectively, and the solutions generated from this initial data represent perturbations of the homogeneous solutions \eqref{Hom-A} and \eqref{Hom-C}, respectively.

\begin{figure}[h]
\centering
\subfigure[Subfigure 1 list of figures text][$t=0.799$]{
\includegraphics[width=0.3\textwidth]{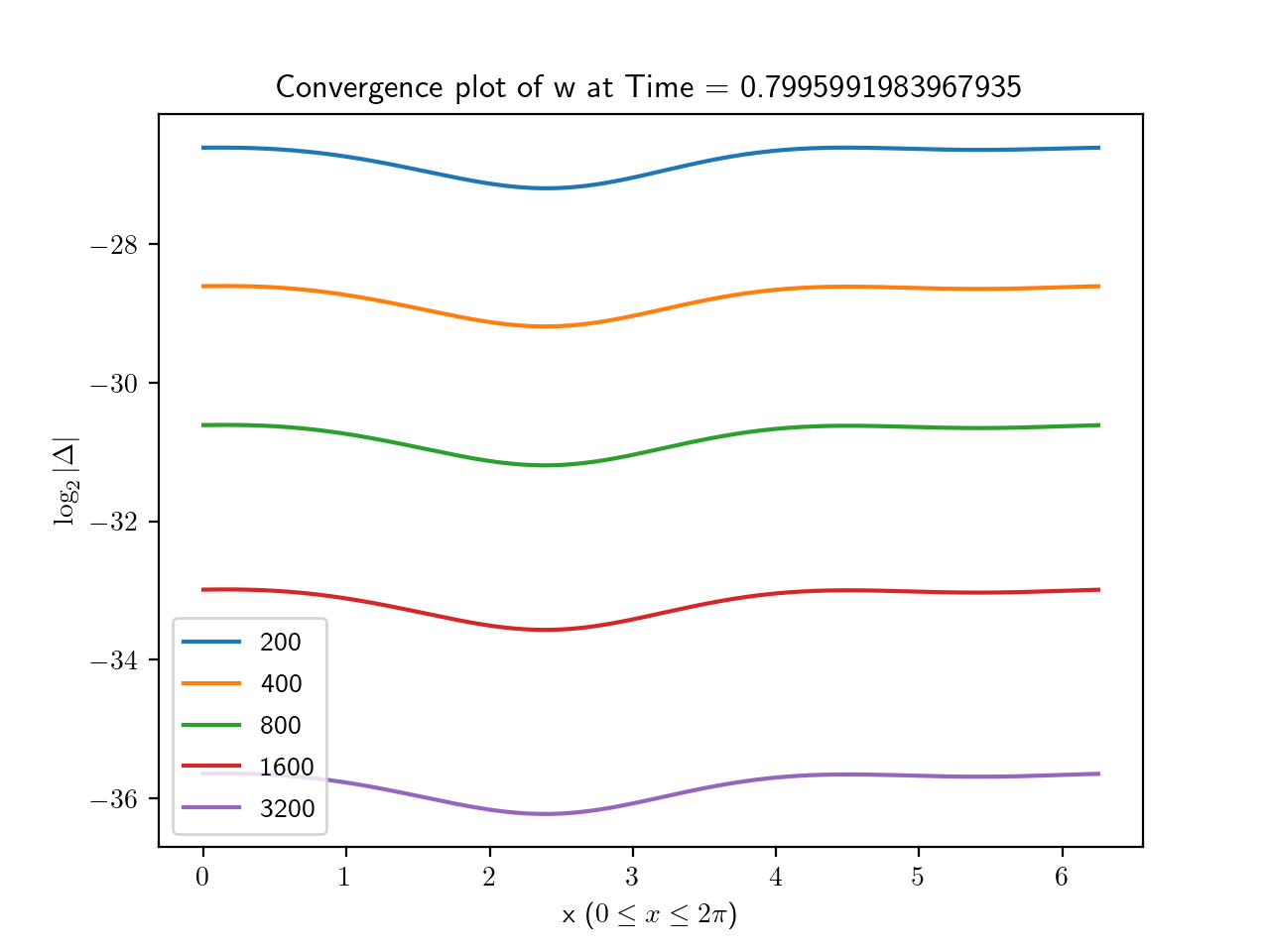}
\label{fig:subfig1}}
\subfigure[Subfigure 2 list of figures text][$t=0.599$]{
\includegraphics[width=0.3\textwidth]{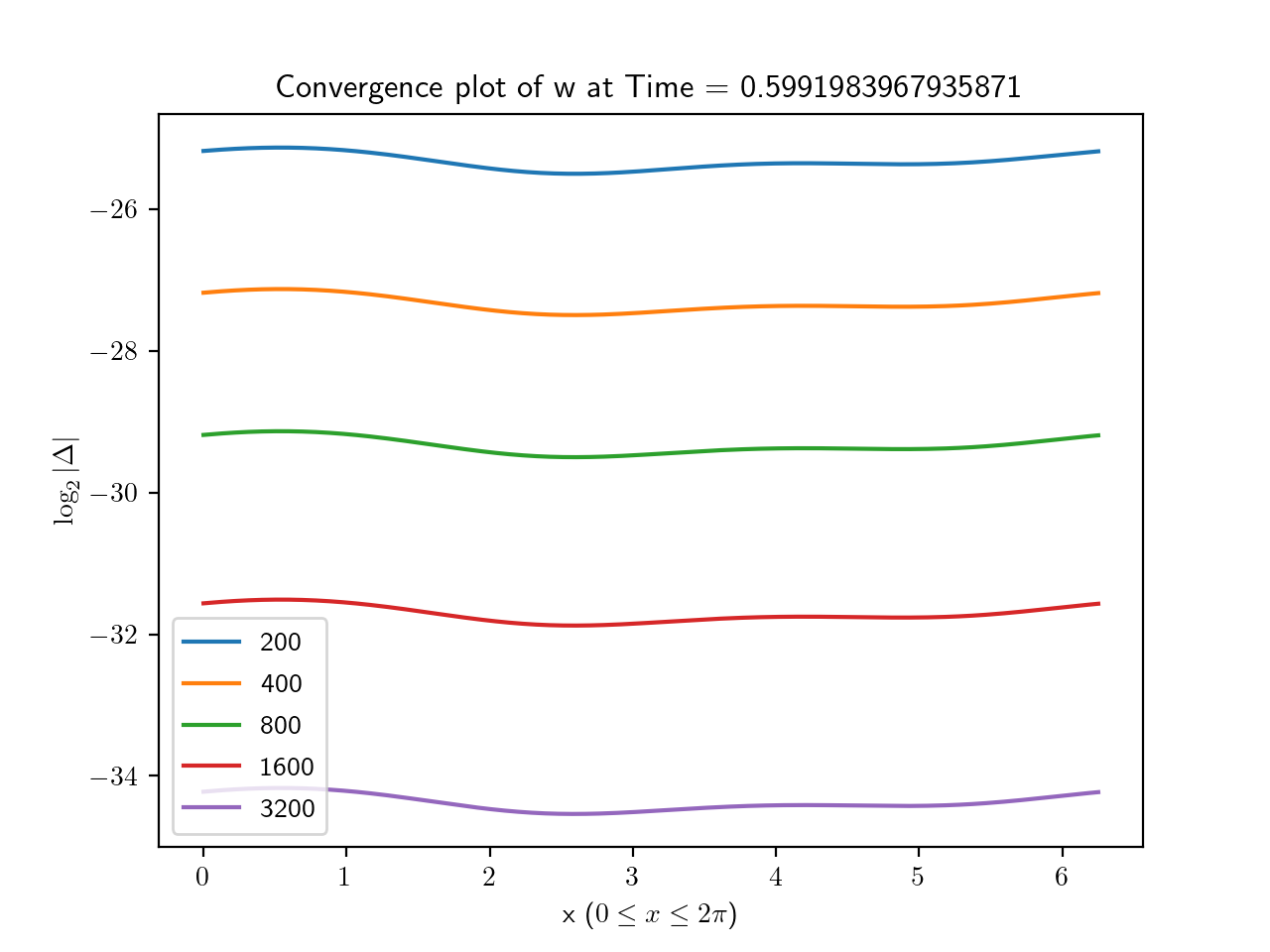}
\label{fig:subfig2}}
\subfigure[Subfigure 5 list of figures text][$t=0.028$]{
\includegraphics[width=0.3\textwidth]{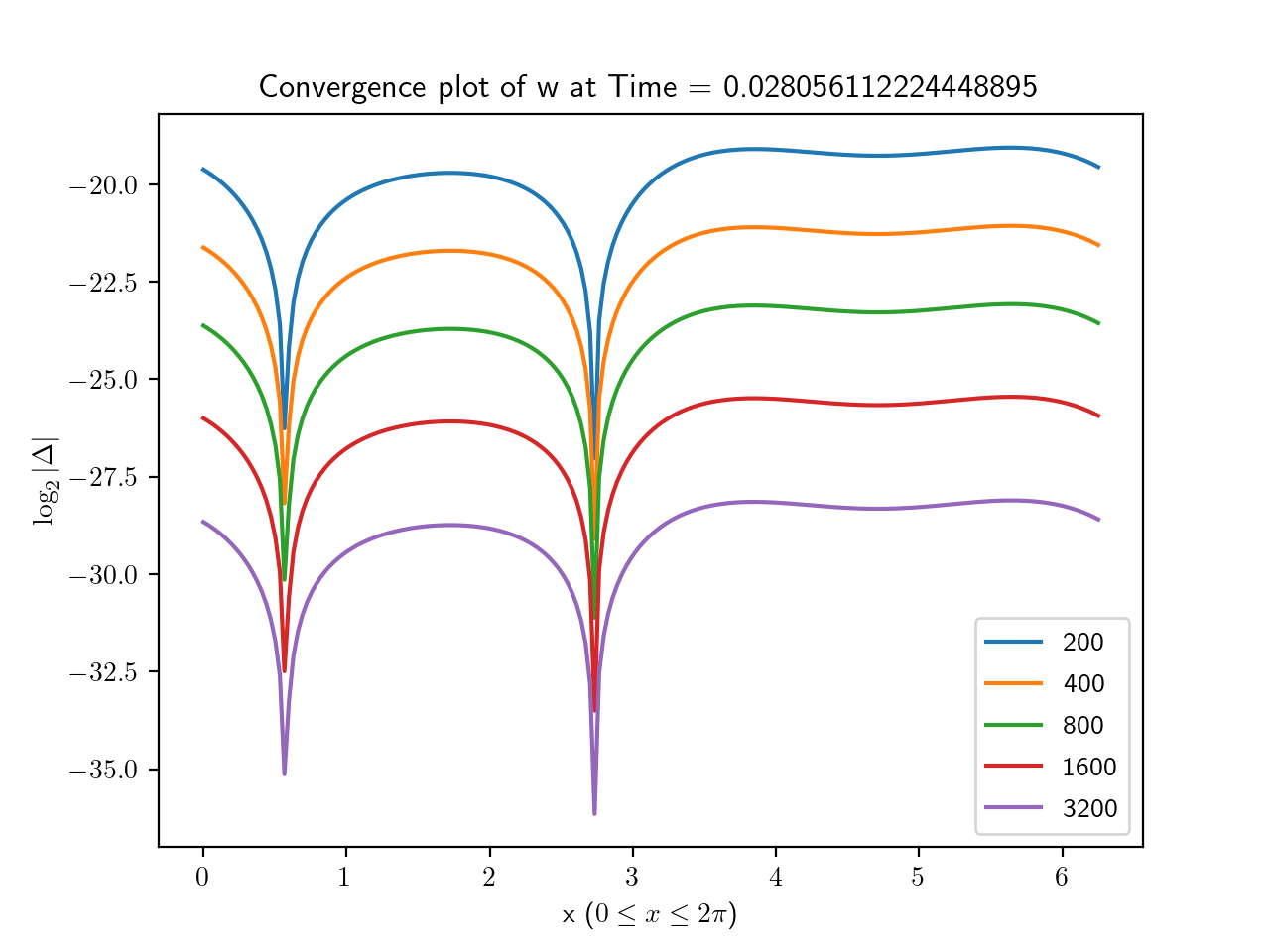}
\label{fig:subfig5}}
\caption{Convergence plots of \texttt{w} at various times. $K=0.5,(\ztt_{0},\wtt_{0}) = (0,0.1\sin(x)+0.15)$. }
\label{fig:W_conv_pos}
\end{figure}

\begin{figure}[h]
\centering
\subfigure[Subfigure 1 list of figures text][$t=0.799$]{
\includegraphics[width=0.3\textwidth]{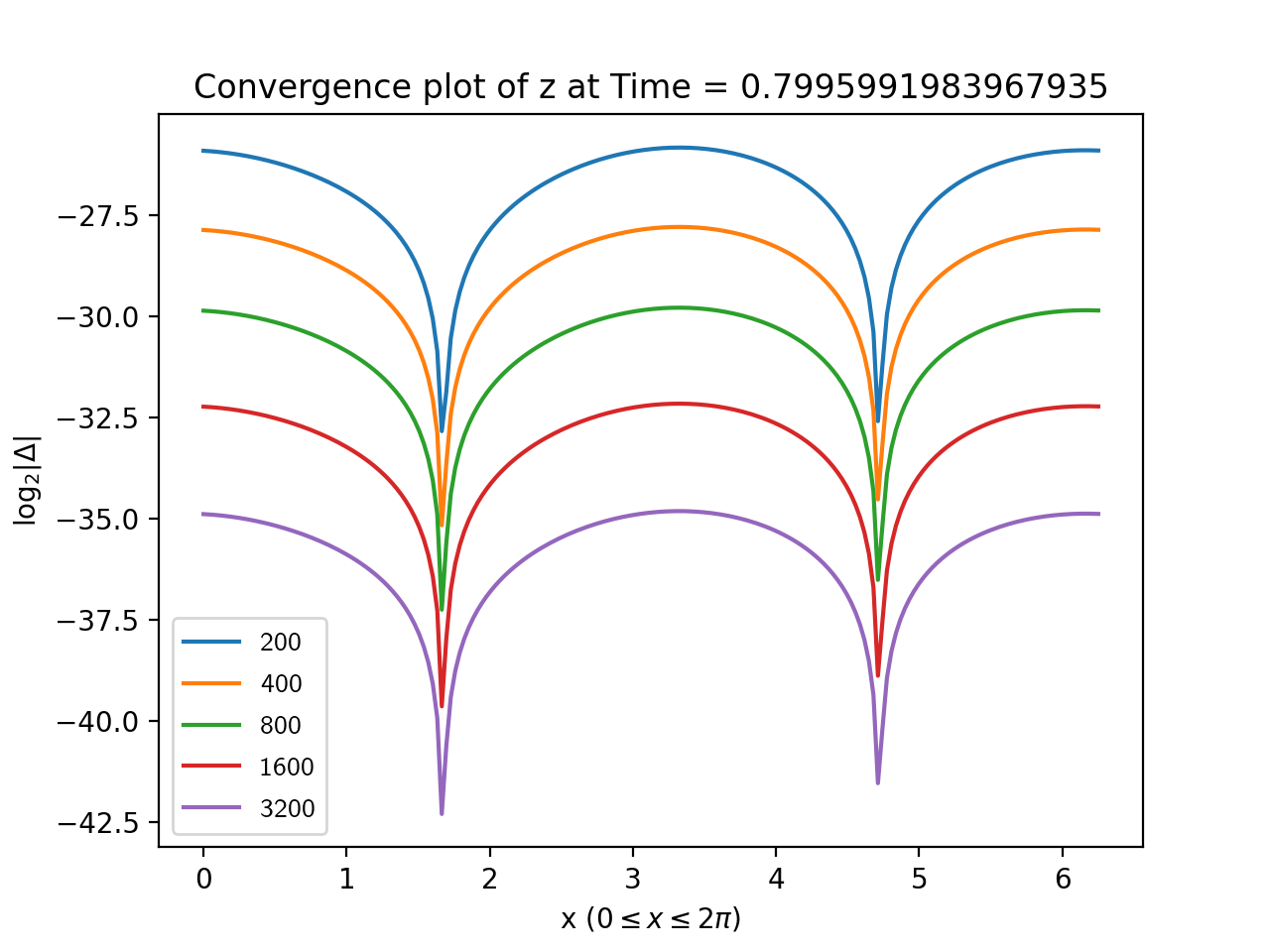}
\label{fig:subfig1}}
\subfigure[Subfigure 2 list of figures text][$t=0.599$]{
\includegraphics[width=0.3\textwidth]{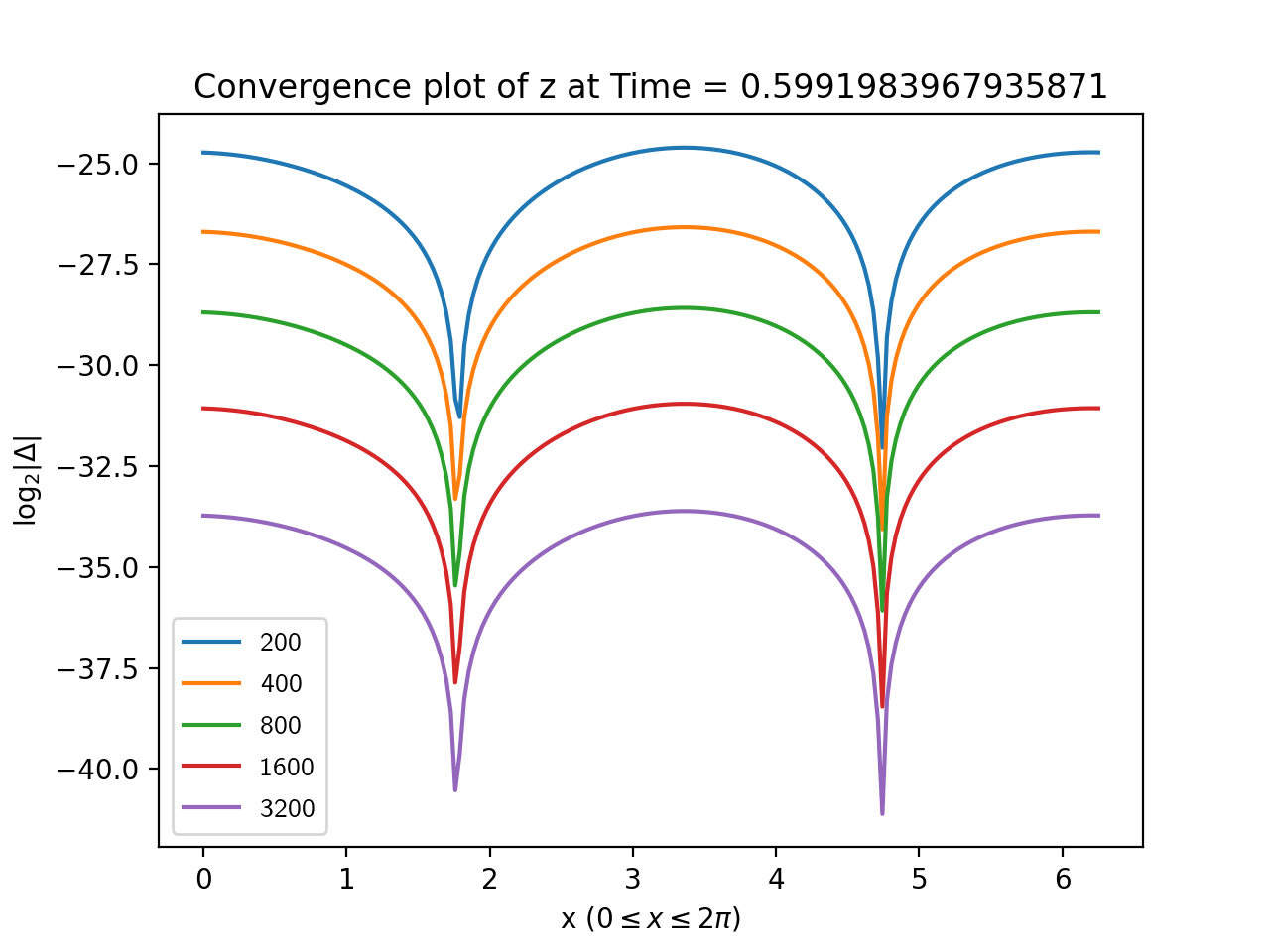}
\label{fig:subfig2}}
\subfigure[Subfigure 5 list of figures text][$t=0.028$]{
\includegraphics[width=0.3\textwidth]{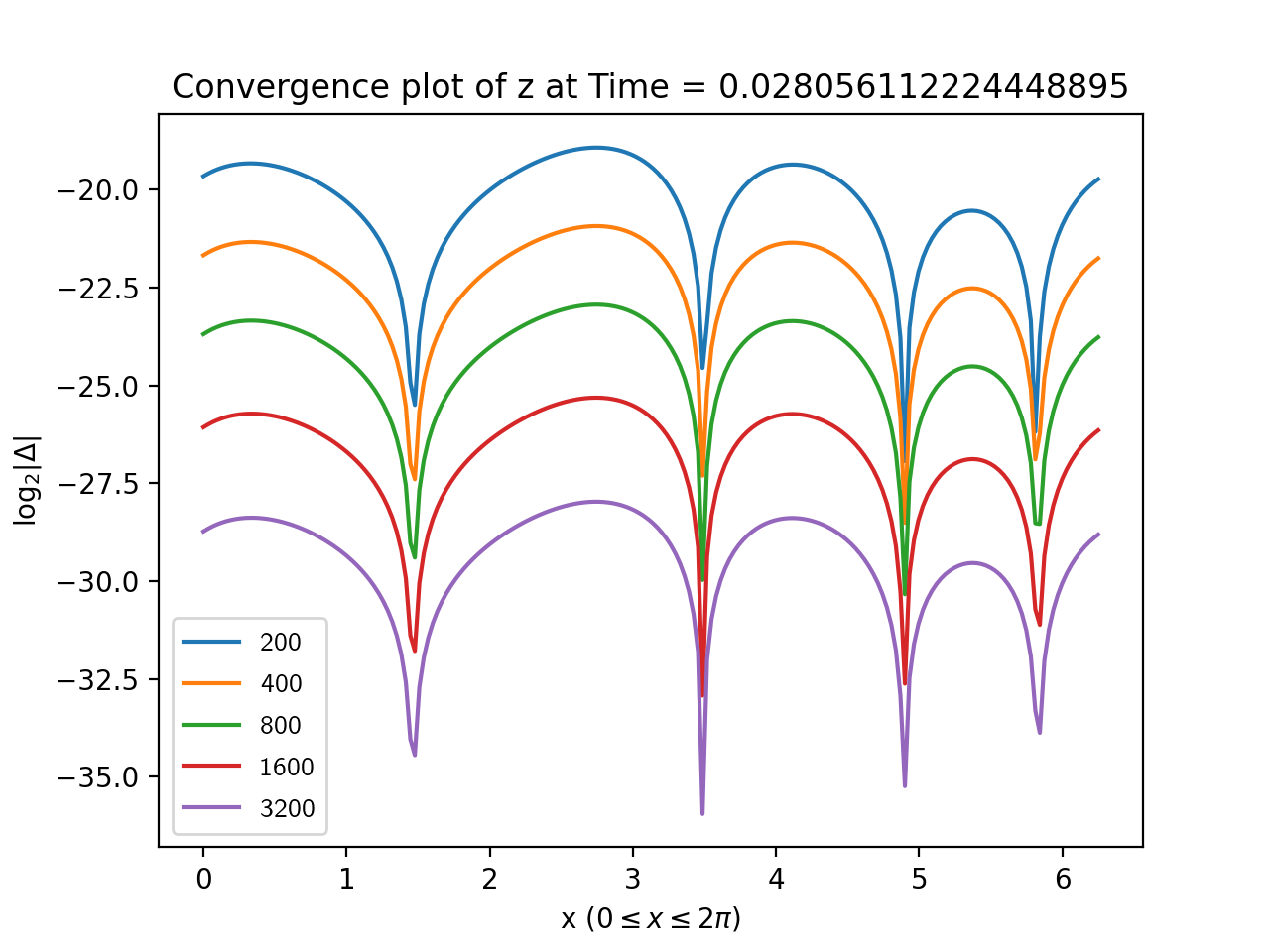}
\label{fig:subfig5}}
\caption{Convergence plots of $\ztt$ at various times. $K=0.5,(\ztt_{0},\wtt_{0}) = (0,0.1\sin(x)+0.15)$.}
\label{fig:Z_conv_pos}
\end{figure}

\begin{figure}[h]
\centering
\subfigure[Subfigure 1 list of figures text][$t=0.799$]{
\includegraphics[width=0.3\textwidth]{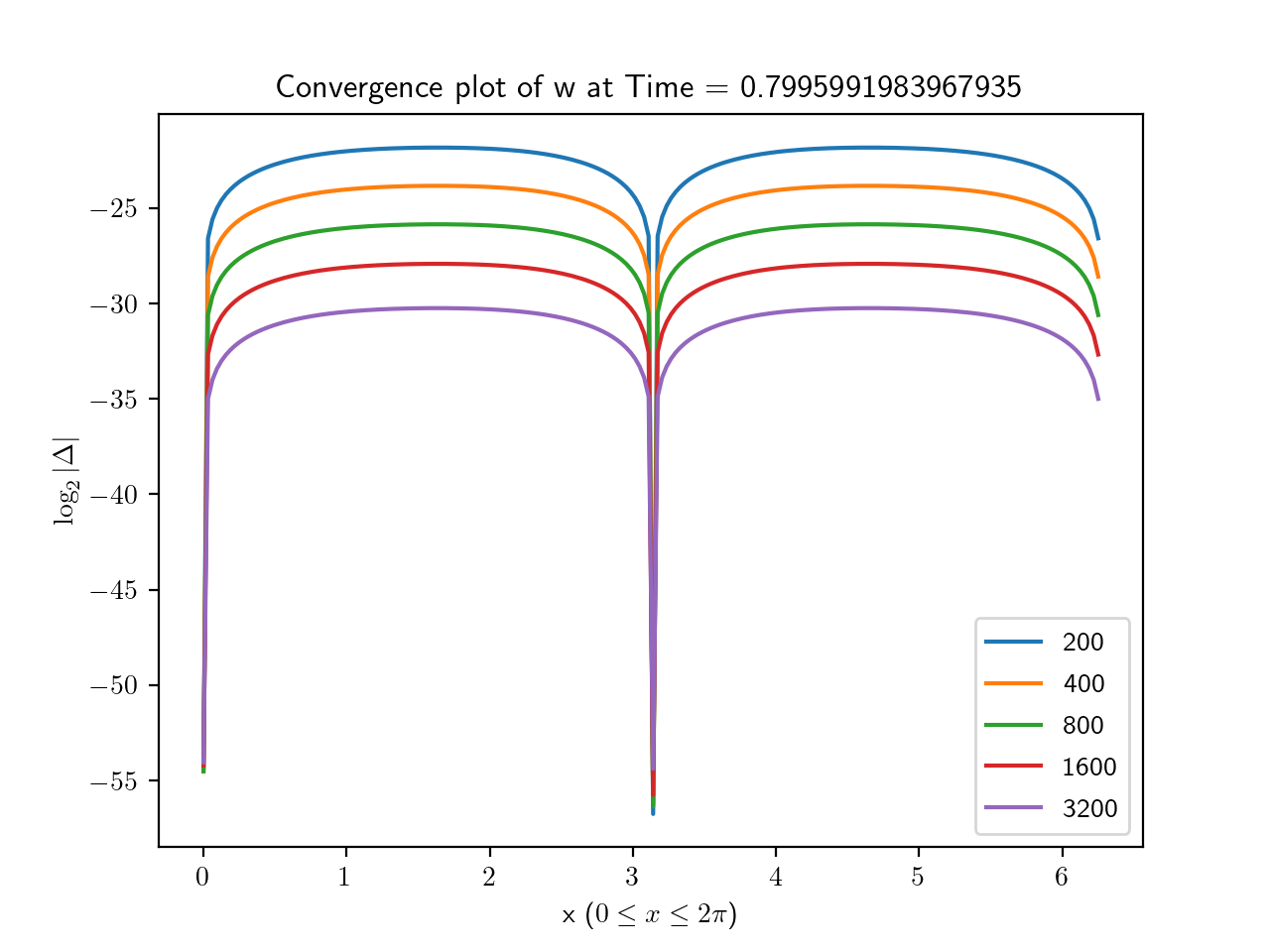}
\label{fig:subfig1}}
\subfigure[Subfigure 4 list of figures text][$t=0.198$]{
\includegraphics[width=0.3\textwidth]{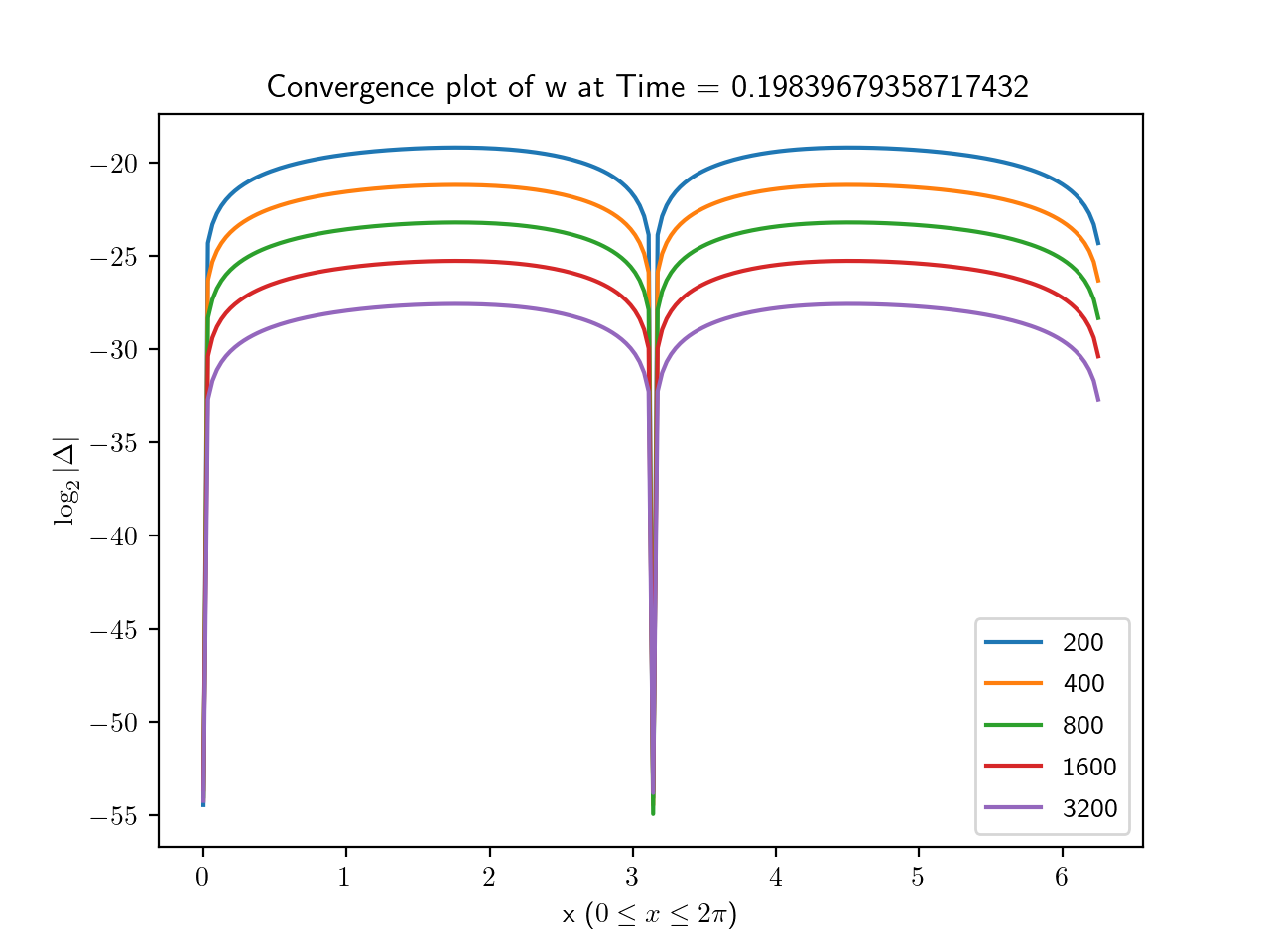}
\label{fig:subfig4}}
\subfigure[Subfigure 5 list of figures text][$t=0.028$]{
\includegraphics[width=0.3\textwidth]{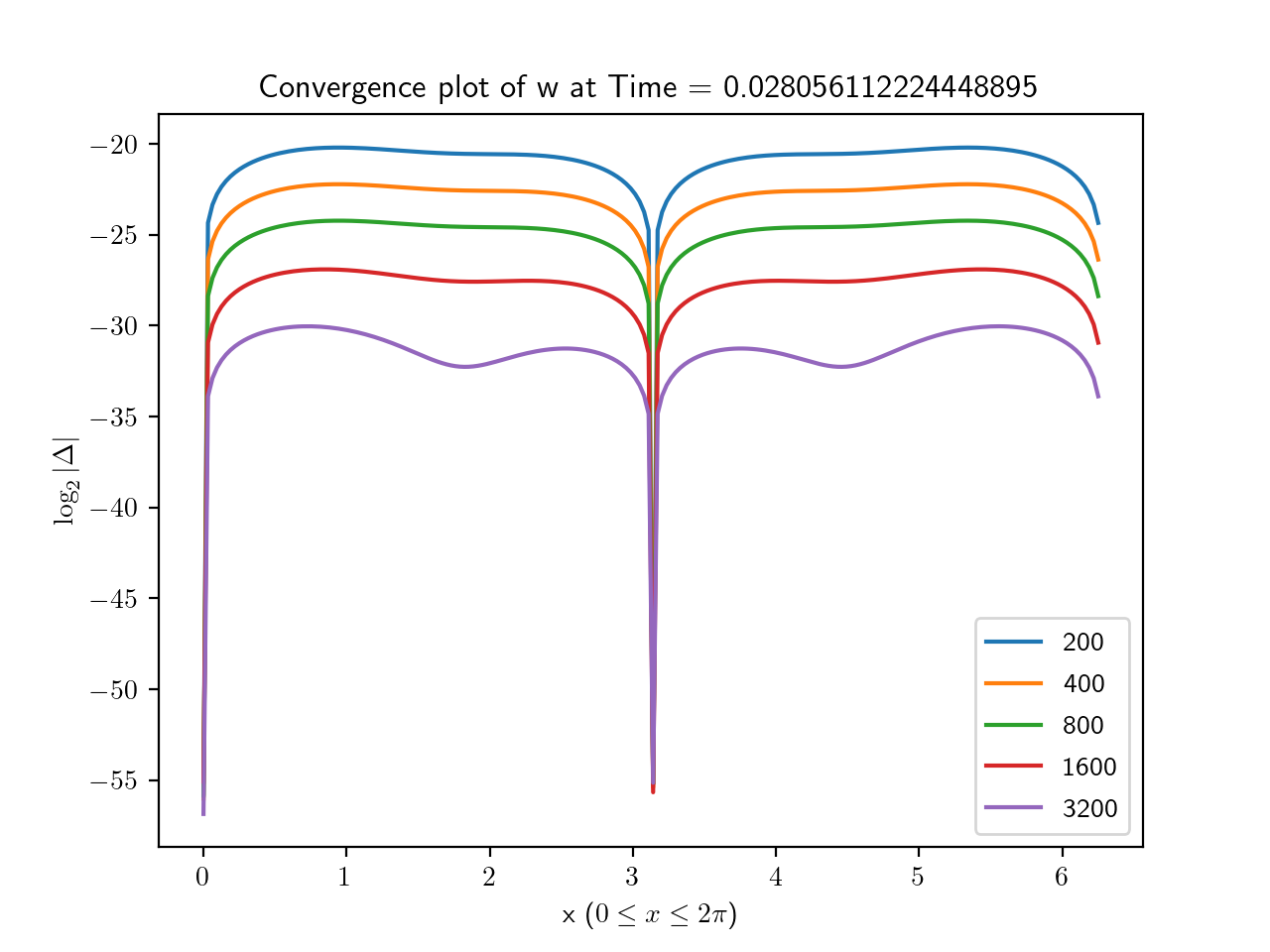}
\label{fig:subfig5}}
\caption{Convergence plots of \texttt{w} at various times. $K=0.5,(\ztt_{0},\wtt_{0}) = (0,0.1\sin(x))$.}
\label{fig:W_conv_cross}
\end{figure}

\begin{figure}[h]
\centering
\subfigure[Subfigure 1 list of figures text][$t=0.799$]{
\includegraphics[width=0.3\textwidth]{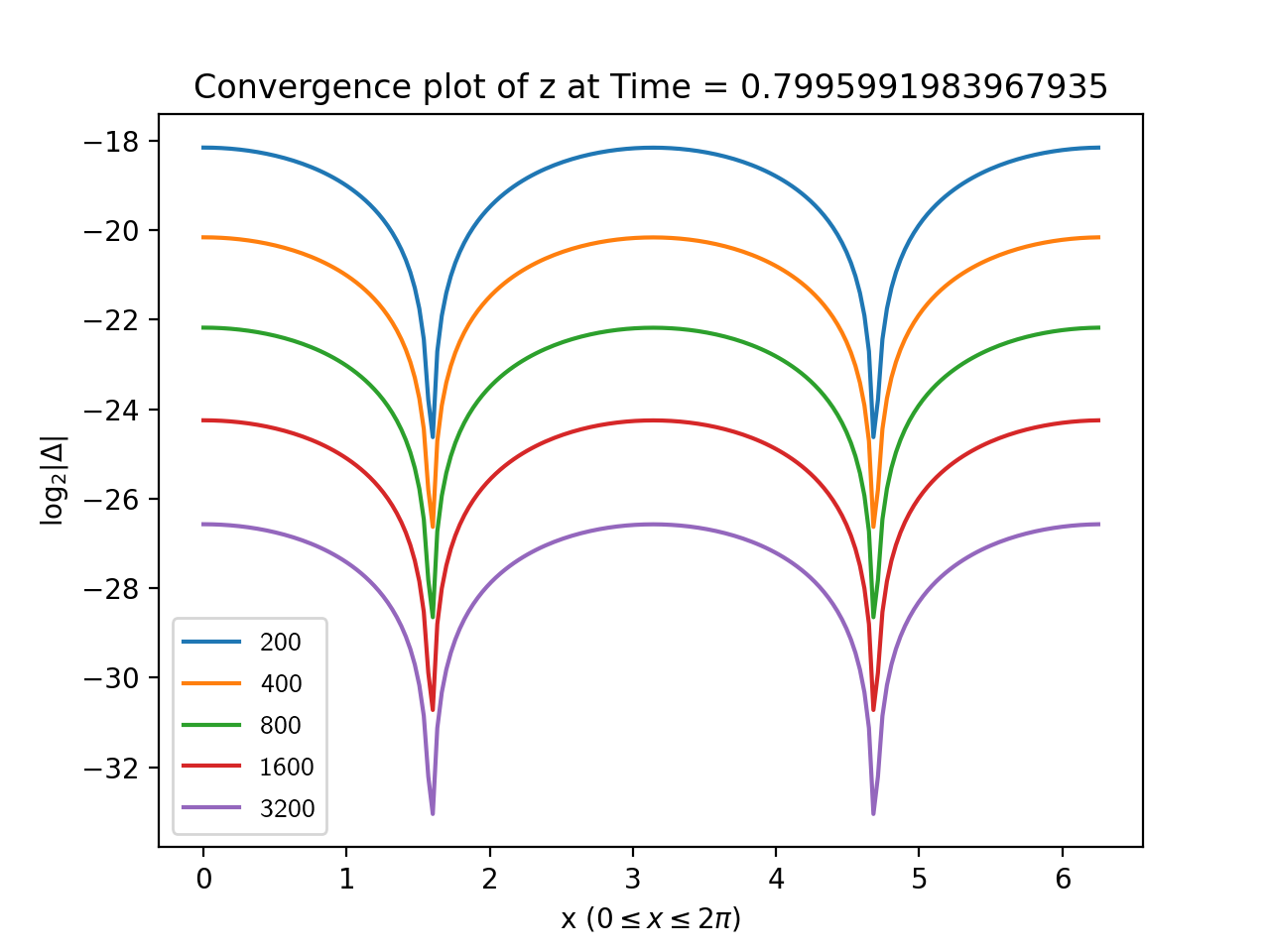}
\label{fig:subfig1}}
\subfigure[Subfigure 4 list of figures text][$t=0.198$]{
\includegraphics[width=0.3\textwidth]{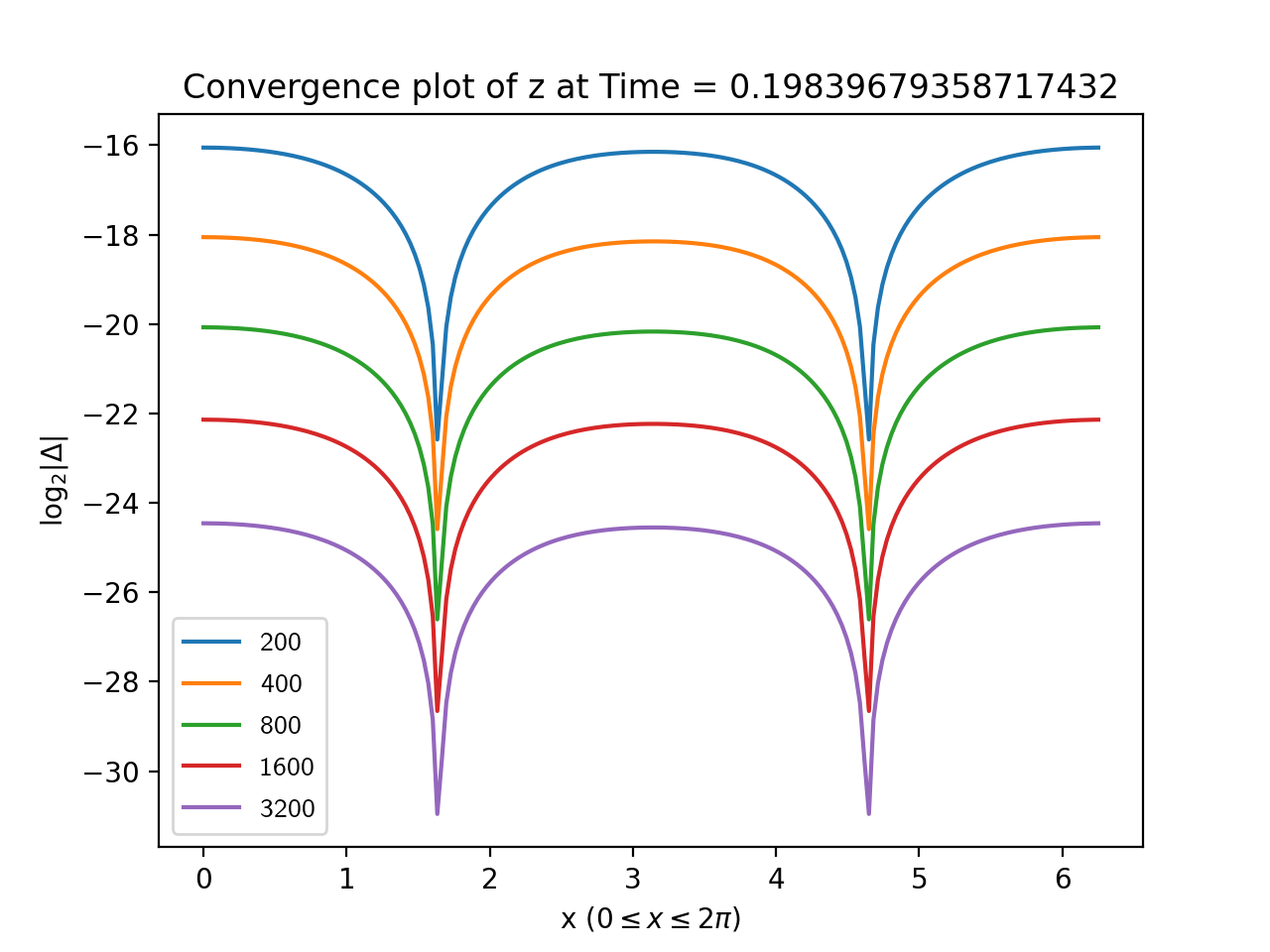}
\label{fig:subfig4}}
\subfigure[Subfigure 5 list of figures text][$t=0.028$]{
\includegraphics[width=0.3\textwidth]{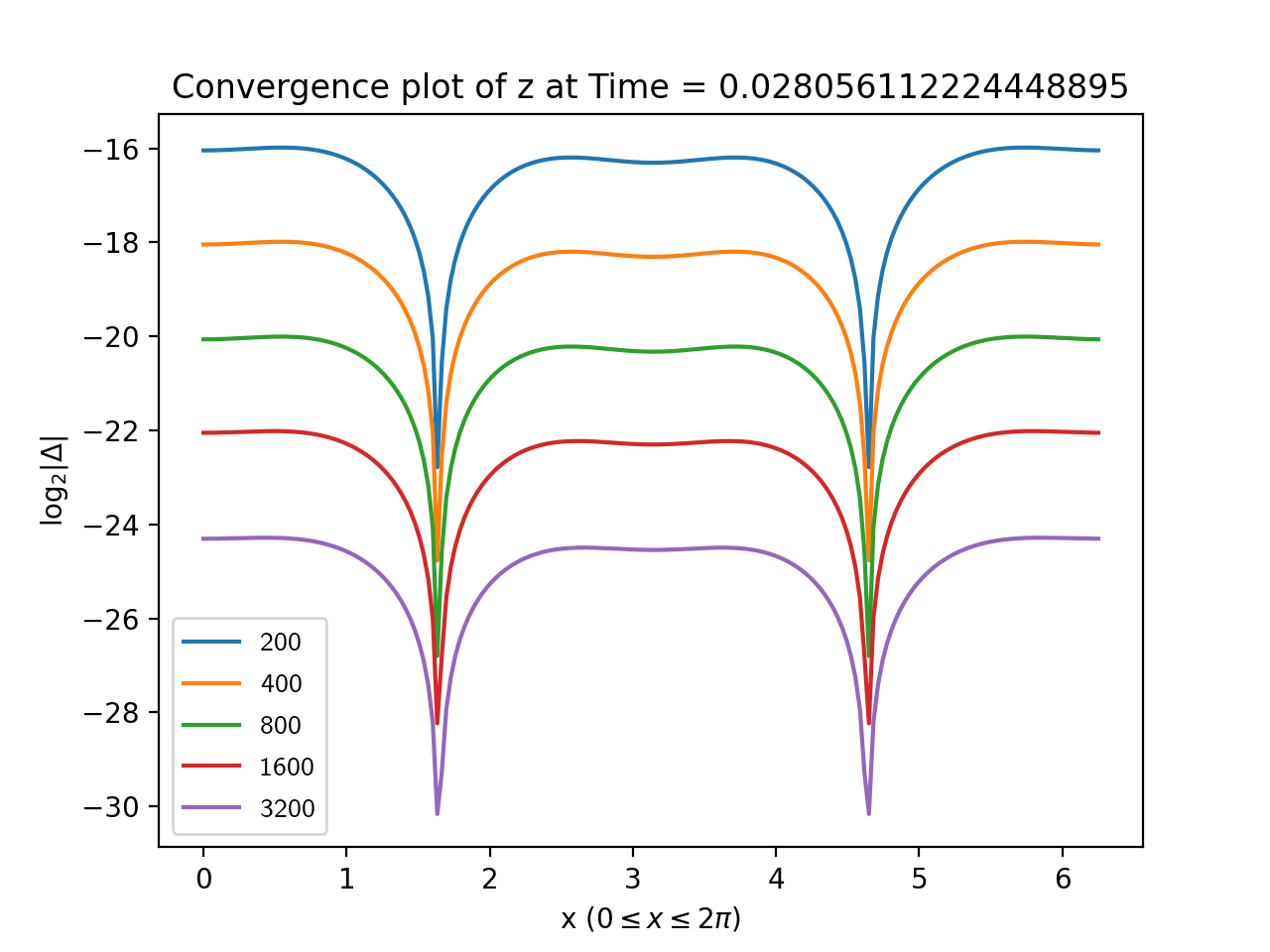}
\label{fig:subfig5}}
\caption{Convergence plots of $\ztt$ at various times. $K=0.5,(\ztt_{0},\wtt_{0}) = (0,0.1\sin(x))$.}
\label{fig:Z_conv_cross}
\end{figure}

To estimate the error, we took the base 2 log of the absolute value of the difference between each simulation and the highest resolution run. The results for are shown in Figures \ref{fig:W_conv_pos}, \ref{fig:Z_conv_pos}, \ref{fig:W_conv_cross} and, \ref{fig:Z_conv_cross} from which the second order convergence is clear. 

\FloatBarrier
\subsubsection{Code validation}
\begin{figure}[h]
\centering
\subfigure[Subfigure 1 list of figures text][Numerical test of \eqref{num_homdecay1}]{
\includegraphics[width=0.3\textwidth]{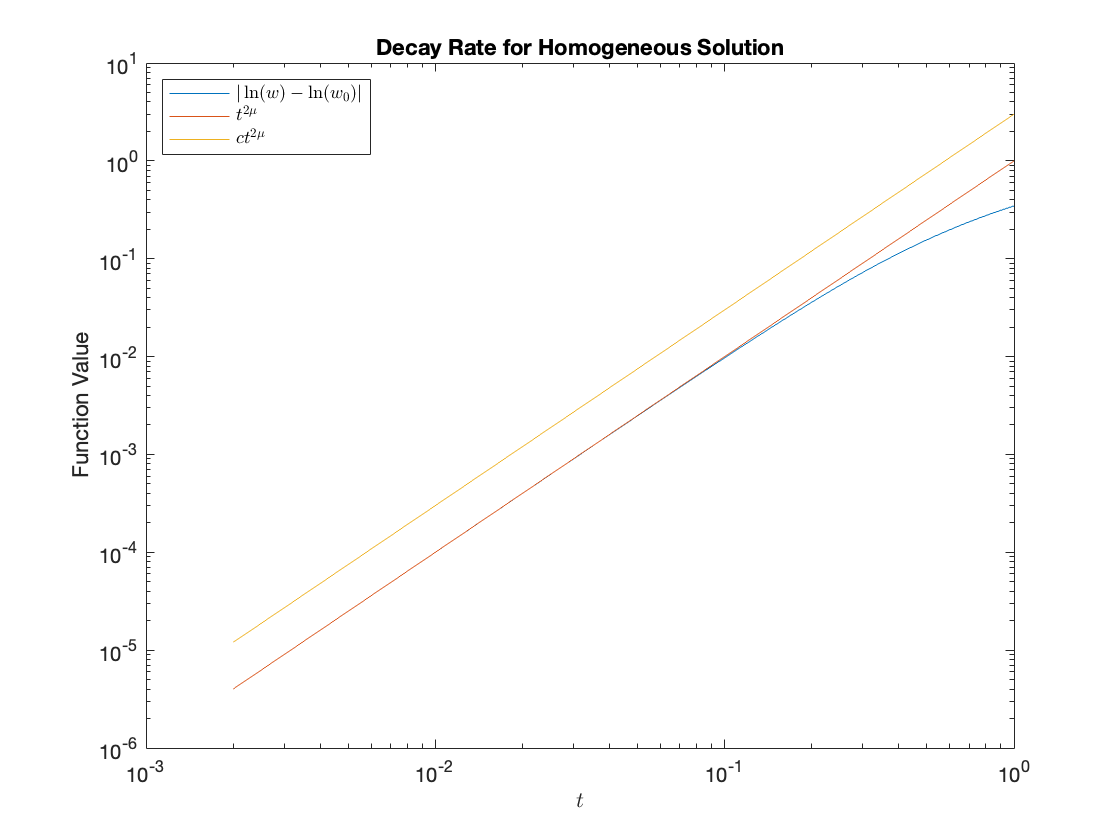}
\label{fig:subfig1}}
\subfigure[Subfigure 4 list of figures text][Numerical test of \eqref{num_homdecay2}]{
\includegraphics[width=0.3\textwidth]{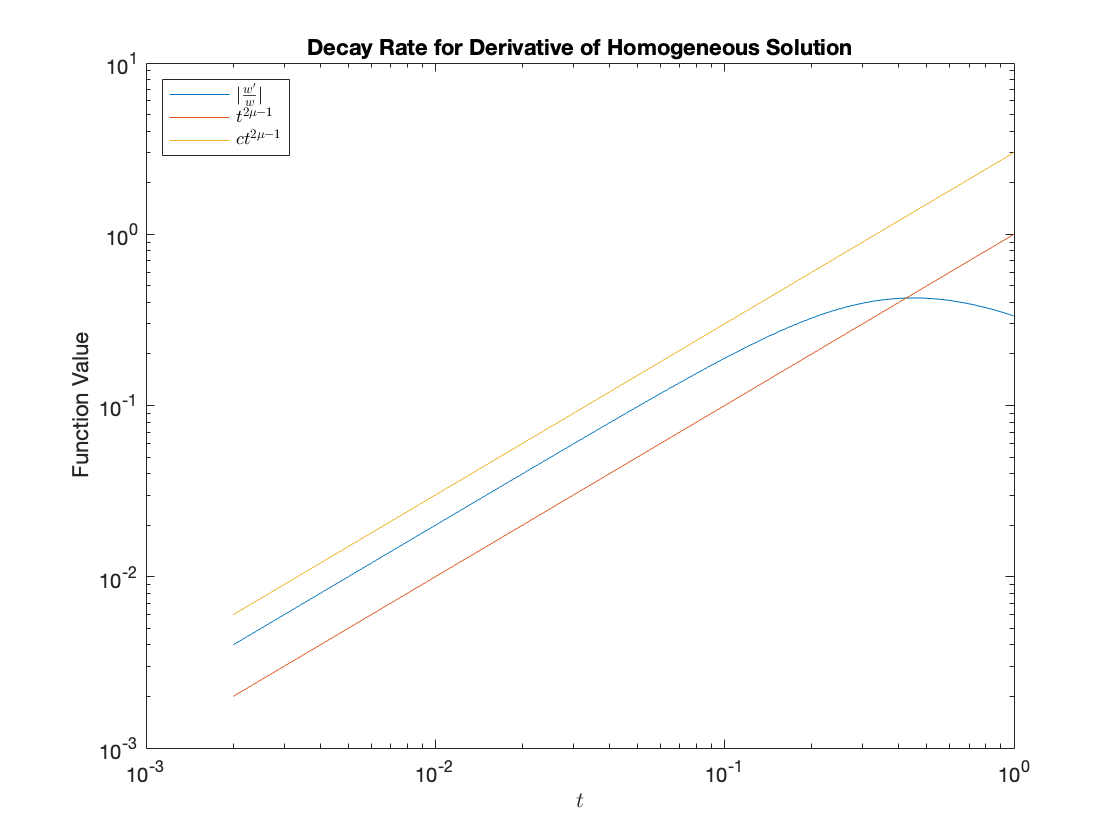}
\label{fig:subfig4}}
\subfigure[Subfigure 5 list of figures text][Numerical test of \eqref{num_nonhomdecay}]{
\includegraphics[width=0.3\textwidth]{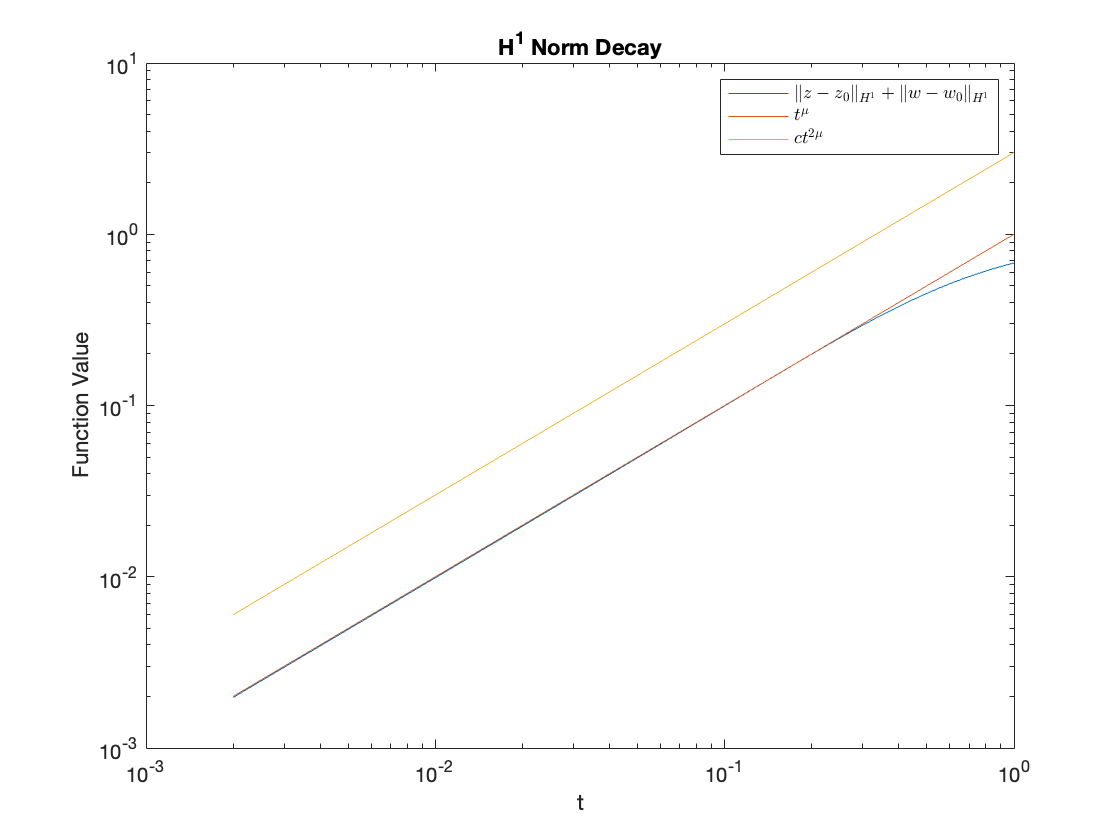}
\label{fig:subfig5}}
\caption{Log-log decay plots of numerical solutions (Blue) against the corresponding bound (Orange) and the bound multiplied by a constant $c$ (Yellow). $K = 0.5, N = 1000$. Initial data for the homogeneous solution is $(\ztt(0,x),\wtt_{H}(0,x)) = (0,1)$. Initial data for the non-homogeneous solution is $(\ztt_{0},\wtt_{0}) = (0,0.1\sin(x)+1)$.}
\label{fig:DecayRates}
\end{figure}

A simple way to test the validity of our code is to verify that numerical solutions to \eqref{eqn:dotzeta}-\eqref{eqn:dotw} that are generated from initial data $(\ztt_0,\wtt_0)$ with $\wtt_0>0$ satisfy the decay rates of Proposition \ref{Homprop} 
\begin{align}
\label{homdecay}
    |u(t)-u(0)| \lesssim t^{2\mu} \AND |u'\!(t)| \lesssim t^{2\mu-1},
\end{align}
and Theorem \ref{mainthm}
\begin{align}
\label{nonhomdecay}
\norm{\zetat(t) - \zetat_*}_{H^{k-1}}+\norm{w_1(t) - w_1^*}_{H^{k-1}}+\norm{t^\mu w_2(t) - \wb_2^*}_{H^{k}}+\norm{t^\mu w_3(t) - \wb_3^*}_{H^{k}} \lesssim t^{\mu-\sigma}, \;\; \sigma>0.
\end{align}
We first note that, by equating \eqref{cov2a} and \eqref{wttt-def} and recalling that $W = 0$ for homogeneous solutions, $u(t)$ can be expressed in terms of a homogeneous solution $\wtt_{H}(t)$ of \eqref{eqn:dotzeta}-\eqref{eqn:dotw} as $u(t) = \ln(\wtt_{H}(t))$.
The decay rates for the homogeneous solution \eqref{homdecay} can then be re-written in terms of $\wtt_{H}$ as
\begin{align}
\label{num_homdecay1}
    |\ln(\wtt_{H}(t))-\ln(\wtt_{H}(0))| &\lesssim t^{2\mu}, \\
\label{num_homdecay2}
    \Bigl|\frac{\wtt^{\prime}_{H}(t)}{\wtt_{H}(t)}\Bigr| &\lesssim t^{2\mu-1}. 
\end{align}
Similarly, for non-homogeneous solutions, we can express $w_{1}$ in terms of $\wtt$ by setting $w_{2}=w_{3}=0$ and equating \eqref{cov2a} and \eqref{wttt-def} to get
$w_{1} = \ln(\wtt(t,x)) - \ln(\wtt_{H}(t))$.
The decay rate \eqref{nonhomdecay}, in the $H^{1}$ norm, is then
\begin{align}
\label{num_nonhomdecay}
    \|\ztt(t,x) -\ztt(0,x)\|_{H^{1}}+\|[\ln(\wtt(t,x))-\ln(\wtt_{H}(t))]-[\ln(\wtt(0,x))-\ln(\wtt_{H}(0)]\|_{H^{1}} \lesssim t^{\mu-\sigma}.
\end{align}
We have estimated $\ztt|_{t=0},\wtt|_{t=0}, \wtt_{H}|_{t=0}$ by taking the values of the functions at a time-step close to $t=0$ and calculated $\wtt^{\prime}_{H}(t)$ using second order central finite differences. As shown in Figure \ref{fig:DecayRates}, the numerical solutions clearly replicate the above decay rates suggesting the code is correctly implemented.

\FloatBarrier

\subsection{Numerical behaviour}

Beyond the convergence tests, we have generated numerical solutions to the system \eqref{eqn:dotzeta}-\eqref{eqn:dotw} from a variety of initial data sets $(\ztt_0,\wtt_0)$ for which $\wtt_0$ satisfies the conditions \eqref{Hom-A-idata} and \eqref{Hom-B-idata}. 
We employed resolutions ranging from 1000 to 160,000 grid points in our simulations. For initial data satisfying \eqref{Hom-A-idata}, we chose functions $\wtt_0$ that cross the $x$-axis at least twice,\footnote{It is necessary to cross the $x$-axis at least twice to enforce the periodic boundary condition.} while for initial data satisfying \eqref{Hom-B-idata}, $\wtt_0$ does not cross the $x$-axis at all.

\begin{figure}[h]
\centering
\subfigure[Subfigure 1 list of figures text][$t=1.0$]{
\includegraphics[width=0.3\textwidth]{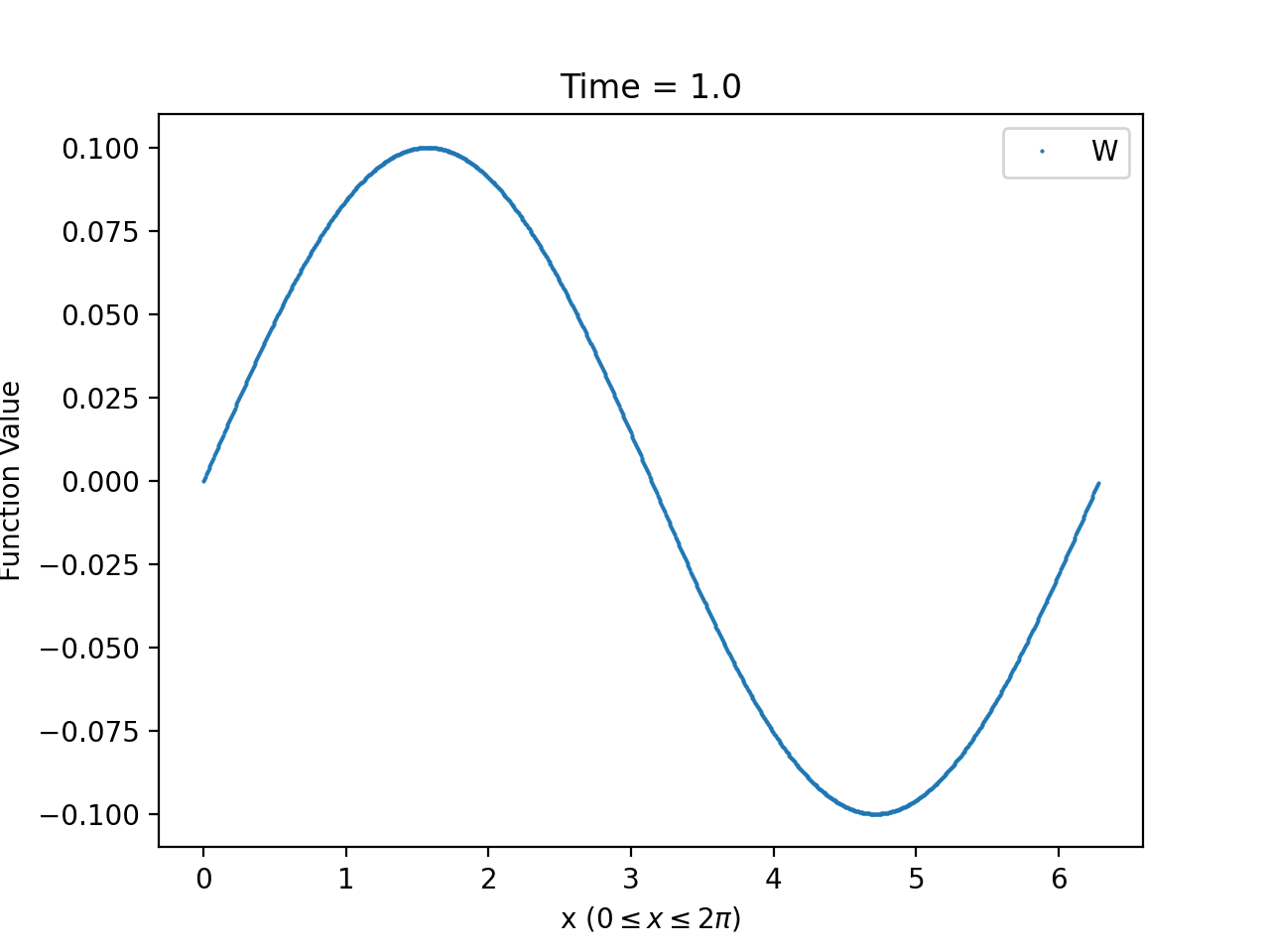}
\label{fig:subfig1}}
\subfigure[Subfigure 3 list of figures text][$t=0.017$]{
\includegraphics[width=0.3\textwidth]{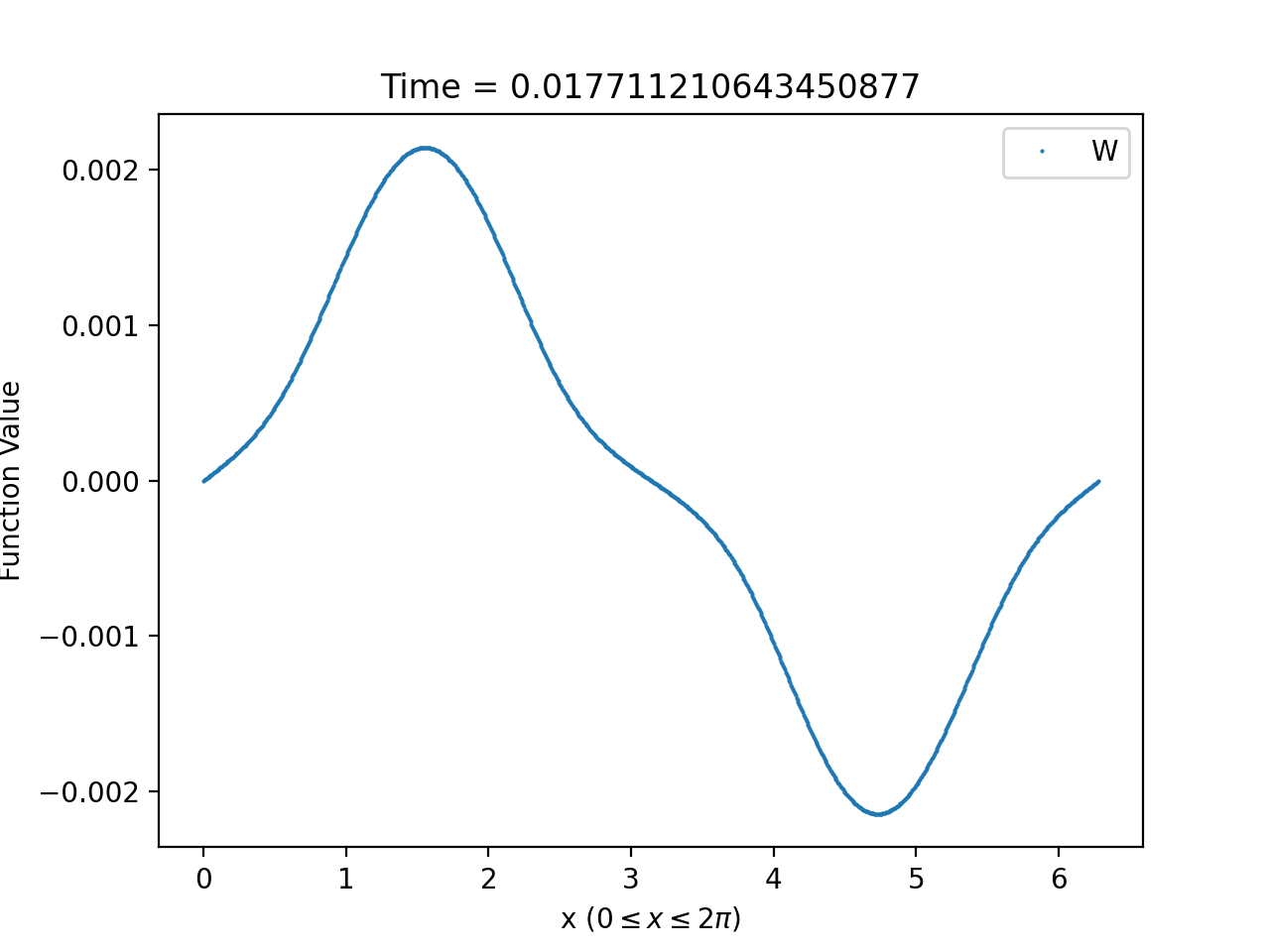}
\label{fig:subfig3}}
\subfigure[Subfigure 5 list of figures text][$t=0.0001$]{
\includegraphics[width=0.3\textwidth]{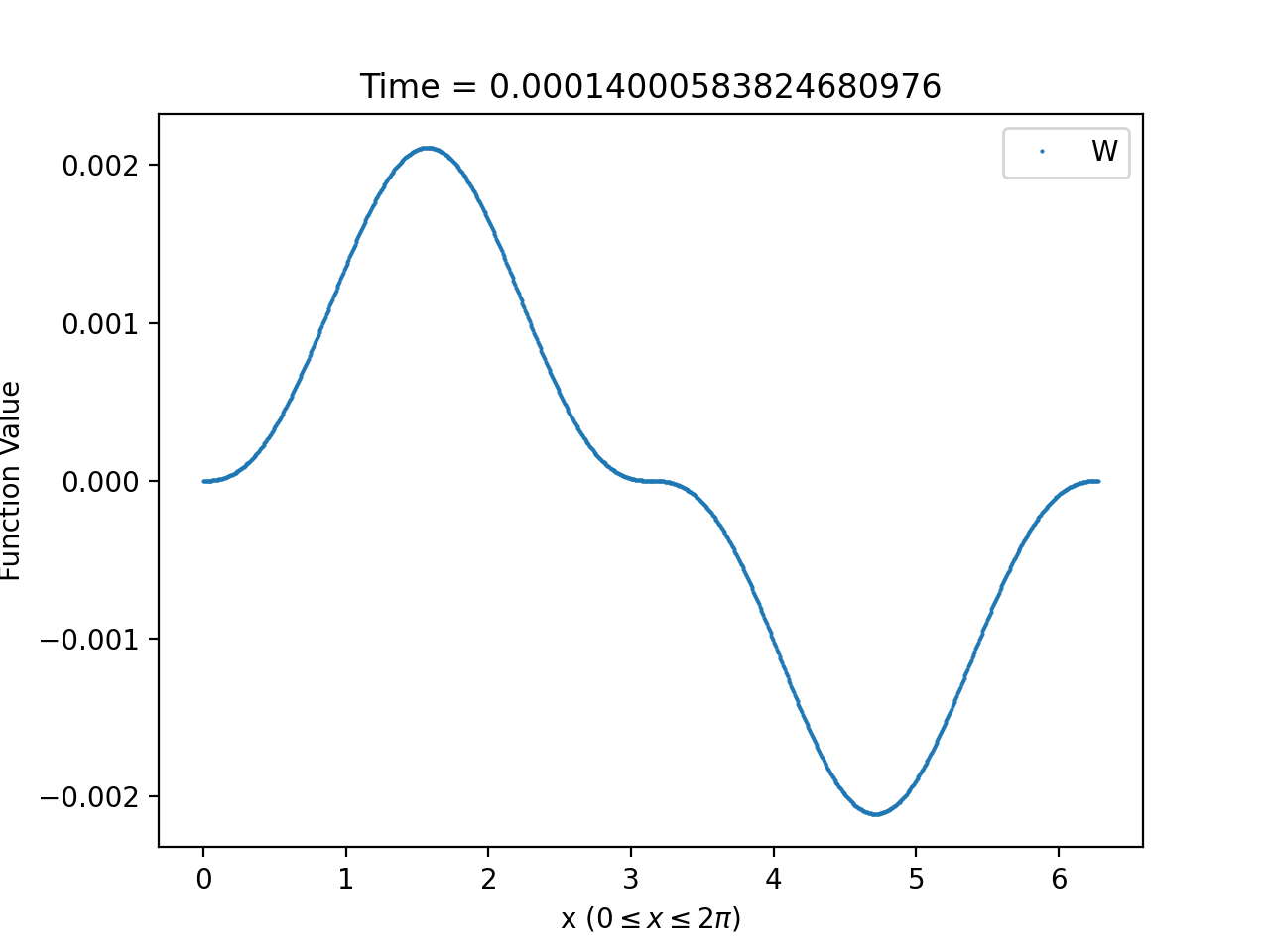}
\label{fig:subfig5}}
\caption{Plots of $\texttt{w}$ at various times. $K=0.6,\;\; N = 1000,(\ztt_{0},\wtt_{0}) = (0,0.1\sin(x))$}
\label{fig:w_evo}
\end{figure}

\begin{figure}[h]
\centering
\subfigure[Subfigure 1 list of figures text][$t=1.0$]{
\includegraphics[width=0.3\textwidth]{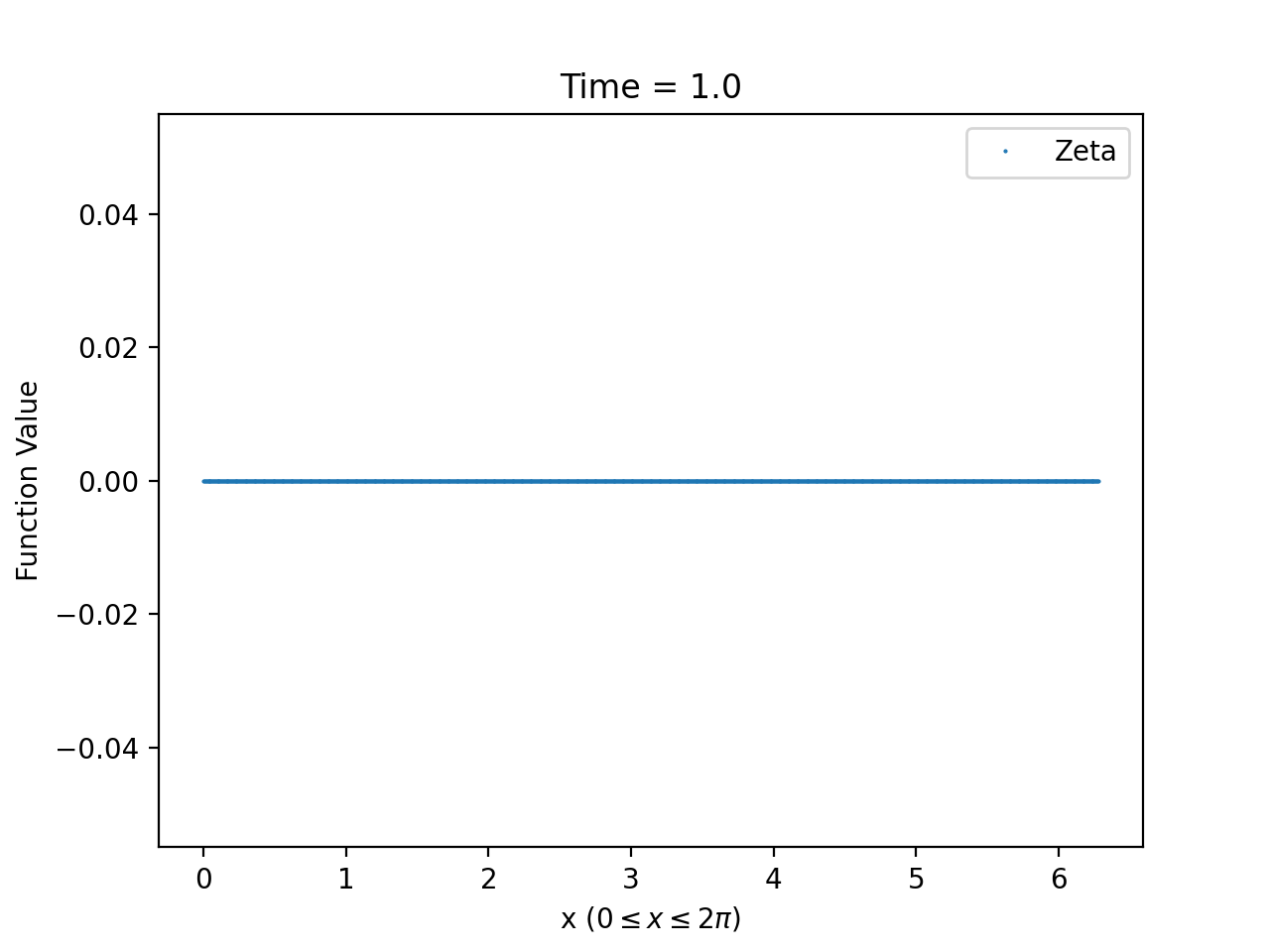}
\label{fig:subfig1}}
\subfigure[Subfigure 2 list of figures text][$t=0.088$]{
\includegraphics[width=0.3\textwidth]{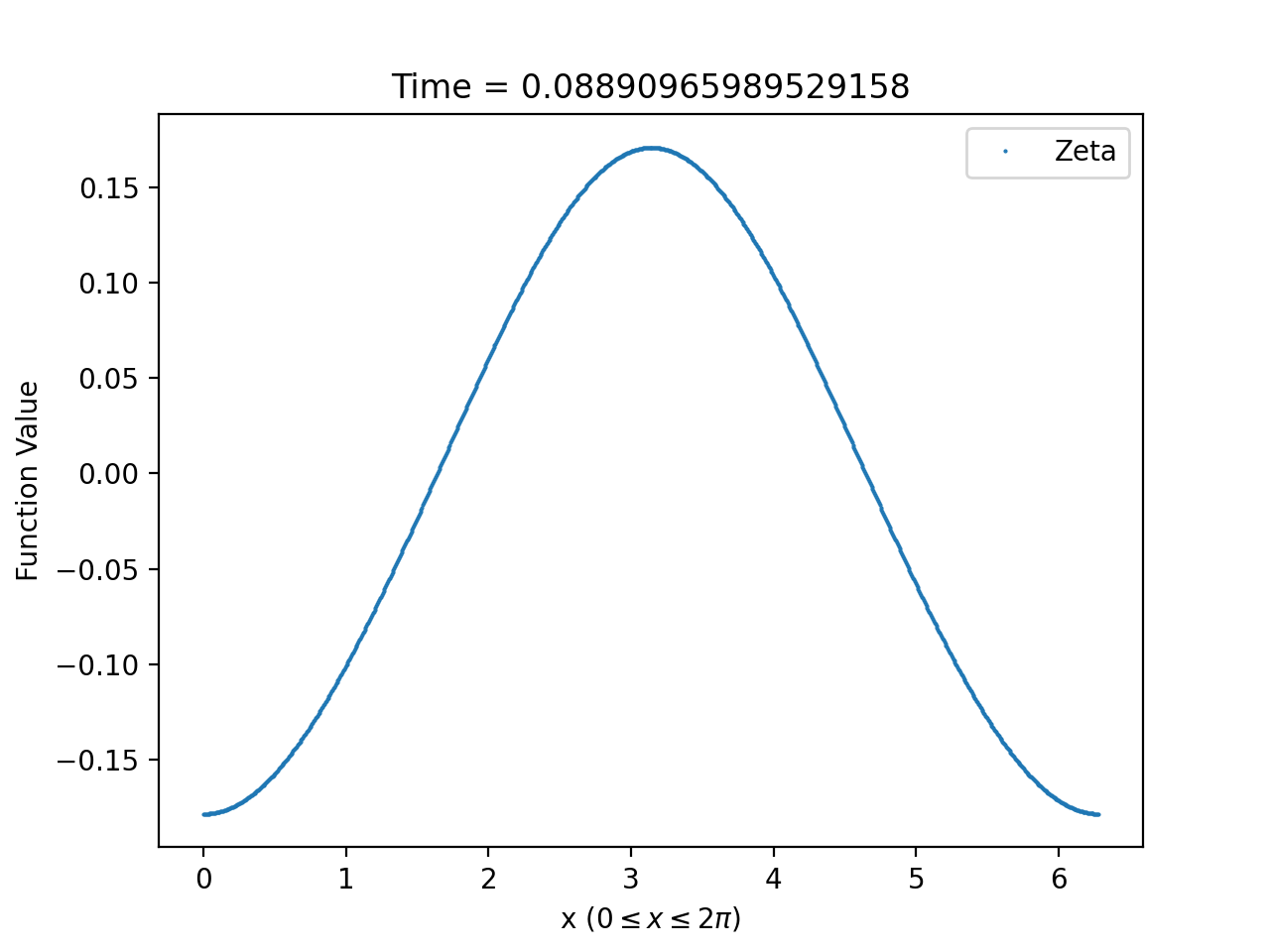}
\label{fig:subfig2}}
\subfigure[Subfigure 5 list of figures text][$t=0.0001$]{
\includegraphics[width=0.3\textwidth]{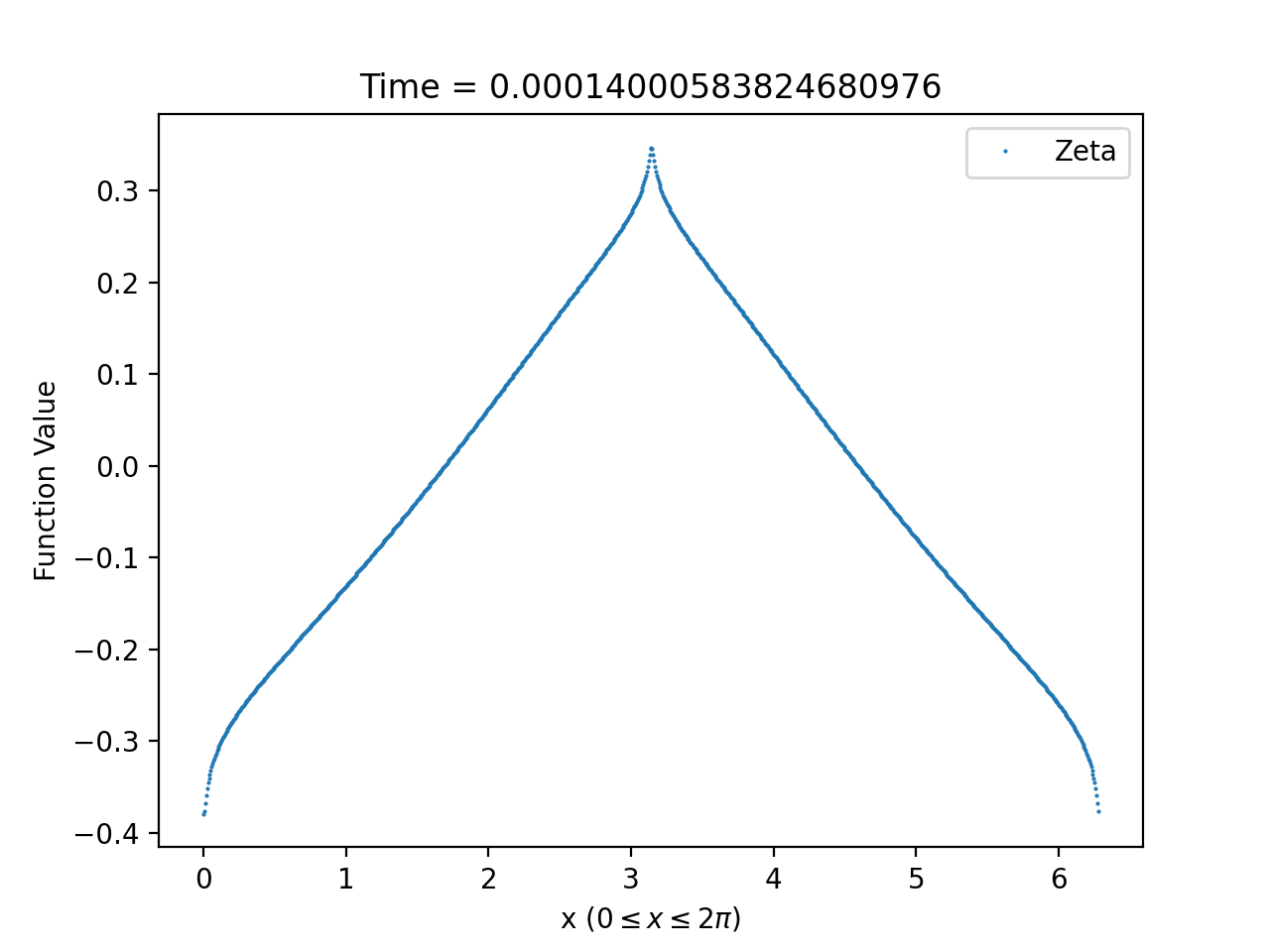}
\label{fig:subfig5}}
\caption{Plots of $\ztt$ at various times. $K=0.6,\;\; N = 1000,(\ztt_{0},\wtt_{0}) = (0,0.1\sin(x))$}
\label{fig:z_evo}
\end{figure}

All of the solutions in this article displayed in the figures are generated from initial data of the form
\begin{align*}
(\ztt_0,\wtt_{0}) = (0, a\sin(x+\theta)+c)
\end{align*}
for some particular choice of the constants $a,c, \theta \in \mathbb{R}$. From our numerical solutions, we observe, for the full parameter range $1/3<K<1$ and all choices of the initial data with $a$ sufficiently small, that $\ztt$ and
$\wtt$ 
remain bounded and converge pointwise as $t\searrow 0$; see Figures \ref{fig:w_evo} and \ref{fig:z_evo}.

\subsubsection{Derivative blow-up at $t=0$}
While $\ztt$ and $\wtt$ remain bounded, our numerical simulations reveal that derivatives of the solutions of sufficiently high order blow-up at $t=0$ 
for the parameter values $1/3<K<1$ and initial data satisfying \eqref{Hom-A-idata}. In Table \ref{ell-table}, we list, for a selection of $K$ values, the corresponding minimum value of $\ell$ for which  $\sup_{x\in \Tbb^1}\bigl(|\del{x}^{\ell} \ztt(t,x)|+|\del{x}^{\ell}\wtt(t,x)|\bigr) \nearrow \infty$ as $t\searrow 0$. 
From these values, it appears that  $\ell$ is a monotonically decreasing function of $K$.
\begin{table}[h] 
\begin{tabular}{|l|l|l|l|l|l|l|l|l|l|}
\hline
$K$ & $0.40$ & 
$0.45$ & $0.50$ & $0.55$ & $0.60$ & $0.65$ & $0.75$ & $0.85$ & $0.95$ \\ \hline
$\ell$ & $4$    & $3$    & $2$    & $1$    & $1$    & $1$ & $1$    & $1$    & $1$    \\ \hline
\end{tabular}
\caption{Observed value of $\ell$ for various $K$}\label{ell-table}
\end{table}
\FloatBarrier

\subsubsection{Asymptotic behaviour and approximations\label{sec:asymp}}
For the full range of parameter $1/3<K<1$ and all choices of initial data, we observe that our numerical solutions display ODE-like behaviour near $t=0$. In particular, these solutions can be approximated by solutions of the asymptotic system \eqref{zttt-asympt}-\eqref{wttt-asympt} at late times using the following procedure:
\begin{enumerate}[(i)]
\item  Generate a numerical solution $(\ztt,\wtt)$ of \eqref{eqn:dotzeta}-\eqref{eqn:dotw} from initial data $(\ztt_{0},\wtt_{0})$ specified at time $t_{0}>0$.
\item Fix a time $\tilde{t}_{0} \in (0,t_{0})$ when the numerical solution $(\ztt,\wtt)$ first appears to be dominated by ODE behaviour.
\item Fix initial data for the asymptotic system \eqref{zttt-asympt}-\eqref{wttt-asympt} at $t=\tilde{t}_{0}$ by setting 
\begin{equation*}
(\zttt_{0},\wttt_{0}) = (\ztt,\wtt)|_{t=\ttld_0}.
\end{equation*}
\item Solve the asymptotic system \eqref{zttt-asympt}-\eqref{wttt-asympt} with initial data as chosen above in (iii) to
obtain the asymptotic solution 
$(\zttt,\wttt)$
where
\begin{equation}\label{zttt-sol}
\zttt = \zttt_0, 
\end{equation}
and $\wttt$ is defined implicitly by
\begin{align}\label{wttt-sol}
\frac{(3 K-\mu -1) \ln \left((3 K-1) t^{2 \mu }-(K-1) \mu \wttt^2\right)}{2 (3 K-1) \mu }-\frac{\ln (|\wttt|(1-3K)}{1-3 K}=c
\end{align}
and
\begin{align*}
c = \frac{(3 K-\mu -1) \ln \left((3 K-1) \tilde{t}_{0}^{2 \mu }-(K-1) \mu \wttt_{0}^2\right)}{2 (3 K-1) \mu }-\frac{\ln (|\wttt_{0}|(1-3K))}{1-3 K}.
\end{align*}
\item Compare the numerical solution $(\ztt,\wtt)$ to the asymptotic solution $(\zttt,\wttt)$ on the region $(0,\ttld_0)\times \Tbb^1$. 
\end{enumerate}

\bigskip

Using this procedure, we find that
numerical solutions $(\ztt,\wtt)$ of the system \eqref{zttt-asympt}-\eqref{wttt-asympt}  can be \textit{remarkably well-approximated} by solutions $(\zttt,\wttt)$ of the asymptotic system. In particular, by setting $t=0$ in \eqref{wttt-sol} and noting that we can solve for $\wttt|_{t=0}$ to get 
\begin{equation}\label{wttf-def}
\wttt_{f} := \wttt|_{t=0}=\frac{\sgn(\wttt_{0})|\wttt_0|^{\frac{1}{1-K}}}{(\ttld_{0}^{2\mu}+\wttt_{0}^{2})^{\frac{K}{2(1-K)}}}  
\end{equation}
where $\sgn(x)$ is the sign function,
we have, with the help of \eqref{zttt-sol}, that
\begin{equation} \label{asymp-sol-t=0}
(\ztt,\wtt)|_{t=0} \approx (\zttt_0, \wttt_f).
\end{equation}
It is worth noting that this ODE-like asymptotic behaviour of solutions generated from initial data satisfying \eqref{Hom-B-idata} is expected by Theorem \ref{mainthm}. What is interesting is that this
behaviour of solutions persists for initial data that violates \eqref{Hom-B-idata}. 

\begin{figure}[h]
\centering
\subfigure[Subfigure 3 list of figures text][$t=0.039$]{
\includegraphics[width=0.3\textwidth]{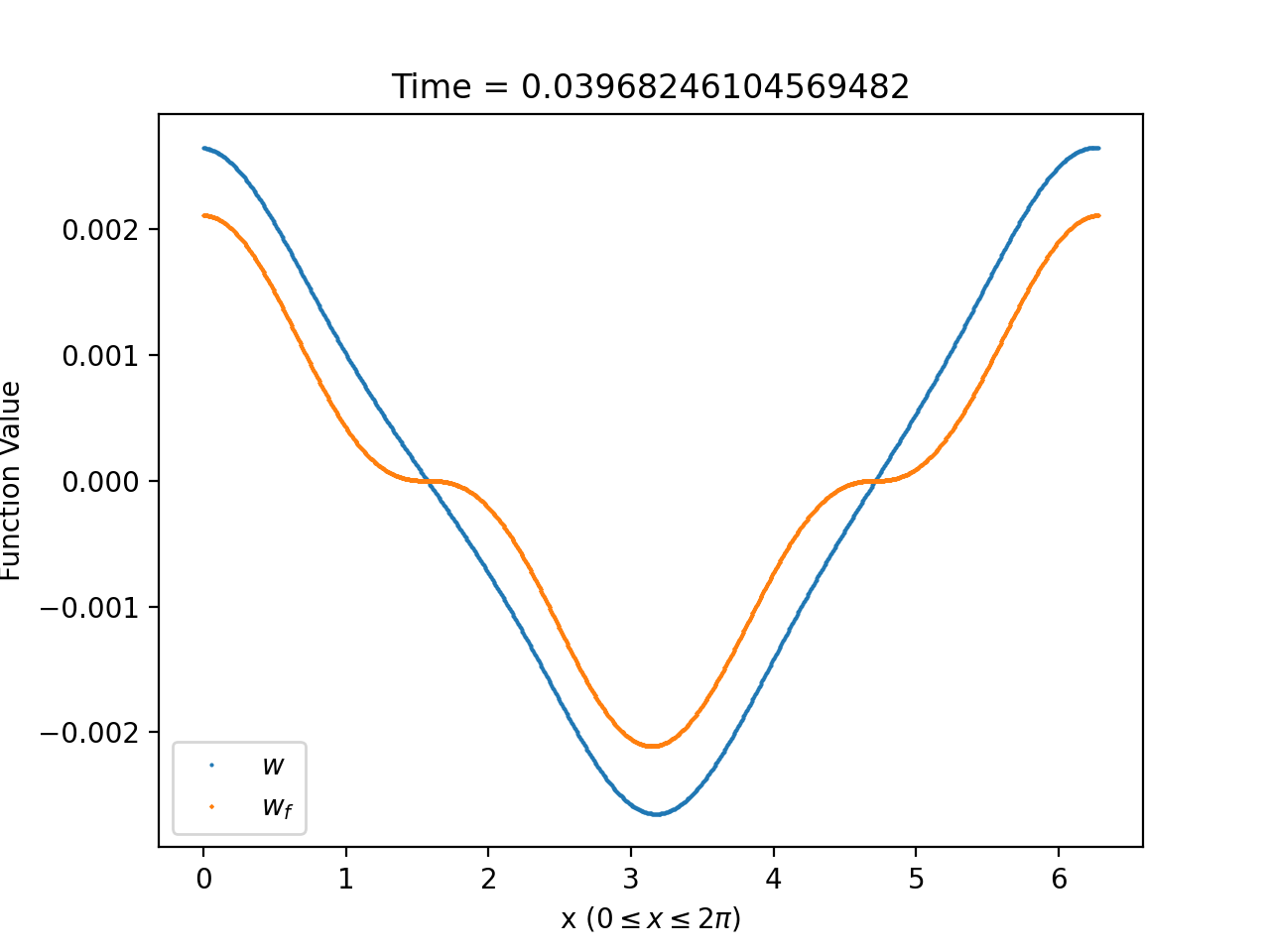}
\label{fig:subfig3}}
\subfigure[Subfigure 4 list of figures text][$t=0.024$]{
\includegraphics[width=0.3\textwidth]{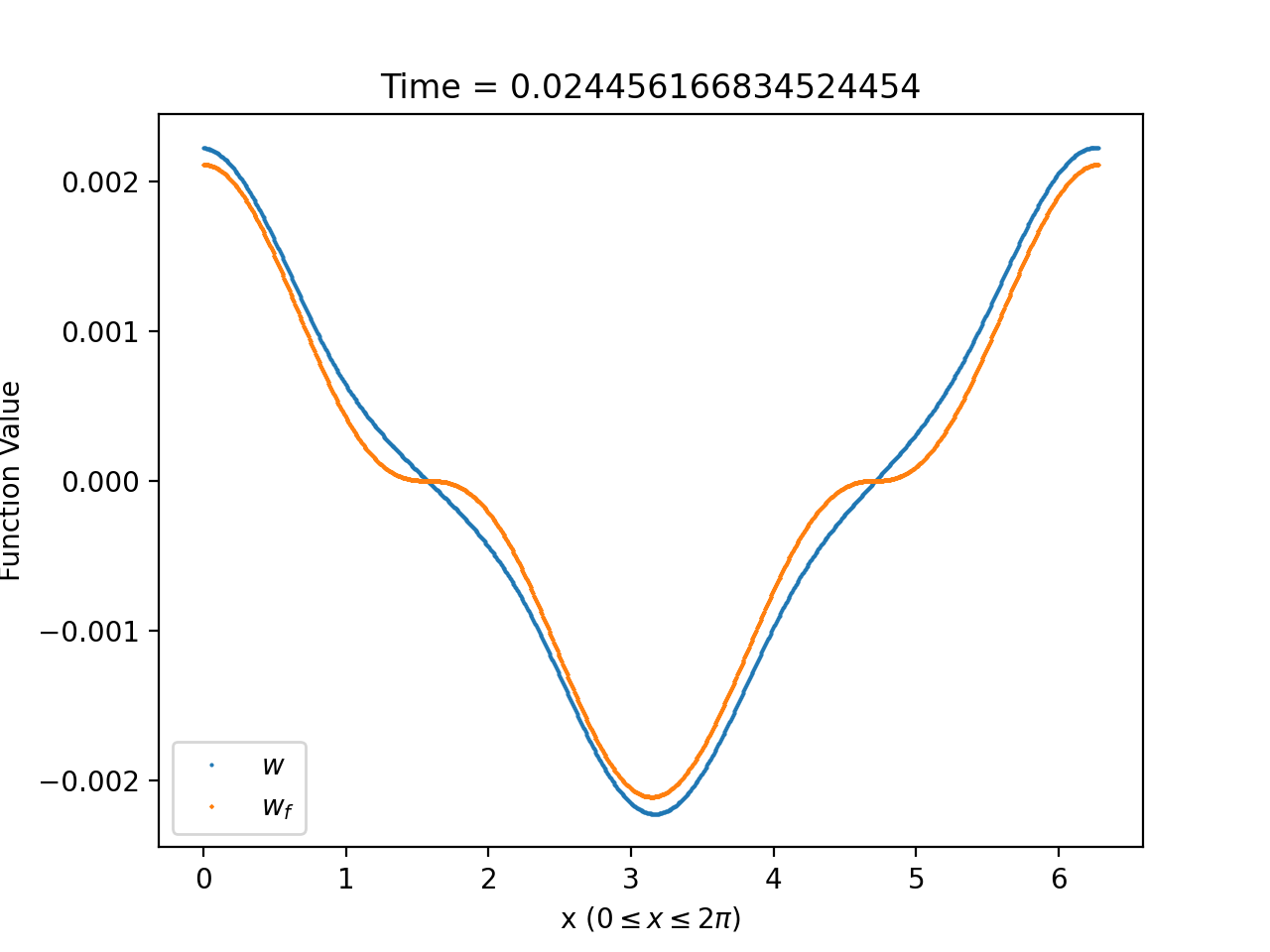}
\label{fig:subfig4}}
\subfigure[Subfigure 5 list of figures text][$t=0.007$]{
\includegraphics[width=0.3\textwidth]{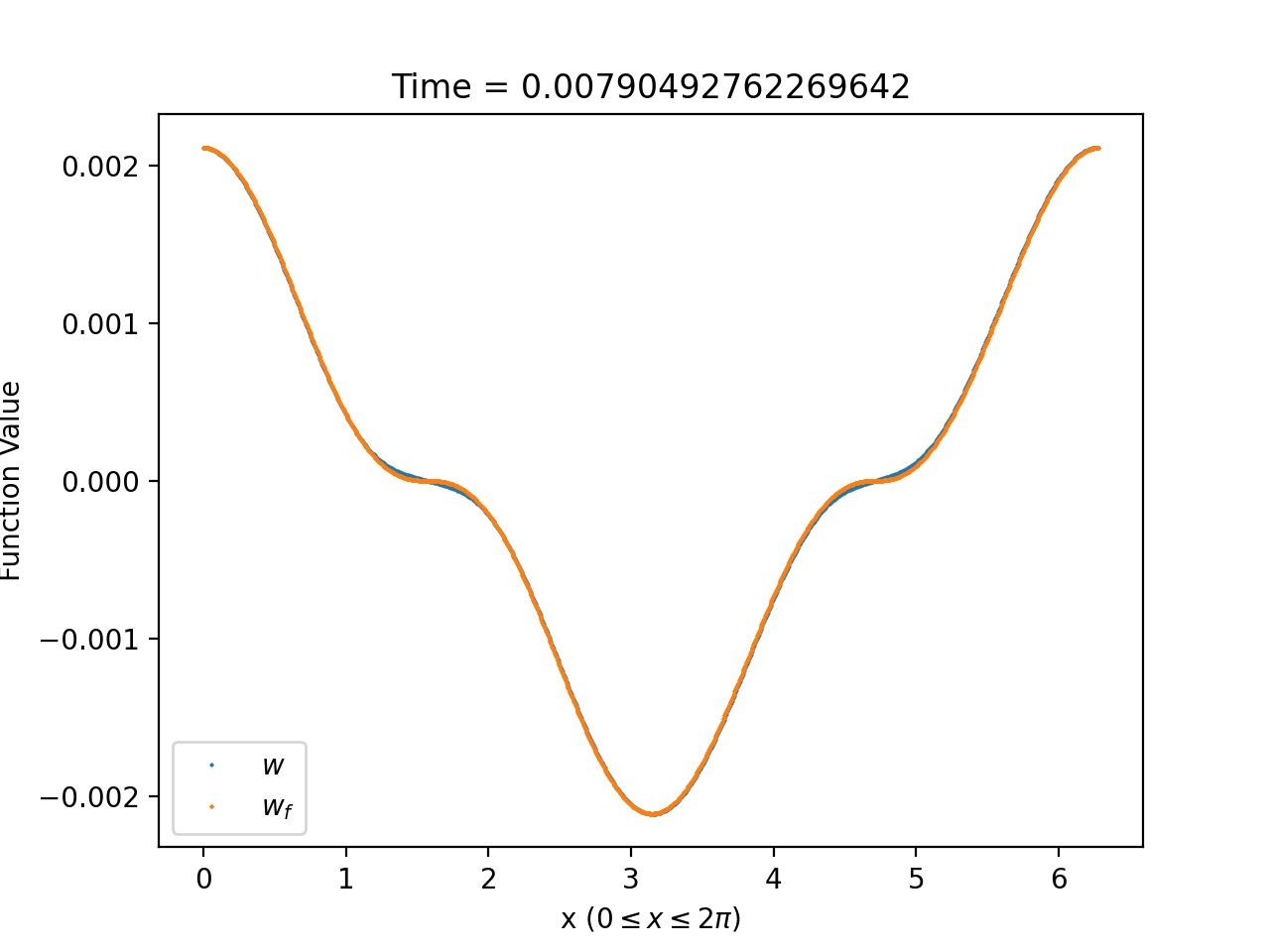}
\label{fig:subfig5}}
\caption{Comparison of numerical solution $\texttt{w}$ (Blue) and $\wttt_{f}$ (Orange). $K=0.6,\;\;N = 1000,\;\;(\ztt_{0},\wtt_{0}) = (0,0.1\cos(x))$, $(\tilde{t}_{0},\wttt_{0}) = (9.93\times 10^{-4},\wtt|_{t=9.93\times 10^{-4}})$.}
\label{fig:HomogFull_Compare}
\end{figure}

To illustrate how well solutions $(\ztt,\wtt)$ of \eqref{eqn:dotzeta}-\eqref{eqn:dotw}
can be approximated by solutions $(\zttt,\wttt)$ of the asymptotic system \eqref{zttt-asympt}-\eqref{wttt-asympt} near $t=0$, we compare in Figure \ref{fig:HomogFull_Compare}
the plot of $\wttt_f=\wttt|_{t=0}$, for a fixed choice of $\ttld_0$ and $\wttt_0$ (see \eqref{wttf-def}), with that of $\wtt(t)$ at times close to zero. From the figure, it is clear that the agreement is almost perfect for times close enough to zero.
\FloatBarrier

\subsubsection{Behaviour of the density contrast}
\begin{figure}[h]
\centering
\subfigure[Subfigure 1 list of figures text][$t=1.0$]{
\includegraphics[width=0.3\textwidth]{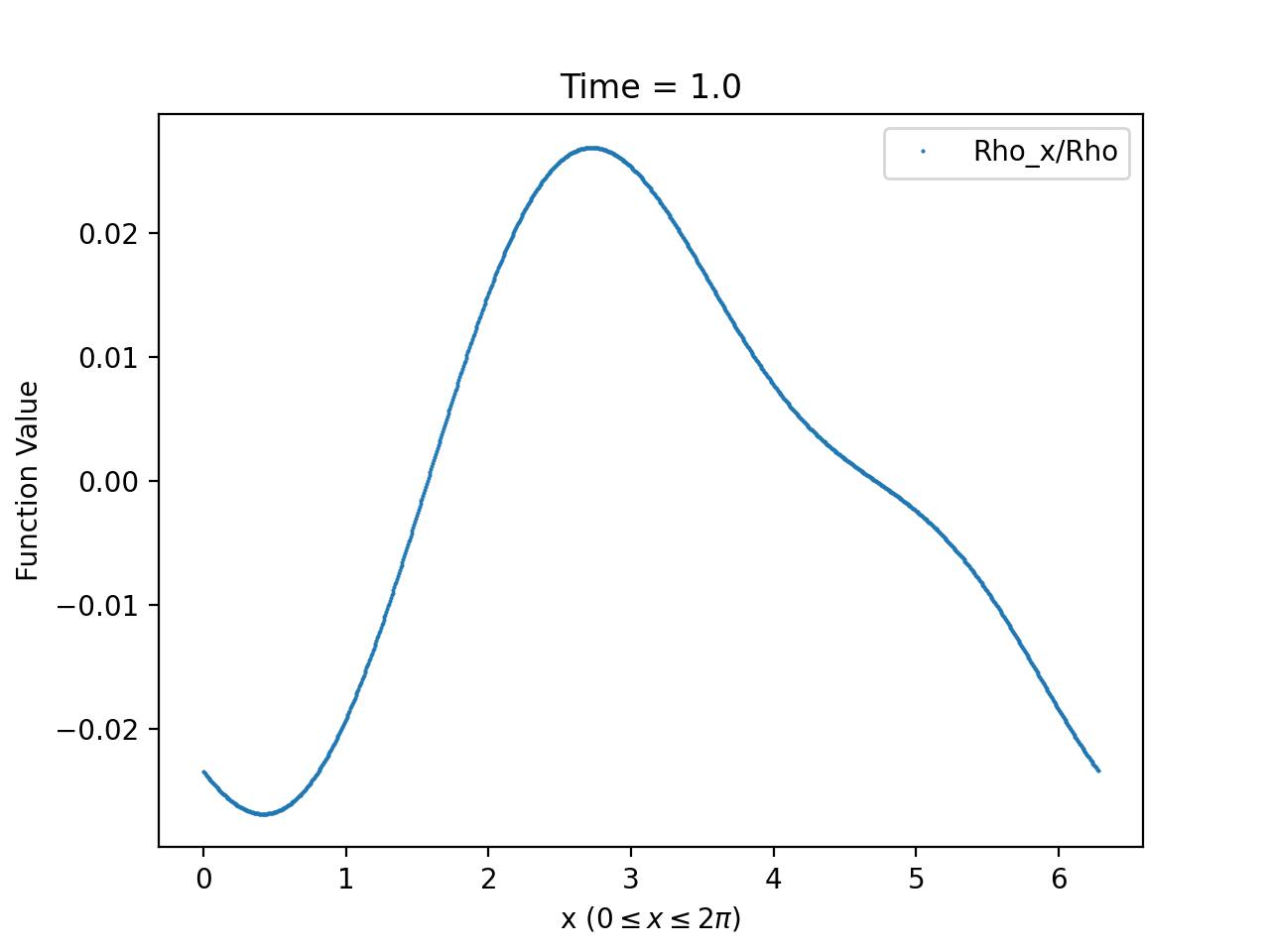}
\label{fig:subfig1}}
\subfigure[Subfigure 3 list of figures text][$t=0.199$]{
\includegraphics[width=0.3\textwidth]{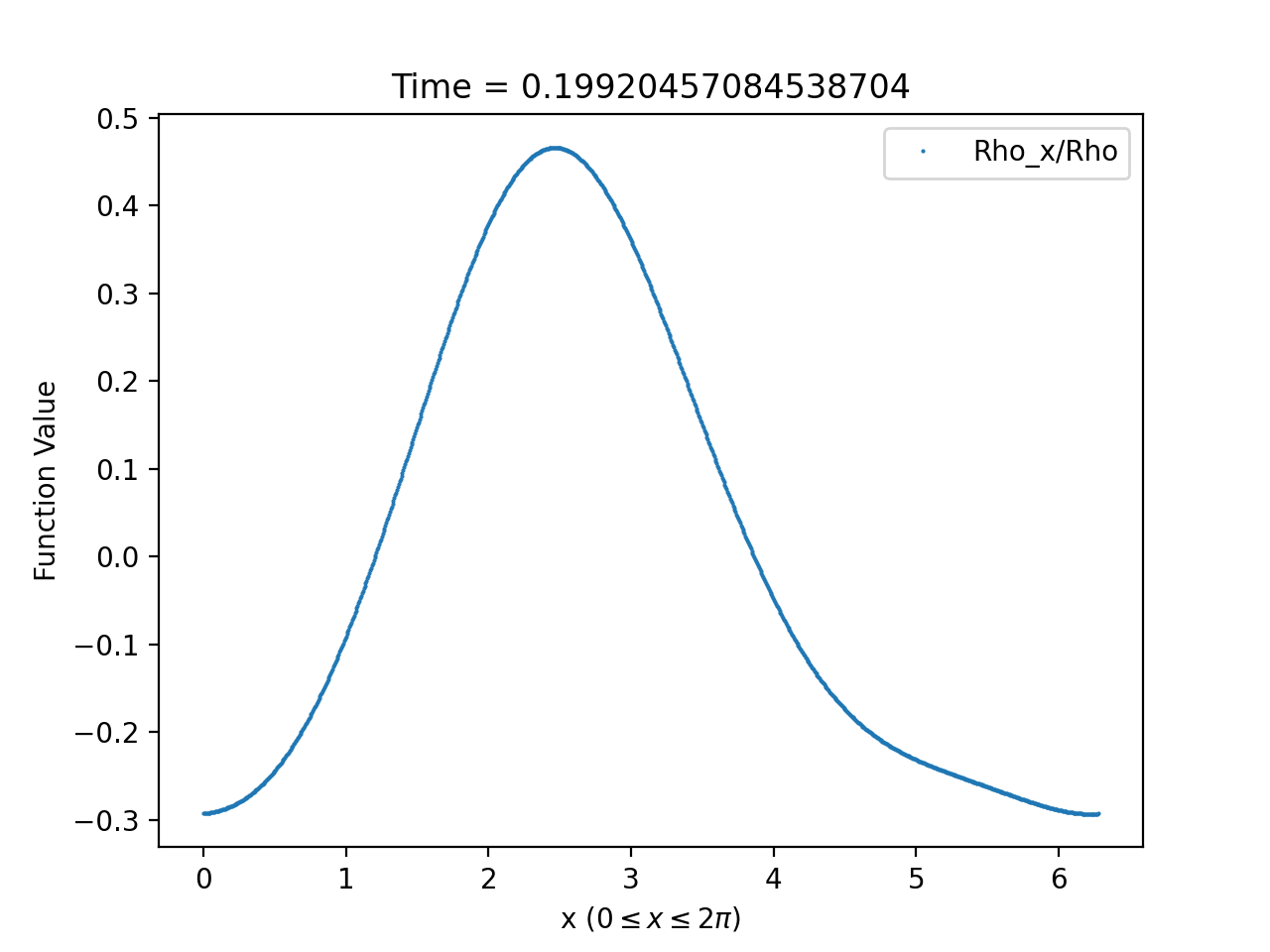}
\label{fig:subfig3}}
\subfigure[Subfigure 5 list of figures text][$t=6.14 \times 10^{-12}$]{
\includegraphics[width=0.3\textwidth]{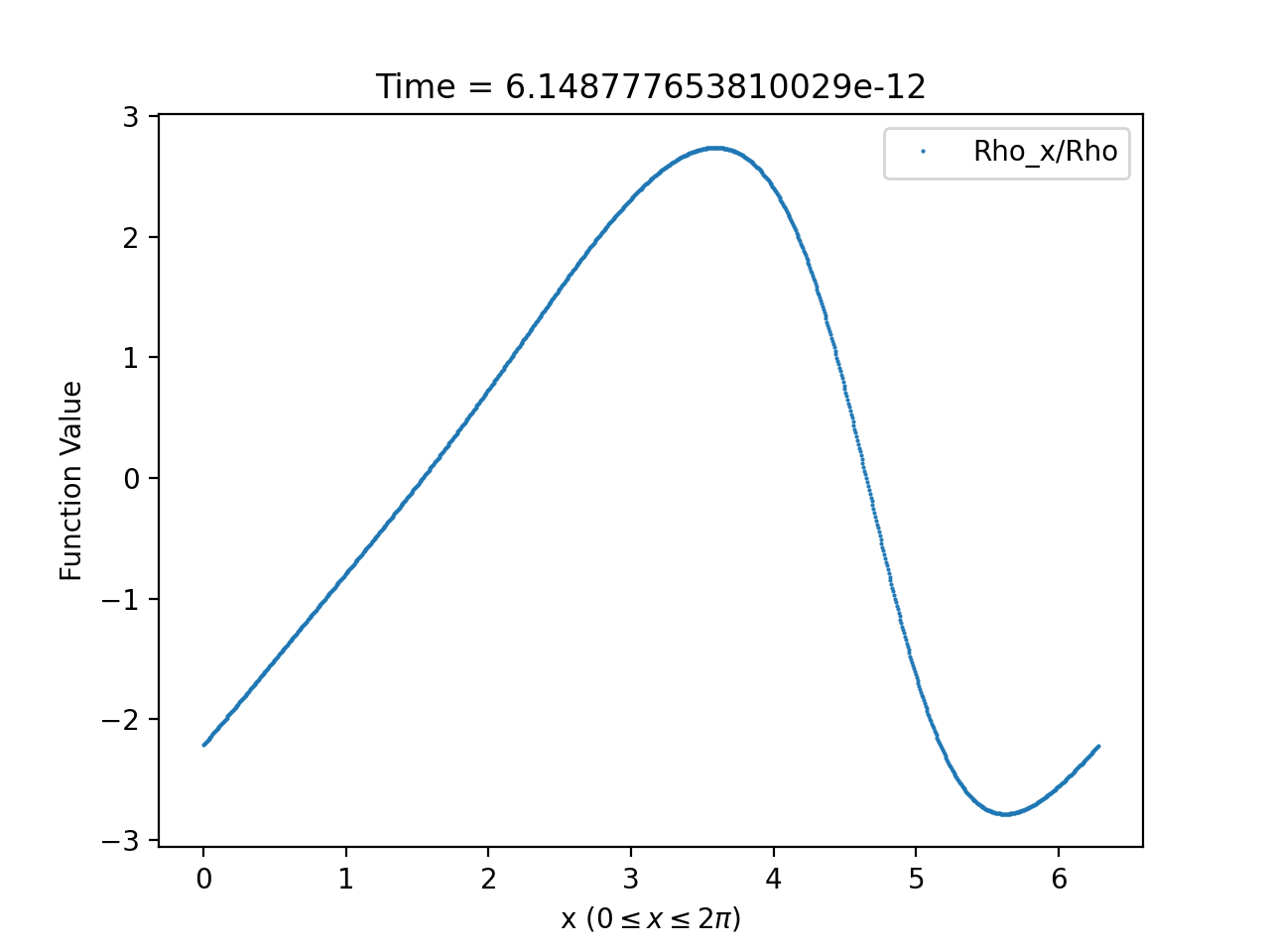}
\label{fig:subfig5}}
\caption{Plots of density contrast, $\frac{\partial_{x}\rho}{\rho}$, at various times. K=0.6, N = 1000, $(\ztt_{0},\wtt_{0}) = (0,0.1\sin(x)+0.15)$}
\label{fig:Rho_x_pos}
\end{figure}

\begin{figure}[h]
\centering
\subfigure[Subfigure 1 list of figures text][$t=1.0$]{
\includegraphics[width=0.2\textwidth]{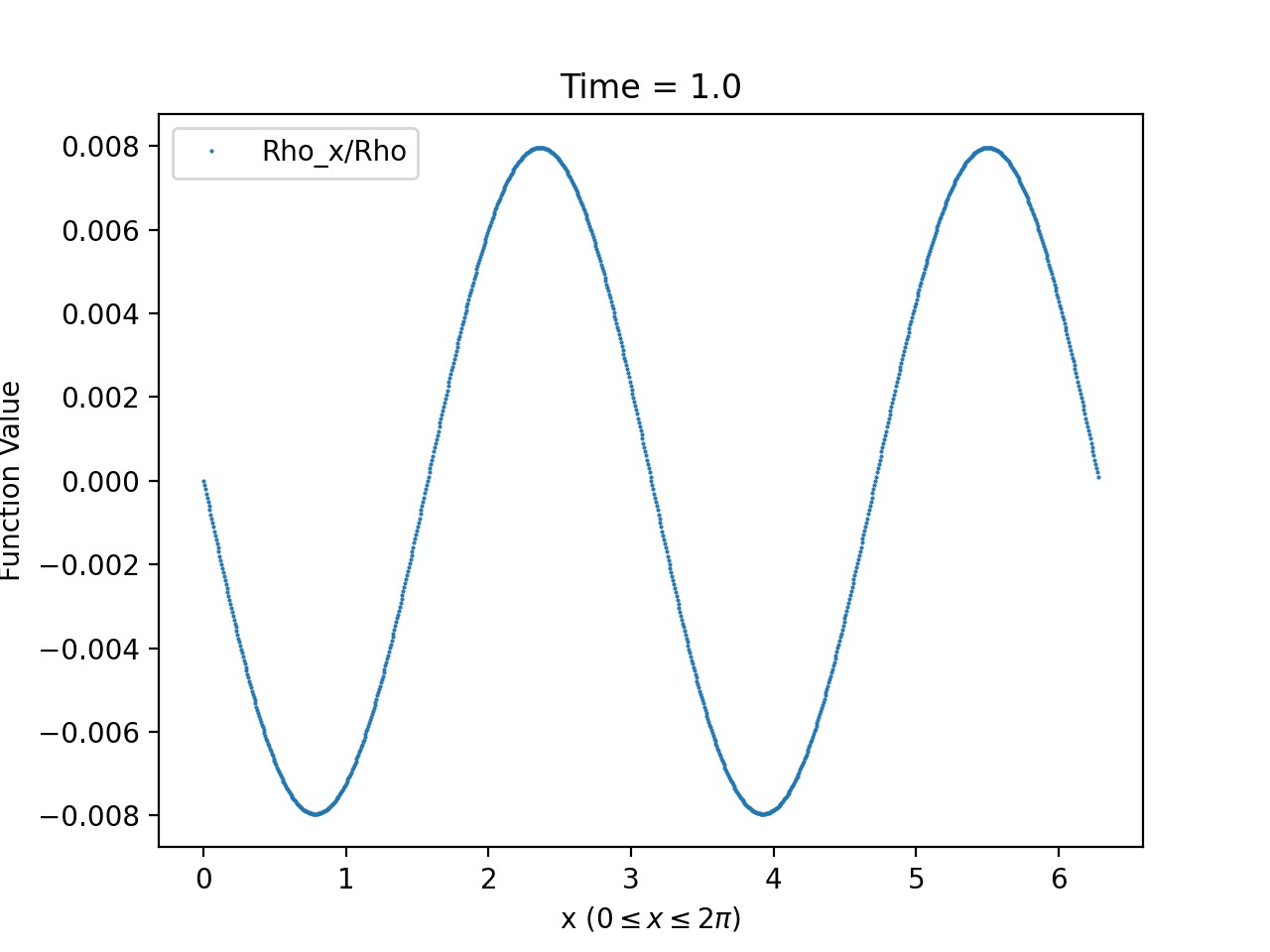}
\label{fig:subfig1}}
\subfigure[Subfigure 4 list of figures text][$t=0.012$]{
\includegraphics[width=0.2\textwidth]{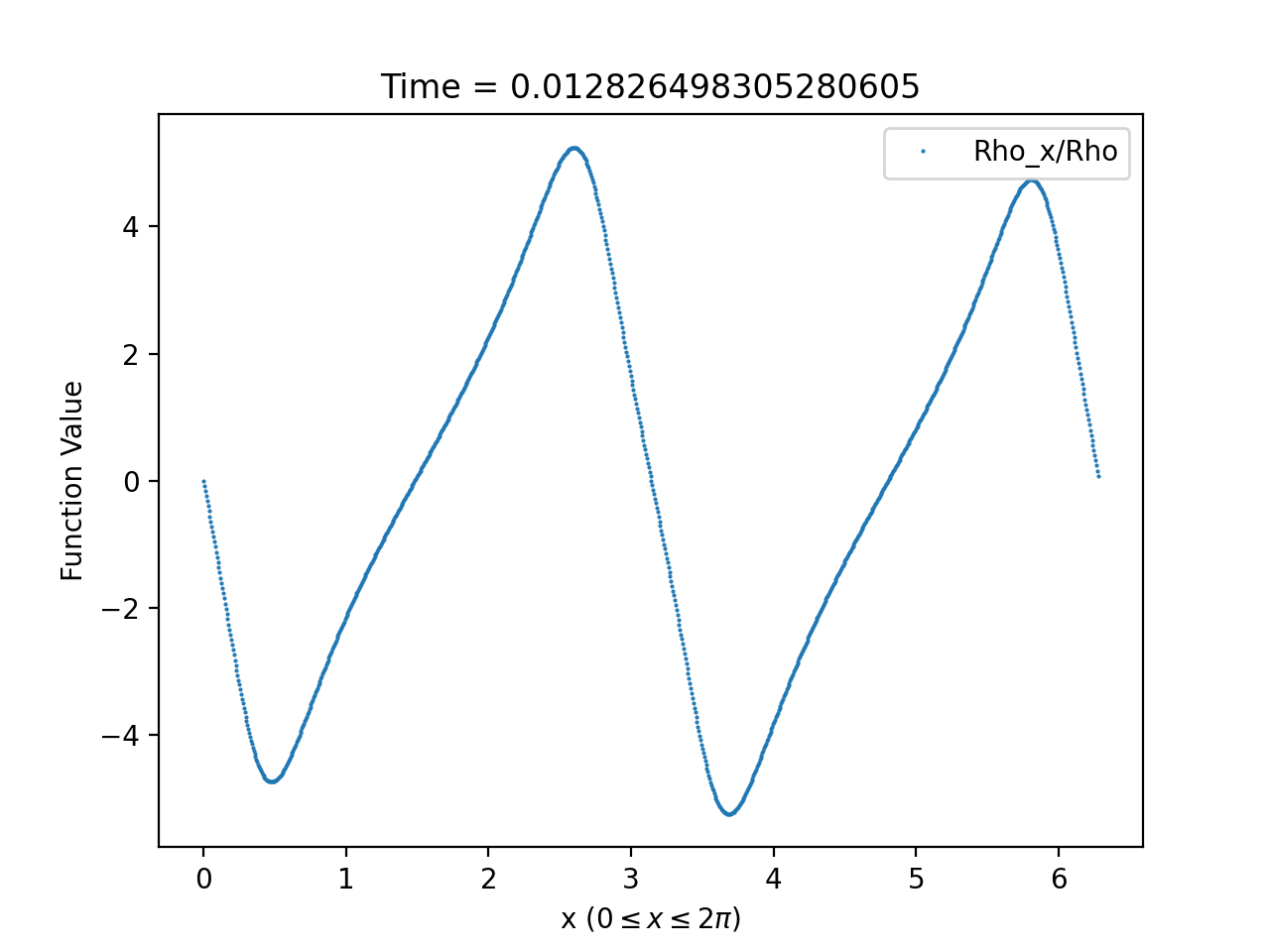}
\label{fig:subfig4}}
\subfigure[Subfigure 5 list of figures text][$t=0.0015$]{
\includegraphics[width=0.2\textwidth]{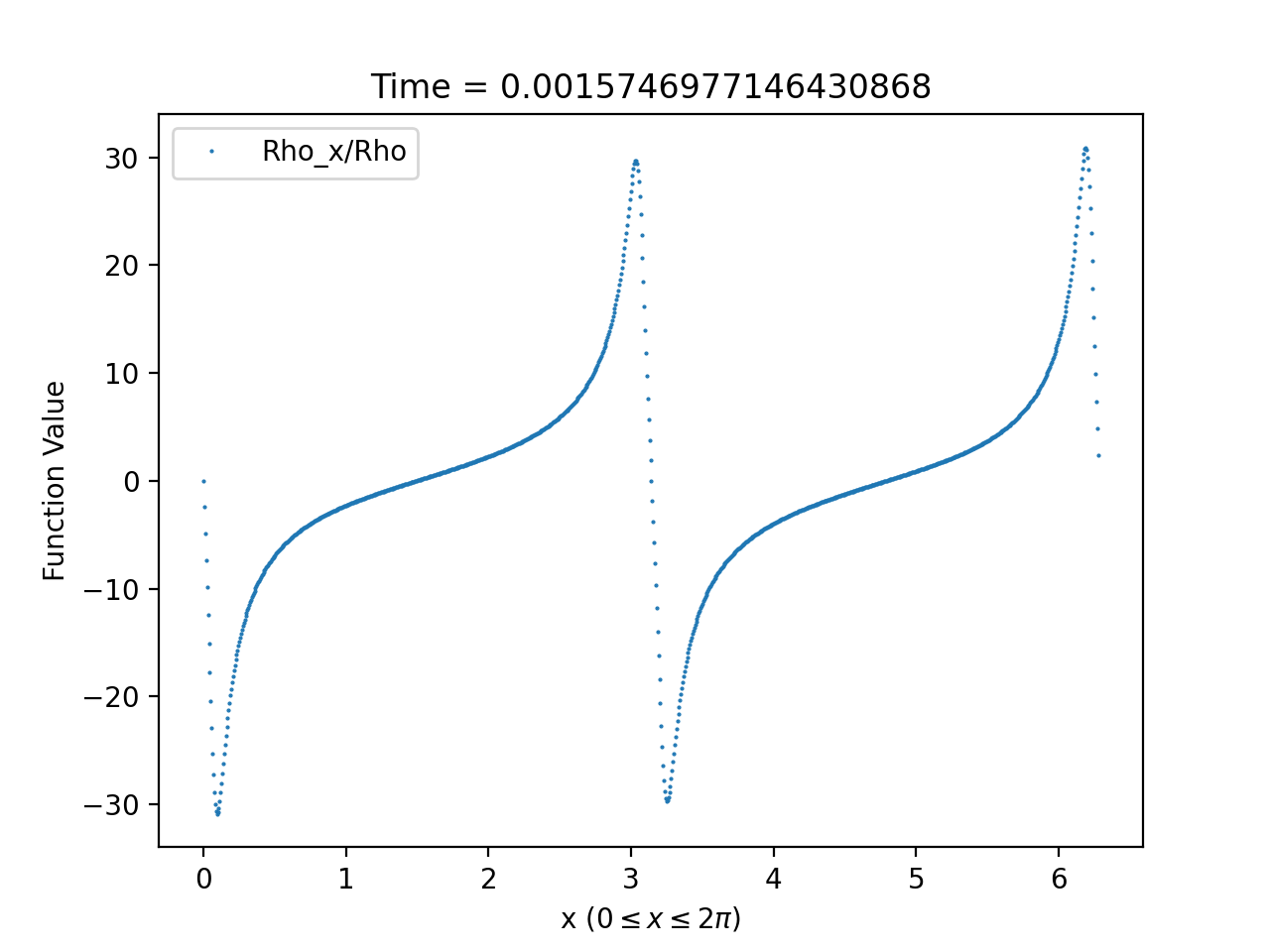}
\label{fig:subfig5}}
\subfigure[Subfigure 5 list of figures text][$t=3.13\times 10^{-4}$]{
\includegraphics[width=0.2\textwidth]{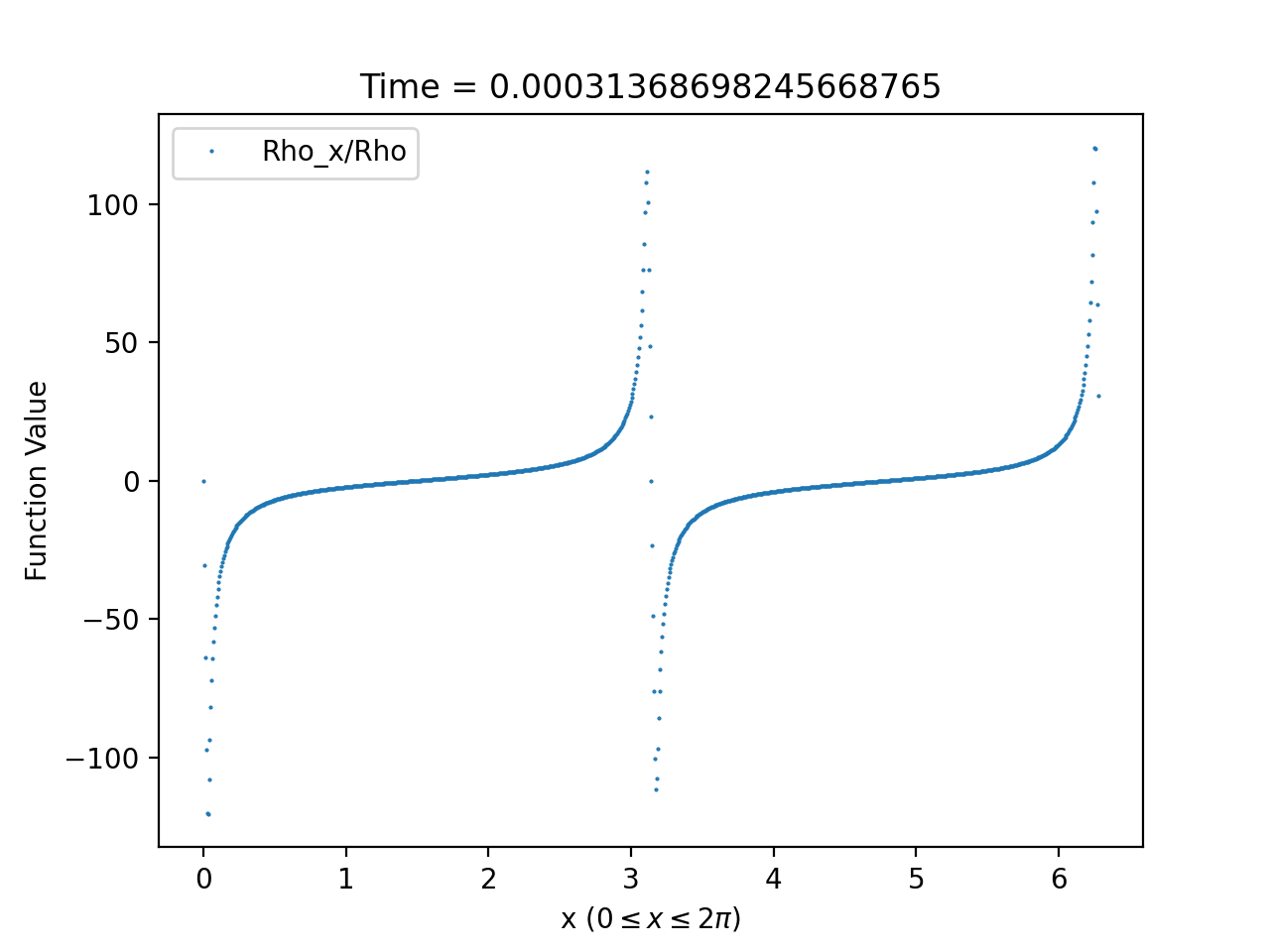}
\label{fig:subfig5}}
\caption{Plots of density contrast, $\frac{\partial_{x}\rho}{\rho}$, at various times. K=0.6, N = 1000, $(\ztt_{0},\wtt_{0}) = (0,0.1\sin(x))$.}
\label{fig:Rho_x_cross}
\end{figure}

By \eqref{mod-den}, \eqref{v0def}, \eqref{zetatdef}
and \eqref{zttt-def}-\eqref{wttt-def}, the density can be written in terms of $\texttt{z}$ and $\texttt{w}$ as $\rho = (\texttt{w}^{2}+t^{2\mu})^{-\frac{K+1}{2}}\rho_{c}t^\frac{2(K+1)}{1-K}e^{(1+K)\texttt{z}}$
where $\rho_c \in (0,\infty)$.
Differentiating this expression, we find after a short calculation that the density contrast is given by
\begin{equation} \label{density-contrast-sym}
\frac{\del{x}\rho}{\rho}= (1+K)\biggl(\del{x}\ztt - \frac{\wtt}{(t^{2\mu}+\wtt^2)}\del{x}\wtt\biggr).
\end{equation}
Using this formula to compute the density contrast for numerical solutions of \eqref{eqn:dotzeta}-\eqref{eqn:dotw},
we observe from our numerical solutions that density contrast displays markedly different behaviour depending on whether or not it is generated from initial data satisfying  \eqref{eqn:numericalID_B}.
For solutions generated from initial data satisfying \eqref{eqn:numericalID_B}, we find that the density contrast remains bounded and converges as $t\searrow 0$ to a fixed function, which is expected by  Theorem \ref{mainthm}. An example of this behaviour is provided in Figure \ref{fig:Rho_x_pos}. On the other hand, the density contrast of solutions generated from initial data violating \eqref{eqn:numericalID_B} develop steep gradients and blows-up at $t=0$ at isolated spatial points; see Figure \ref{fig:Rho_x_cross} for an example of this behaviour.

\begin{figure}[h]
\centering
\subfigure[Subfigure 1 list of figures text][$t=0.001$]{
\includegraphics[width=0.2\textwidth]{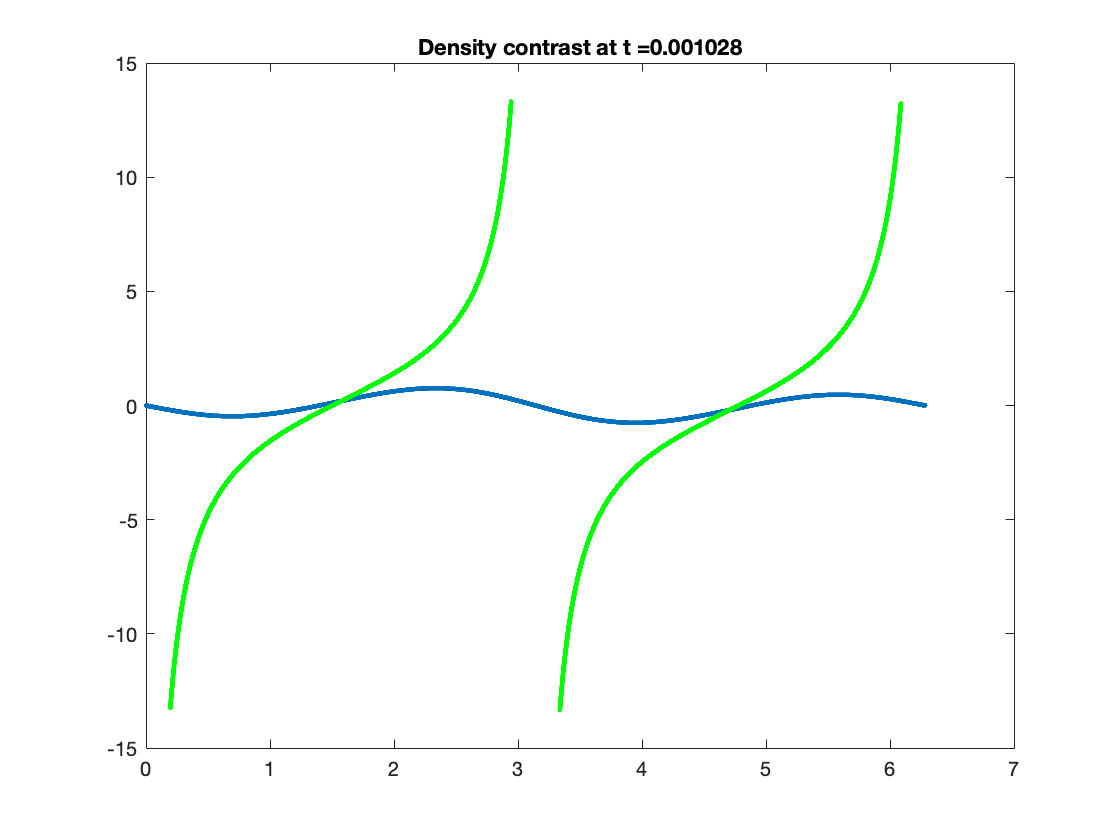}
\label{fig:subfig1}}
\subfigure[Subfigure 2 list of figures text][$t=3.23\times 10^{-5}$]{
\includegraphics[width=0.2\textwidth]{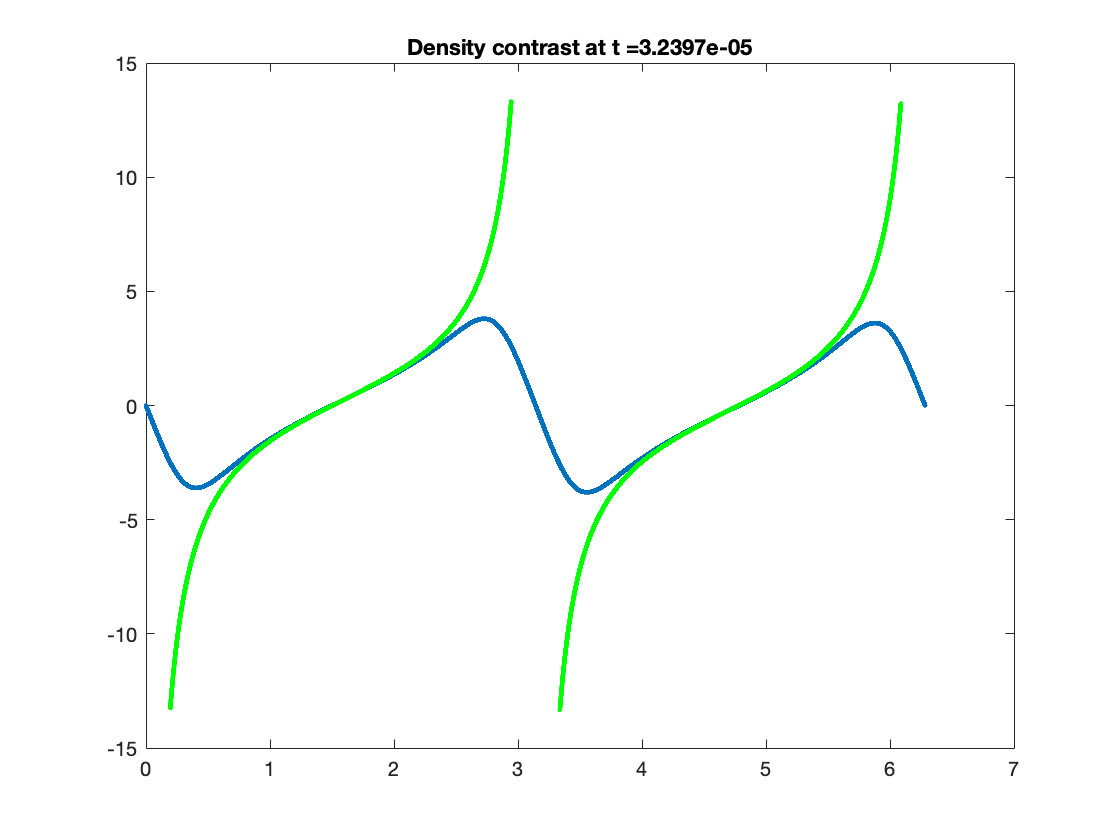}
\label{fig:subfig2}}
\subfigure[Subfigure 3 list of figures text][$t=1.02\times 10^{-6}$]{
\includegraphics[width=0.2\textwidth]{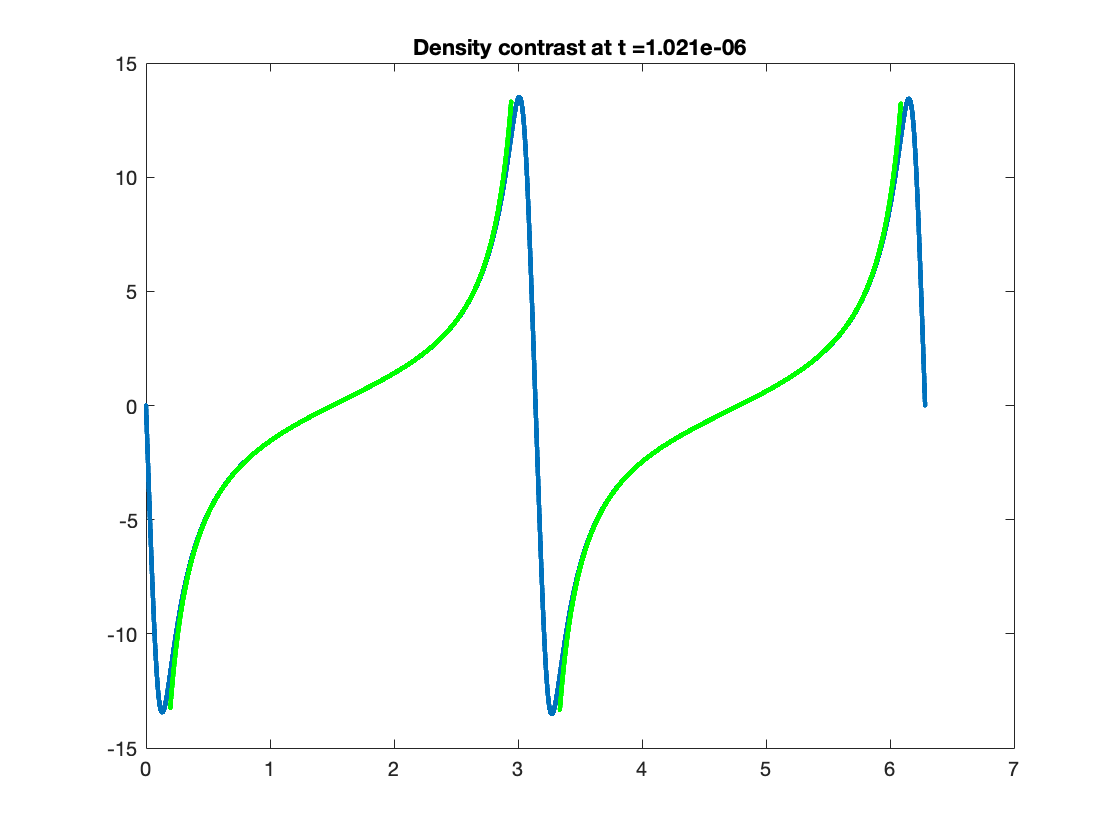}
\label{fig:subfig3}}
\subfigure[Subfigure 4 list of figures text][$t=3.21\times 10^{-8}$]{
\includegraphics[width=0.2\textwidth]{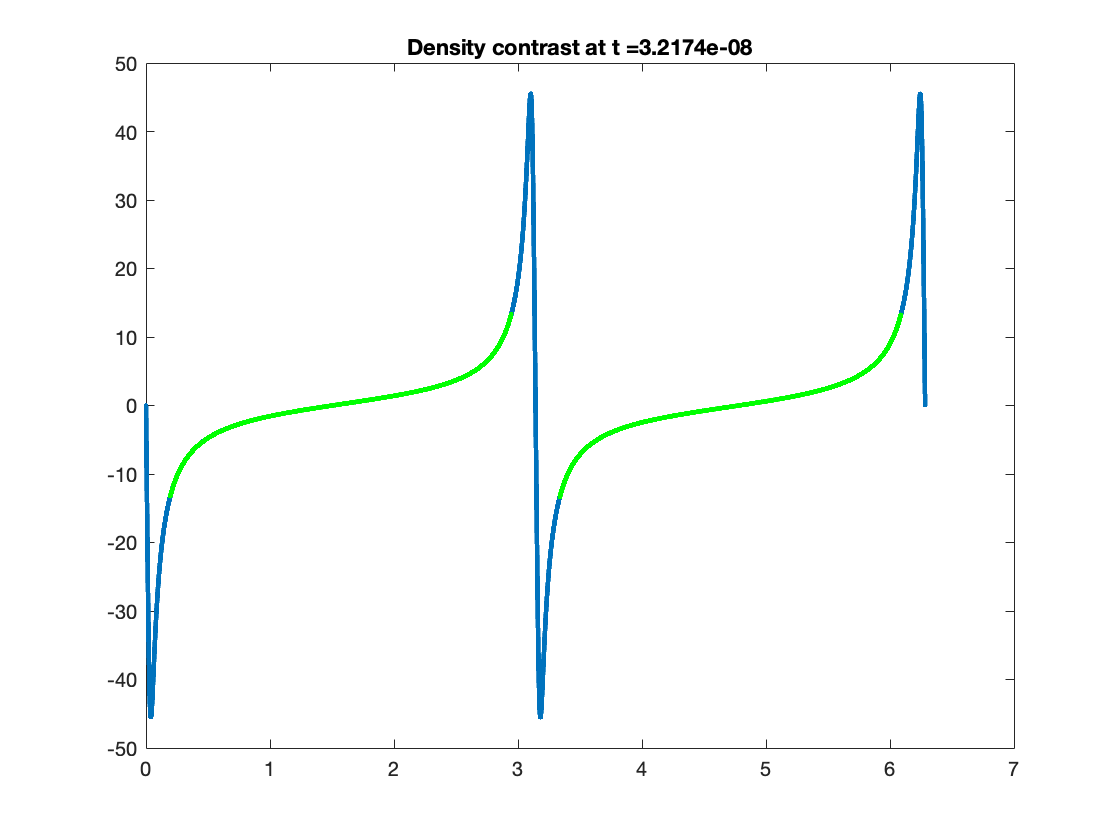}
\label{fig:subfig4}}
\caption{Plots of density contrast, $\frac{\partial_{x}\rho}{\rho}$, calculated from numerical results (Blue) and the asymptotic map (Green). K=0.45, N = 160000, $(\ztt_{0},\wtt_{0}) = (0,0.1\sin(x))$. Points near $\wtt_{0} = 0$ in the asymptotic map have been removed to emphasise agreement of the plots away from the singularities.}
\label{fig:Rho_x_asymptotic}
\end{figure}

As in Section \ref{sec:asymp}, we can compare  the density contrast of the full numerical solutions with the density contrast computed from a solutions of the asymptotic equation.  We do this by evaluating \eqref{density-contrast-sym} at $t=0$ and using \eqref{asymp-sol-t=0}
to approximate the density contrast at $t=0$ by
\begin{equation*}
\frac{\del{x}\rho}{\rho}\biggl|_{t=0}\approx  (1+K)\biggl(\del{x}\zttt_0 - \frac{(\ttld_0^{2\mu}+(1-K)\wttt_0^2)}{(1-K)(\ttld_0^{2\mu}+\wttt_0^2)\wttt_0}\del{x}\wttt_0\biggr).
\end{equation*}
This formula identifies, at least heuristically, that the blow-up at $t=0$ in the density density contrast is due the vanishing of $\wtt$.
Once again the agreement between the numerical and asymptotic plots is close enough that the two are practically indistinguishable as can be seen from  Figure \ref{fig:Rho_x_asymptotic}.

\FloatBarrier
\subsection*{Acknowledgements}
We thank Florian Beyer for helpful discussions and suggestions regarding the numerical aspects of this article. 

\bibliographystyle{amsplain}
\bibliography{K-gth-one-third_v2}

\end{document}